\documentclass[preprint2]{aastex}
\usepackage{natbib}
\slugcomment{Submitted to ApJ}

\shorttitle{X-ray nuclear activity in nearby galaxies} 
\shortauthors{Zhang et al.}

\begin{document}


\title{A census of X-ray nuclear activity in nearby galaxies}


\author{Wei Ming Zhang\altaffilmark{1}, Roberto Soria\altaffilmark{2}, Shuang Nan Zhang\altaffilmark{1}, 
Douglas A. Swartz\altaffilmark{3} and JiFeng Liu\altaffilmark{4}}


\altaffiltext{1}{Physics Department and Center for Astrophysics,
Tsinghua University, Beijing 100084, China; email: 
zhangweiming@mails.thu.edu.cn.} \altaffiltext{2}{MSSL, University
College London, Holmbury House, Dorking RH5 6NT, UK.}
\altaffiltext{3}{Universities
Space Research Association, NASA Marshall Space Flight Center,
VP62, Huntsville, AL, USA.}
\altaffiltext{4}{Harvard-Smithsonian Center for Astrophysics,
60 Garden st., Cambridge, MA 02138, USA.}


\begin{abstract}
We have studied the X-ray nuclear activity of 187 nearby
(distance $< 15$ Mpc) galaxies observed with {\it{Chandra}}/ACIS.
We found that 86 of them have a point-like X-ray core, consistent
with an accreting BH. We argue that the majority of them are
nuclear BHs, rather than X-ray binaries.
The fraction of galaxies with an X-ray detected nuclear BH
is higher ($\approx 60$ per cent) for ellipticals and
early-type spirals (E to Sb), and lower ($\approx 30$ per cent)
for late-type spirals (Sc to Sm). There is no preferential association
of X-ray cores with the presence of a large-scale bar; in fact,
strongly barred galaxies appear to have slighly lower detection
fraction and luminosity for their nuclear X-ray sources, compared 
with non-barred or weakly barred galaxies of similar Hubble types.
The cumulative luminosiy distribution of the nuclear sources 
in the $0.3$--$8$ keV band is a power-law 
with slope $\approx -0.5$, from $\approx 2 \times 10^{38}$
erg s$^{-1}$ to $\approx 10^{42}$ erg s$^{-1}$. The Eddington ratio
is lower for ellipticals ($L_{\rm X}/L_{\rm Edd} \sim 10^{-8}$) and higher for
late-type spirals (up to $L_{\rm X}/L_{\rm Edd} \sim 10^{-4}$), but in all cases,
the accretion rate is low enough to be in the radiatively-inefficient regime.
The intrinsic absorbing column density is generally low, especially 
for the less luminous sources: there appear to be
no Type-2 nuclear BHs at luminosities $\la 10^{39}$ erg s$^{-1}$.
The lack of a dusty torus or of other sources of intrinsic absorption
(e.g., an optically-thick disk wind) may be directly related to the lack of
a standard accretion disk around those faint nuclear BHs.
The fraction of obscured sources increases with the nuclear BH luminosity: 
two thirds of the sources with $L_{\rm X} > 10^{40}$ erg s$^{-1}$ 
have a fitted column density $> 10^{22}$ cm$^{-2}$. 
This is contrary to the declining trend of the obscured fraction 
with increasing luminosities, observed in more luminous AGN and quasars.

\end{abstract}




\section{INTRODUCTION}
%An AGN is a compact region that at the centre of a galaxy which
%has a much higher than normal luminosity over some or all of the
%spectrum. Now it's believed that the radiation form AGN is a
%result of accretion of mass by the supermassive black holes (BHs)
%at the center of the host galaxy.

Supermassive black holes (BHs) are now believed to exist in all massive
galaxies with a spheroidal component \citep{mago98,mf01b,kor04}.
Low-mass galaxies tend to harbour a nuclear star cluster,
whose mass is also related to the mass of the spheroidal
component \citep{feet06}. But in some low-mass galaxies,
nuclear BHs have also been identified
\citep{set08,fh03,baet05,gh04,gh07,doet07,ghb08}.
Thus, it is still a topic of active investigation whether
there is a mass threshold between spiral galaxies containing
a nuclear BH or a nuclear star cluster, to what extent
they coexist, and what the mass relation is between
the two nuclear systems when they are both present.
It is also still debated whether there is a Hubble-type
threshold for the presence of a nuclear BH,
that is whether late-type spiral galaxies without
a spheroidal component can have nuclear BHs, and if so, whether
the BH formation mechanisms and growth processes are different
from those of nuclear BHs in massive spheroidals \citep{set08,wan08}.

The main difficulty in resolving these questions is
determining which galaxies host nuclear BHs. In particular,
kinematic mass determinations require prohibitively high
spatial resolution at the low-mass end. 
When a kinematic mass determination of the central dark mass
is not available (i.e., in all but a few cases), the strongest
evidence of the presence of a nuclear BH comes from
its accretion-powered activity \citep{sal64,lyn69}. AGN activity
can be identified either from an emission-line optical/UV spectrum
from the nuclear region, or from a point-like X-ray nuclear source,
or from a flat-spectrum radio core; surveys in different bands 
lead to different selection biases.
Statistical studies of the AGN population provide direct information
on what fraction and what types of galaxies contain a nuclear BH,
and indirect constraints on their masses and accretion rates.

Optical spectroscopic studies such as the Palomar Survey
\citep{hfs97a,hfs97b} and the Sloan Digital Sky Survey \citep{hao05a}
suggest \citep{ho08} that $\approx 10$\% of nearby galaxies are Seyferts,
$\approx 20$\% are ``pure'' Low-Ionization Nuclear Emission-line Regions
(LINERs), and another $\approx 10$\% are ``transition objects'' whose spectra
are intermediate between those of pure LINERs and H{\footnotesize{II}} regions
\citep{hec80}. 
%This means that overall, 1/10 of nearby galaxies are fully-fledged AGN,
%and at least another 3/10 are low-luminosity AGN. 
This means that overall, at least 4/10 of nearby galaxies
contain a currently active nuclear BH. This is only a
lower limit to the population of currently active nuclear BHs:
another $40$\% of galaxies have an emission-line nucleus
(H{\footnotesize{II}} nucleus). H{\footnotesize{II}} nuclei are powered
by a compact star-forming region, but in some cases they may also
harbour a weakly accreting BH. Optical surveys also show that
Seyferts, LINERs and transition objects are mostly found in
early-type galaxies: $\approx 60$\% of E to Sb galaxies have such
nuclear signatures; conversely, almost all late-type disk galaxies
are dominated by H{\footnotesize{II}} nuclei, and $\la 20$\% of them
have evidence of an accreting nuclear BH \citep{ho08}.
Thus, it is more difficult to obtain reliable BH demographics
in late-type galaxies (containing smaller BHs) using optical
spectroscopy, due to confusion from star formation.
{\it Spitzer}'s mid-infrared spectroscopic studies \citep{saty08}
have recently proved more effective at finding nuclear BHs
in late-type galaxies, with an AGN detection rate possibly
4 times larger than suggested by optical spectroscopic
observations.

Radio-band surveys are not yet as complete or as sensitive
as the Palomar or Sloan optical surveys. Nonetheless, arcsec-resolution
VLA surveys at 5 GHz \citep{wh91}, 8.4 GHz \citep*{nfw05}
and 15 GHz \citep*{fbh06} have confirmed the presence of
AGN signatures (non-thermal radio cores) in $\approx 30$--$40$\%
of early-type galaxies, mostly Seyferts and LINERs.
The radio emission in LINERs is mainly confined to a compact
core or the base of a jet (sub-arcsec size); Seyferts are more
often accompanied by extended (arcsec-size) jet-like features
\citep{nfw05}. Somewhat in contrast with the optical classification,
radio detections of compact cores in transition objects
are much rarer; this suggests that perhaps $\approx 50$\% of transition
objects are not AGN.

X-ray surveys are an under-utilized and very promising tool to further
our understanding of low-level nuclear activity in the
local Universe. X-ray emission probes regions much closer
to the accreting BH than, for example, optical emission lines.
And a direct study of the nuclear X-ray source allows
better constraints on its mass accretion rate and output power,
stripped of the often messy or ambiguous optical-line phenomenology
of the host galaxy.
Sub-arcsec spatial resolution and astrometric accuracy
are required for the study of low-luminosity AGN,
for at least two reasons: to confirm
that an X-ray source is point-like and coincident with
the optical/radio nucleus; and to separate
the point-like, nuclear X-ray source from the surrounding
unresolved emission (mostly from diffuse hot gas
and faint X-ray binaries), which is often present
in galactic cores. Only {\it Chandra} can provide
such resolution and astrometric accuracy.
There are still no complete, unbiased {\it Chandra}
X-ray surveys of galaxies in the local universe, mostly because
time-allocation constraints lead to an implicit
bias towards X-ray luminous targets.
However, a number of {\it Chandra} studies
have targeted a sizeable fraction of nearby LINERs
(about half of the LINERs in the Palomar survey; \citet{ho08}).
They have confirmed that most LINERs and transition objects
contain an X-ray luminous,
accreting BH \citep{ssd04,duet05,pel05,saet05,flet06,goet06}.
More specifically, $\approx 75$\% of LINERs have an X-ray nucleus,
and this fraction goes to $100$\% for the subsample
of LINERs that also have a detected radio core (\citet{ho08}
and references therein).
In such {\it Chandra} studies, the typical detection limit
for point-like nuclear X-ray sources is $\sim 10^{38}$ erg s$^{-1}$.
This is directly comparable to the detection limit of nuclear H$\alpha$
emission in the Palomar survey, $\sim 10^{37}$ erg s$^{-1}$,
because, in typical Seyferts, $L_{\rm X} \sim 10 L_{\rm{H}\alpha}$
\citep{ho08}. For such low-luminosity AGN, radio and X-ray studies
are in principle less affected by contamination from possible
surrounding star-forming regions than studies based on photoionized
H$\alpha$ emission. It is still not known whether or what fraction of
H{\footnotesize{II}} nuclei in late-type spirals
also have an X-ray active nuclear BH, and whether there is any
correlation with the presence and strength of a large-scale bar.
Pioneering studies \citep{gho08,des09} suggest that a few nearby
late-type spirals with an optically-classified
H{\footnotesize{II}} nucleus may harbour a low-luminosity AGN
with $L_{\rm X} \approx$ a few $10^{37}$ erg s$^{-1}$.
The drawback of X-ray studies is that at such low luminosities,
it becomes extremely difficult
to determine for any individual galaxy whether the nuclear
source is the accreting supermassive BH or an unrelated
X-ray binary in the dense nuclear region or in a nuclear
star cluster.

X-ray surveys, in combination with other bands,
can do more than just refine our statistical census
of active nuclei across the Hubble sequence: they can be used
to obtain a physical understanding of the process of accretion
across the whole range of mass accretion rates.
A fundamental issue that deserves investigation is whether
there is a continuous distribution of nuclear X-ray activity
from the most luminous quasars ($L_{\rm bol} \sim L_{\rm Edd}$)
to run-of-the-mill Seyferts
($L_{\rm bol} \sim (10^{-4}$--$10^{-2}) L_{\rm Edd}$),
LINERs ($L_{\rm bol} \sim (10^{-6}$--$10^{-4}) L_{\rm Edd}$)
transition objects ($L_{\rm bol} \sim (10^{-7}$--$10^{-6}) L_{\rm Edd}$)
all the way down to ``quiescent'' galactic nuclei such as
the BH in the Milky Way or in nearby elliptical galaxies
($L_{\rm bol} \la 10^{-9} L_{\rm Edd}$);
or, instead, whether there is a series of clear thresholds along this
sequence, corresponding to fundamental changes in the accretion mode
and inflow structure. Such transitions may or may not
coincide with the ``canonical'' accretion states found
in stellar-mass BHs. Moreover, thresholds in the X-ray
properties may or may not coincide with optically-classified
classes of AGN.

One such physical transition may be between accreting BHs
with and without an optically-thick accretion disk.
The presence or absence of disk signatures such as Compton
reflection and a broad Fe line
in the X-ray spectra can provide strong constraints,
in parallel with the presence or absence of a UV bump
in the spectral energy distribution.
It was suggested \citep{ho08} that the disappearance of
the inner disk marks the transition between Seyferts
and LINERs. A related issue yet to be properly understood
in low-luminosity AGN is the redistribution of the accretion
power between a radiative component (UV/X-rays),
a mechanical component (radio jets) and an advected component;
AGN become radio louder at lower accretion rates, in agreement
with the trend seen in Galactic BHs \citep*{fen04,mel03}.
Order-of-magnitude estimates of the accretion rate
based on the mass-loss rate from evolved stars
and the gravitational capture rate of hot gas
from the interstellar medium typically lead to large
overestimates of the nuclear BH luminosity, especially
in LINERs and transition objects \citep{ho08}. Therefore, such objects
are an ideal test for radiatively inefficient accretion models 
such as Advection-Dominated Accretion Flows (ADAF)
or Radiatively Inefficient Accretion Flows (RIAF) 
\citep{ny95a,qua99,nar02,yua04}.
X-ray studies also allow quantitative comparisons between
the amount of gas that is used for star formation
in the nuclear region of late-type spirals, and that
used for fuelling a possible accreting BH (or an upper limit
to that accretion rate), and how this ratio depends
on Hubble type and bar structure.

Another physical problem that can be addressed with
X-ray surveys is the validity of the unified model
\citep{ant93} at low accretion rates and low luminosities
($L_{\rm X} \sim 10^{38}$--$10^{41}$ erg s$^{-1}$).
The standard unified model is based on the presence
of a geometrically and optically thick parsec-scale
structure (``dusty torus'')
that blocks our direct view of the innermost disk
in high-inclination sources (Type-2 AGN).
The nature and physical structure of the obscuring material
is still controversial. In alternative to the parsec-scale torus
scenario, it was suggested that the absorption could instead be due
to an optically-thick disk wind, launched from
a few 100 Schwarzschild radii \citep{elv00,elv04,mur95,mur98}.
X-ray spectral surveys of low-luminosity AGN are crucial
to determine whether the thick absorber disappears along with
the disk signatures (which would point to an intimate connection
between the two components), and at what luminosities.
We already know that there are no dusty tori around
extremely faint BHs such as the one in our Galaxy.
{\it Chandra} and {\it XMM-Newton} studies show
low absorbing column densities and weak or undetected
narrow Fe K$\alpha$ emission in a large fraction of nearby
LINERs, which suggests a direct, unobstructed view of
the nucleus \citep{ho08}. In contrast, strong Fe K$\alpha$
emission \citep{caet06} and a nearly continuous
distribution of absorbing columns \citep{paet06}
suggest that Seyferts are more gas-rich than LINERs.

To address those issues, we have collected and analyzed
the archival {\it Chandra} data available for nearby galaxies
(see Section 2 for the selection criteria of our sample).
In this paper, we present preliminary results of our X-ray
population study, focussing in particular on the luminosity
distribution and on the dependence of the absorbing column
density on the nuclear luminosity.

\begin{figure}[h]
%\plotone{XClassNew.eps}
\hspace{-0.9cm}\includegraphics[width=9.35cm]{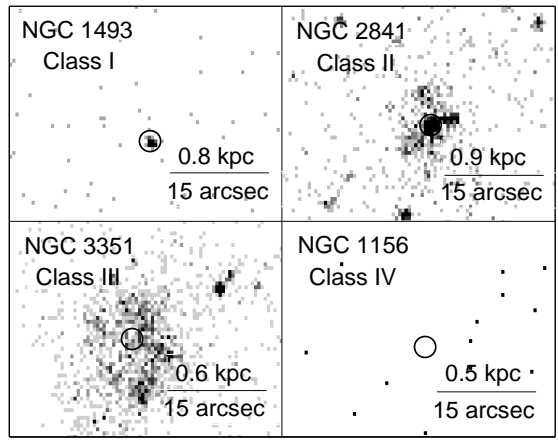}
\caption{Our X-ray classification
of the nuclear regions (see also Table 1),
based on the {\it Chandra}/ACIS images:
I = dominant point-like X-ray nucleus;
II = point-like nuclear source embedded in diffuse or unresolved
emission (from hot gas and fainter X-ray binaries);
III = diffuse X-ray emission without a clearly-idenfied point-like core;
IV = no detectable X-ray emission at the nuclear position.
\label{fig1}}
\end{figure}

\section{SAMPLE SELECTION}

In general, we have two choices when selecting a sample of nearby galaxies
for X-ray population studies of their nuclei. We can select
all targets observed by {\it Chandra} within a certain distance
or flux detection limit; this sample will probably be biased in favour
of X-ray luminous galaxies (or galaxies with starbursts, or with
some other X-ray peculiarities), which are more likely to be selected
as targets. Alternatively, we may select a limited sub-sample
of such targets, which may be regarded as complete (or at least
unbiased) according to some optical criteria.

For our work, we have started from an optically/IR selected sample
previously used by \citet*{swa08} and Swartz et al.~(2009, in prep.)
for a statistical X-ray study of ultraluminous X-ray sources (ULXs)
population. This sample is defined as a volume-limited set
of galaxies within $14.5$ Mpc that are both contained in the Uppsala
Galaxy Catalog (UGC; \citet{nil73}) with photographic magnitude
$m_p<14.5$~mag, and in the Infrared Astronomical Satellite ({\sl IRAS})
catalogs \citep{ful89,mos93}
with a flux $f_{\rm FIR}\ge10^{-10.3}$ erg cm$^{-2}$ s$^{-1}$,
where $f_{\rm FIR}/10^{-11}=3.25S_{60}+1.26S_{100}$ erg cm$^{-2}$ s$^{-1}$ \citep{ric88}.
Here, $S_{60}$ and $S_{100}$ are the total flux densities, expressed in Jy,
at 60 and 100~$\mu$m respectively.
The {\sl IRAS} catalogs are complete to approximately 1.5~Jy for point-like sources.
The UGC contains all galaxies north of B1950 $\delta=-2^{\circ}30^{\prime}$.
This combined selection criteria favor nearby, predominantly
optically-bright galaxies with at least a modest amount
of recent star formation. The sample is known to exclude
smaller dwarf galaxies in the neighborhood \citep{swa08}. 
Most ellipticals would be included by the selection criteria; 
however, there are not many Northern-hemisphere nearby ellipticals 
(this bias is further discussed in Swartz et al.~2009, in prep.).
%and most ellipticals; it is more suitable for a study of nuclear
%BH activity in spiral galaxies. 
There are 140 galaxies in this complete sample. However, only 116 have been observed
with {\it Chandra}; the rest only have {\it XMM-Newton} and {\it ROSAT} data,
which are less suitable for our study due to their larger point-spread-functions.
Nonetheless, we estimate that the bias introduced by the lack
of {\it Chandra} coverage for those 24 galaxies out of 140 is minimal.
Henceforth, we will refer to this subsample of 116 galaxies
as the optical/IR sample, for simplicity.
The morphological classification of those 116 galaxies is:
3\% ellipticals; 44\% early-type spirals (S0 through Sb);
42\% late-type spirals (Sc through Sm);
11\% dwarfs/irregulars. The optical line classification
is: 9\% Seyfert;  11\% LINERs; 11\% transition objects;
29\% {H\footnotesize{II}} nuclei; 40\% unknown.

We have then used a larger sample, defined by Liu (2009, in prep.): it includes
all {\it Chandra}/ACIS targets within $15$ Mpc, contained in the Third Reference 
Catalog of Galaxies \citep[RC3;][]{dev91}, which is complete 
for nearby galaxies having apparent diameters
$\ge 1\arcmin$~at the D25 isophotal level and total B-band magnitudes $m_B <15.5$ mag,
in both the Northern and Southern sky.
%, and redshifts $z\le 15,000$ km s$^{-1}$.
We have taken all the {\it Chandra}/ACIS targets within this sample
(as of the end of 2007); the exposure times range from 500 s to 120 ks.
This selection adds another 71 galaxies to those already contained
in the optical/IR sample. The Liu sample is slightly biased in favour
of the brighter and larger galaxies, as expected; it is essentially
an X-ray selected sample, because only $\sim 10$ per cent of the RC3 galaxies
have been {\it Chandra} targets. From the morphological
point of view, the Liu sample has an overabundance of early-type spirals,
which somewhat compensates the bias of the optical/IR sample
in favour of late-type galaxies; moreover, it includes 
a few Southern-hemisphere ellipticals, lacking from the optical/IR sample.

In summary, there are altogether 187 nearby galaxies with publicly-available
{\it Chandra} observations (as of December 2007) included either in the
optically/IR-selected sample or in the X-ray-selected sample,
within 15 Mpc (full list in Table 1)\footnote{We excluded M\,31-Andromeda from the sample. 
Its nuclear BH has a $0.3$--$8$ keV luminosity $\approx 10^{36}$ erg s$^{-1}$ 
\citep{gar05}; this would be below the detection limit for all other 
galaxies in our sample.}. More than half (104/187) of those galaxies 
are also included in the optical Palomar sample.
The morphological classification of those 187 galaxies is:
8\% ellipticals; 43\% early-type spirals (S0 through Sb);
36\% late-type spirals (Sc through Sm); 13\% dwarfs/irregulars. 
The optical line classification is: 8\% Seyfert; 11\% LINERs;
10\% transition objects; 40\% {H\footnotesize{II}} nuclei; 31\% unknown.
For simplicity, we will call this sample of 187 galaxies 
the ``extended sample''.

\section{DATA ANALYSIS AND RESULTS}

For each of the 187 {\it Chandra} target galaxies in our extended sample,
we searched for a point-like X-ray source at or close to the nuclear position.
For the nuclear coordinates, we generally referred
to the positions listed in the NASA/IPAC Extragalactic Database 
(NED\footnote{http://nedwww.ipac.caltech.edu/}).
In most cases, they come from the Two Micron All Sky Survey (2MASS) 
catalog \citep{skr06}; the semi-major axes of the 95\% confidence 
ellipse vary between $\approx 1\farcs0$ and $1\farcs5$. 
In a few cases, the NED positions come from the Sloan Digital Sky Survey 
Data Release 6\footnote{http://www.sdss.org/dr6} 
(semi-major axes $\approx 0\farcs5$), or from 
VLA radio observations (for example in the case of NGC\,4203, 
with semi-major axes $\approx 0\farcs1$).
%defined by VLA radio observations
%when available, or by the Two Micron All Sky Survey (2MASS) catalog \citep{skr06}; 
%we generally referred
%to the positions listed in the NASA/IPAC Extragalactic Database
%(NED\footnote{http://nedwww.ipac.caltech.edu/}).
%Sloan Digital Sky Survey Data Release 6
%the semi-major axes of the 95\% confidence ellipse

We used the Chandra Interactive Analysis of Observations software
({\footnotesize{CIAO}}) V4.1 for filtering and analyzing the event files.
For each {\it Chandra} observation, we checked and screened out
exposure intervals corresponding to background flares.
We used the standard task {\it wavdetect} to identify sources
in the nuclear region of each galaxy. The morphologies of the nuclear regions
in the {\it Chandra} images can be loosely grouped into four
classes (Figure 1): (I) a dominant nuclear source; (II) a nuclear
source embedded in extended, unresolved emission; (III) unresolved 
emission in the nuclear region but no point-like source; (IV) no nuclear 
source at all\footnote{This classification is reminiscent of, but 
not identical to the X-ray morphological classification in \citet{hoet01}.}.
Eighty-six of the 187 galaxies in our sample belong to class I and II,
that is have a point-like X-ray source within $1\arcsec$ of the
independently-identified nuclear position (details in Section 3.1).
%HERE YOU SHOULD ALSO GIVE THE FRACTION OF DETECTED SOURCES
%IN THE SWARTZ SAMPLE ALONE.\\
For all the nuclear sources, we extracted spectra (in the $0.3$--$8$ keV
band) from circular regions of radius $1\farcs5$ (which include
$\approx 90$\% of the counts), and corresponding background and 
response files using the {\footnotesize{CIAO}} task {\it psextract}.
We used {\footnotesize{XSPEC}} Version 12.1 \citep{arn96}
for spectral modelling.
We used the Cash statistics \citep{cas79}
for sources with $\la 200$ counts,
and the $\chi^2$ statistics (with suitably binned data)
in the other cases. In addition, some galaxies have been 
the target of detailed {\it Chandra} studies in the literature: 
in those cases, we have used their spectral results to complement 
our analysis.

\subsection{Census of nuclear X-ray sources}

For the optical/IR sample, $\approx 50$ per cent (27 out of 53) of
early-type galaxies (Hubble types E to Sb) have a point-like nuclear
X-ray core (Table 2). By contrast, this proportion drops to $\approx 30$ 
per cent (20 out of 63) for later Hubble types (Sc to Sm and
Irr). Most Seyferts (7 out of 9) and LINERs (12 out of 13) 
have a detected X-ray nucleus; this
proportion drops to less than half for transition objects (5 out of 12)
and H{\footnotesize{II}} nuclei (16 out of 46).

For the extended sample, $\approx 60$ per cent (57 out of 94) of early
type galaxies have an X-ray nucleus, but only $\approx 30$ per cent
(29 out of 93) of later Hubble types. More than $90$ per cent (33 out of 36) 
of Seyferts and LINERs are X-ray detected, while an X-ray core is found 
in $\approx 60$ per cent of transition objects (12 out of 20) 
and $\approx 30$ per cent of H{\footnotesize{II}}
nuclei (18 out of 54). Thus, the optical/IR sample and the extended sample 
are consistent with each other.

The numbers above are roughly consistent with the AGN fraction inferred from optical
spectroscopic studies \citep{ho08} in early-type ($\approx 50$--$70$ per cent)
and late-type ($\approx 15$ per cent) galaxies respectively.
There is a slight overabundance of X-ray core detections in late-type 
spirals compared with the optical AGN fraction in the Palomar sample, 
and with the estimates of \citet{hao05b} from
the Sloan Digital Sky Survey (based on H$\alpha$
and [OIII] emission lines).
This might be due to contamination from high-mass X-ray binaries, but 
we do not think this is a very significant effect (see
Section 3.3). It is more likely due to the negligible
effect, through heating and photo-ionization,
that a faint X-ray core ($L_{\rm X} \sim 10^{38}$ erg s$^{-1}$)
would have on the optical/UV nuclear spectrum; 
as a result, it may not appear as an active nucleus 
in the Palomar sample. See also \citet{hop09} for a discussion 
of how optical surveys may miss low-luminosity AGN, or erroneously 
classify them as optically-obscured, because of dilution effects.
Another reason why nuclear activity is more often
detected in the X-ray band is that optical emission lines
from the nuclear BH may be more severely obscured
by dust in the surrounding star-forming environment.

We then searched for possible correlations between the presence of a
nuclear X-ray source and of a large-scale bar. In the optical/IR sample,
X-ray cores are found in $\approx 60$ per cent of
non-barred (SA) galaxies (21 out of 34) and
weakly barred (SAB) galaxies (16 out of 27), and in $\approx 20$ per cent 
of strongly barred (SB) galaxies (6 out of 26). For the extended sample,
nuclear sources are detected in $\approx 60$ per cent (30 out of 48)
of SA, $\approx 60$ per cent (24 out of 41) of SAB and $\approx 40$
per cent (16 out of 42) of SB galaxies. 
Part of this effect is due to the larger presence of early-type spirals 
among the SA and SAB classes. To remove this bias, we compared 
the effect of a bar separately within the early-type and late-type spiral subsamples 
(Table 3). We still find a slightly higher fraction of active nuclei 
in the SA/SAB classes compared with the SB class. In early-type spirals, 
$\approx 70$ per cent of non-barred/weakly-barred galaxies have an active nucleus, 
compared with about half for strongly-barred galaxies; in late-type spirals, 
about half of non-barred/weakly-barred galaxies have an X-ray core, 
but only about 1/4 of strongly-barred galaxies. This confirms that
bar structures have no positive influence on nuclear BH feeding
\citep{hfs97d,sak99}. If anything, there may be a slight
anticorrelation (see also Section 4), but it may be still be 
attributed to small number statistics.
%Sakamoto, K.; Okumura, S. K.; Ishizuki, S.; Scoville, N. Z.1999ApJ...525..691S

\begin{figure*}
\includegraphics[width=8.6cm]{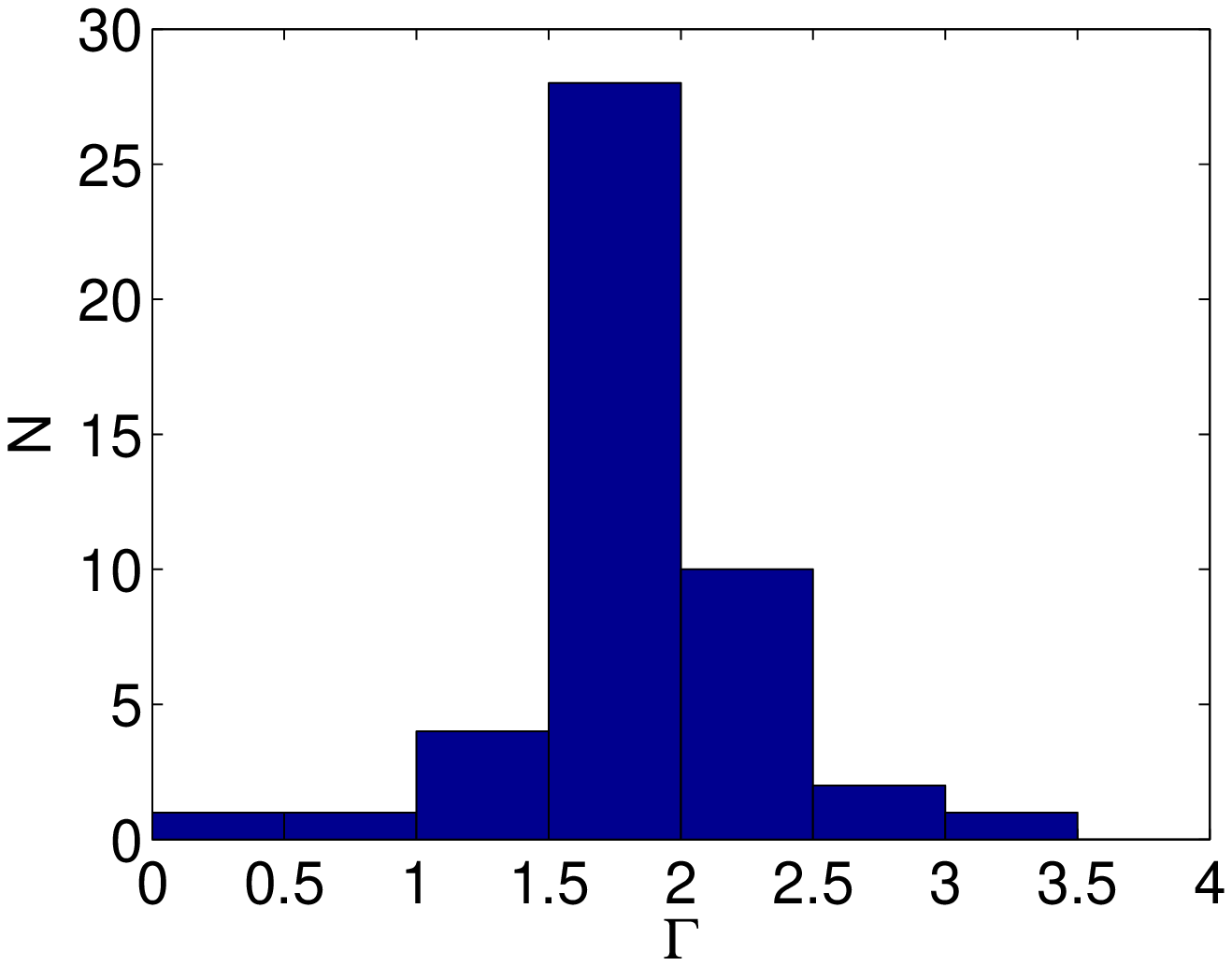}
\includegraphics[width=8.6cm]{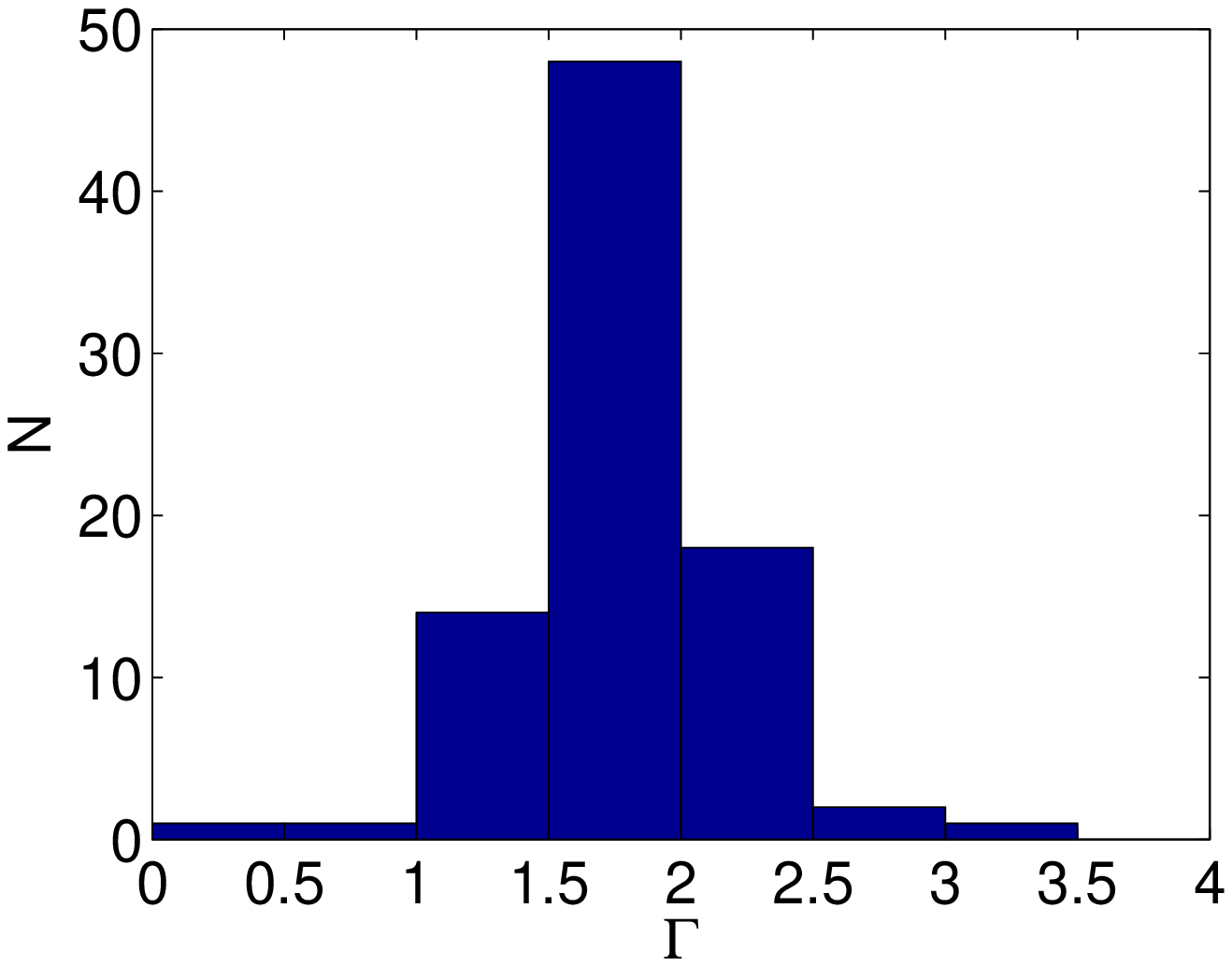}
%\plottwo{SWPI.eps}{PI.eps} 
\caption{Left panel: photon index
distribution for the nuclear X-ray sources in the optically/IR-selected
sample of galaxies. Right panel: photon index distribution for
the extended (X-ray selected) sample. \label{fig2}}
\end{figure*}

\subsection{Spectral modelling and luminosity distribution}

There are about 15 sources with $\approx 4$--$10$ net counts: these
are significant detections because the expected positions are
independently known, the background level is very low, and there are
no other point-like sources nearby. We used the Bayesian method of
\citet*{kbn91} to confirm the significance of those detections. For
sources with such a low number of counts, we fitted their spectra
with a simple power-law model with fixed galactic absorption (no
intrinsic absorption). For other sources with more counts, we used
an absorbed power-law model with free intrinsic column density
$N_{\rm H}$. A power-law with cold absorption generally provides a
good fit for most of the sources. However, about 20 nuclear sources
clearly require an additional ionized absorber and/or an
optically-thin thermal-plasma emission component, which we have
included in our spectral fits. 
%We summarize our spectral modelling results in Table 4.
% $N_{\rm H}$ values are listed in column 3 of Table 3.
Figure 2 shows the distribution of the fitted photon indices for the
optical/IR sample and the extended sample. Most of the nuclear sources
have a photon index $\approx 1.5$--$2.0$, which is the typical range
for AGN. Figure 3 shows the distribution of the intrinsic $N_{\rm
H}$ for the two samples (which are once
again consistent with each other). It is clear that most nuclear
sources are not heavily obscured: very few of them can be classified
as Type-2, in the unified scheme (see Section 3.4).

We then used the fitted values of photon index and absorption to
estimate the intrinsic $0.3$--$8$ keV isotropic luminosities of all
86 nuclear sources in the extended sample (Table 4 and Figure 4).
The cumulative luminosity distribution is consistent with a
power-law with a slope of $\approx -0.45$ above $\approx 2 \times
10^{38}$ ergs s$^{-1}$, which we estimate as the completeness limit,
based on the exposure times and detection thresholds for the
shortest observations in our sample. Because of the different
exposure times and distances of our galaxies, in addition to
non-constant background levels in the nuclear regions, it is not
meaningful to give a detection threshold for the whole sample.

\begin{figure*}
\includegraphics[width=8.6cm]{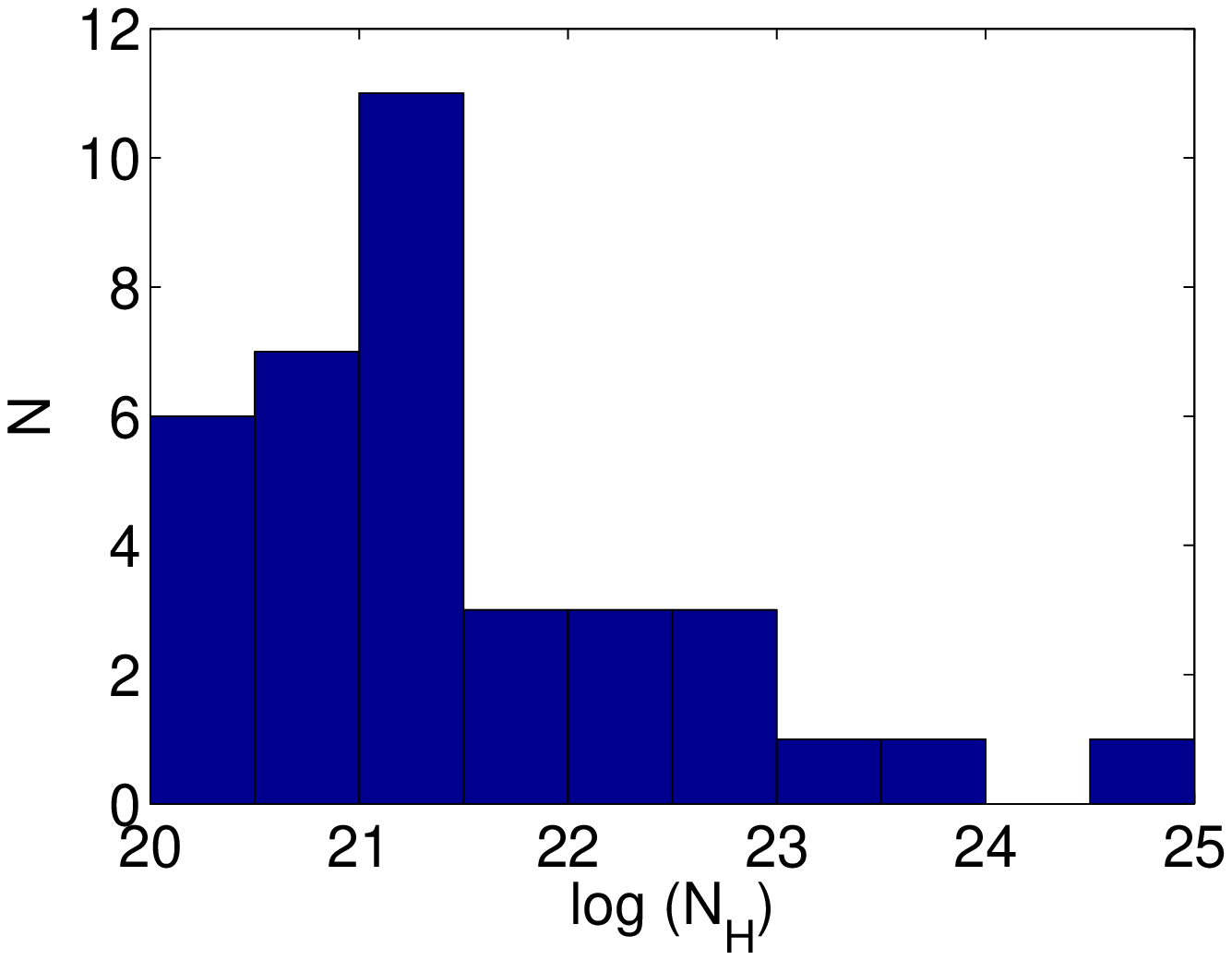}
\includegraphics[width=8.6cm]{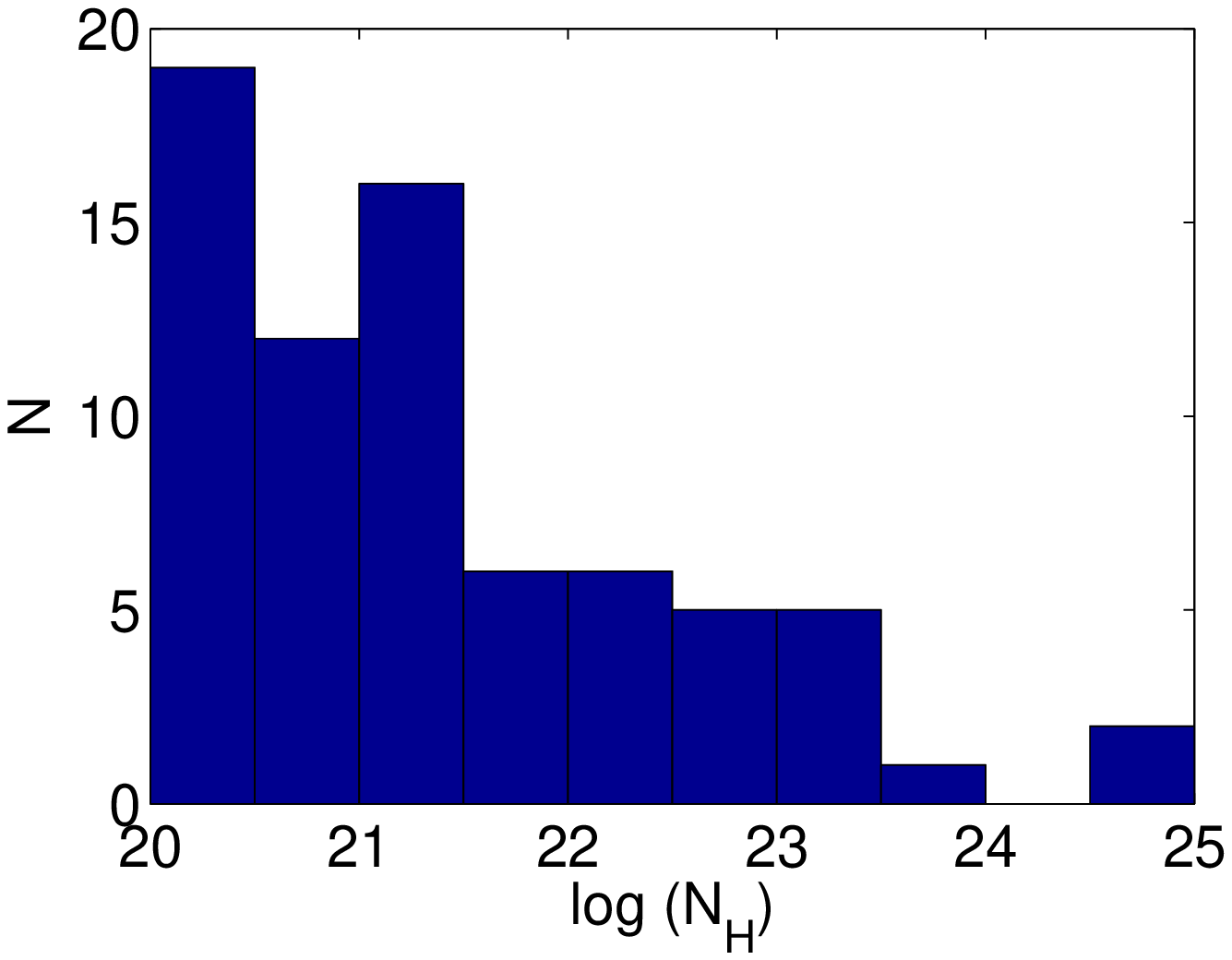}
%\plottwo{SWNH.eps}{NH.eps} 
\caption{Left panel: intrinsic
$N_{\rm H}$ distribution for the nuclear sources
in the optical/IR sample. Right panel: intrinsic $N_{\rm H}$ distribution
for the extended sample. \label{fig3}}
\end{figure*}

\subsection{Physical identification of the nuclear X-ray source population}

The main problem affecting X-ray surveys of low-luminosity ($L_{\rm
X} \la 10^{39}$ erg s$^{-1}$) nuclear BH activity is the possibility
of confusion with X-ray binaries, in the overlapping range of
luminosities. A typical example is the nuclear X-ray source in M\,33
\citep{geb01,dcl04}. Detailed investigations of individual galactic
nuclei \citep{gho08} have highlighted the difficulties and
ambiguities of any such identifications. On a statistical basis, the
slope and normalization of the cumulative luminosity distribution in
our extended sample (Figure 4) is also consistent with the
luminosity distribution of a population of high-mass X-ray binaries,
in a galaxy or ensemble of galaxies with a total star formation rate
of $\approx 25 M_{\odot}$ yr$^{-1}$ \citep*[Eq. 5]{ggs03}, for
which we would also expect an integrated H$\alpha$ luminosity
$L_{\rm H\alpha} \approx 3 \times 10^{42}$ erg s$^{-1}$. This could
be a priori consistent with our sample, if the majority of our 187
galaxies were in the high-luminosity tail of the
H{\footnotesize{II}} nuclei distribution \citep{hfs97c}. However,
there are a few arguments in support of a nuclear BH identification
for the majority of our 86 nuclear X-ray sources.

\begin{figure}
\includegraphics[width=8.1cm]{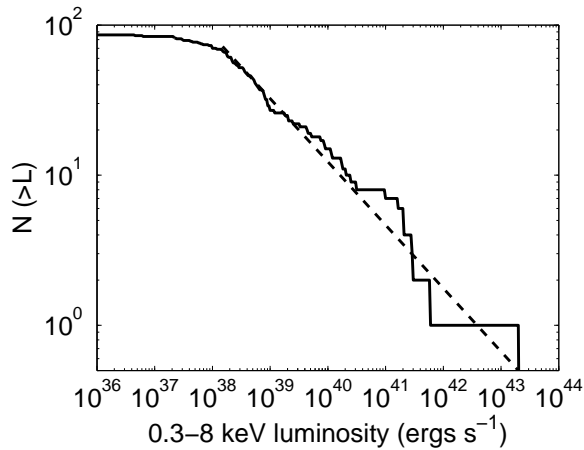} 
\caption{Cumulative luminosity distribution ($0.3$--$8$ keV band)
for all the 86 nuclear X-ray sources in our extended sample.
The completeness limit (estimated from the
shortest exposure times in the survey) is $\approx 2 \times 10^{38}$ erg s$^{-1}$.
A power-law fit has a slope $d\log N/d\log L_{\rm X} \approx -0.5$.
\label{fig4}}
\end{figure}

Firstly, we limited our sample to X-ray sources located
within $1\arcsec$ (typically, $\approx 30$--$70$ pc
for the majority of our galaxies) of the independently known nuclear position.
When we considered the annulus between $1\arcsec$ and
$5\arcsec$ around the nuclear position (a projected area
24 times larger), we found only $\approx 50$ sources,
identified with the same criteria used for the 86 nuclear sources.
Such a concentration of point-like X-ray sources exactly
at the nucleus but not in its immediate surroundings
is much cuspier than typical projected stellar densities (except
when the galaxy has a nuclear star clusters) or densities of
star formation rate. And the argument is even stronger
if we consider that high-mass X-ray binaries may easily disperse
over $\ga 100$ pc in their lifetime, owing to proper motion: thus,
it would be even more unlikely to find most of them exactly
at the nuclear position.

Secondly, our cumulative luminosity distribution is consistent with
an unbroken power-law up to $\sim 10^{42}$ erg s$^{-1}$, where we
run out of galaxies in our sample (Figure 4). Instead, the
luminosity distribution of high-mass X-ray binaries (including
ultraluminous X-ray sources) shows a downturn at $\approx 10^{40}$
erg s$^{-1}$ \citep{ggs03,swartz04}. A scenario where our
nuclear-source population is composed of low-luminosity AGN for
$L_{\rm X} \ga 10^{40}$ erg s$^{-1}$ and high-mass X-ray binaries
for $L_{\rm X} \la 10^{40}$ erg s$^{-1}$, coincidentally with the
same normalization, appears highly contrived.

Thirdly, the majority of the nuclear sources (57 out of 86) are
detected in the earlier-type galaxies of our sample (E to Sb), that
is in galaxies where the nuclear region is part of a massive, old
spheroidal component (Section 3.1). If the contamination from
high-mass X-ray binaries
%in nuclear star clusters and compact starburst nuclei
were significant, we would expect to find more sources in late-type spirals. 
Ellipticals and massive spheroidals do have a high stellar density 
in their cores, with old populations, so they may have 
low-mass X-ray binaries near the nuclear position. 
But a significant contribution of low-mass X-ray binaries 
to our source population above $2 \times 10^{38}$ erg s$^{-1}$ 
is also ruled out, because their cumulative luminosity distribution 
would be much steeper \citep{gil04,swartz04}.
In addition, we used the following argument to estimate
the possible contribution of low-mass X-ray binaries
in the nuclear region of ellipticals and spheroidal bulges.
From surface brightness profiles \citep{gebet03,lau95}, 
we can estimate the total luminosity
and hence the total stellar mass of characteristic
galaxy types within, say, $1\farcs5$ ($\sim 50$--$100$ pc
for our distance range), that is, the radius of our source
extraction region around each nuclear position. This is of course
a function of Hubble type and galaxy history, among other
things, but $M \sim 10^8 M_{\odot}$ is an order-of-magnitude
estimate for massive ellipticals, and a conservative upper
limit for spheroidal bulges. From the population studies
of \citet{gil04}, we expect $\sim 0.02$ sources with X-ray
luminosities $\ga 2 \times 10^{38}$ erg s$^{-1}$, from
an old stellar population with a stellar mass $\sim 10^8 M_{\odot}$.
A similar estimate can be obtained directly from the growth
curves plotted in Fig.~3 of \citet{gil04} for a sample
of nearby elliptical galaxies and spheroidal bulges: those plots
also suggest the presence of $\la 0.1$ sources
above $10^{38}$ erg s$^{-1}$ within $1\farcs5$, scaled
to the range of distances of our sample.
Since our main statistical results are based on sources
above a completeness limit $\approx 2 \times 10^{38}$ erg s$^{-1}$,
the contamination of at most 2 or 3 X-ray binaries
in our whole sample of early-type galaxies is not 
a significant problem.

Taken together, the previous arguments strongly support
our identification of the X-ray source population as
active nuclear BHs---without ruling out the possibility of
stellar-mass interlopers in some limited cases, such as M\,33, 
where stellar kinematics suggest a BH mass $< 1500 M_{\odot}$ 
\citep{geb01}. The distribution of X-ray photon indices $\approx 1.5$--$2.0$
is also in the typical range for AGN, and rules out,
for example, thermal emission from compact starburst nuclei.

\begin{figure}[h]
%\plotone{halpha.ps}
\includegraphics[angle=-90,scale=0.32]{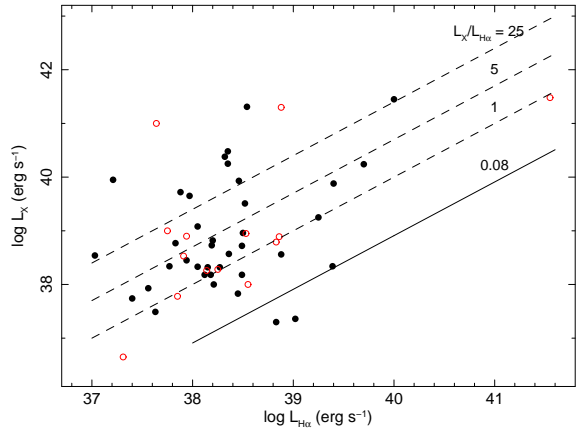}
\caption{Correlation between H$\alpha$
luminosity and X-ray luminosity ($0.3$--$8$ keV) from the nuclei
with point-like X-ray emission. The H$\alpha$ luminosities are from
the Palomar survey \citep{hfs97a}; open circles denote H$\alpha$
measurements during non-photometric nights, which may be taken
as lower limits. The observed range
$1 \la L_{\rm X}/L_{\rm H\alpha} \la 100$ is typical
of accreting nuclear BHs (compare also with Fig.~10 in \citet{ho08}
and Fig.~7 in \citet{flet06});
instead, we would expect $L_{\rm X}/L_{\rm H\alpha} \la 0.1$
for a compact star-forming nucleus. \label{fig5}}
\end{figure}

\begin{figure*}
\includegraphics[width=8.6cm]{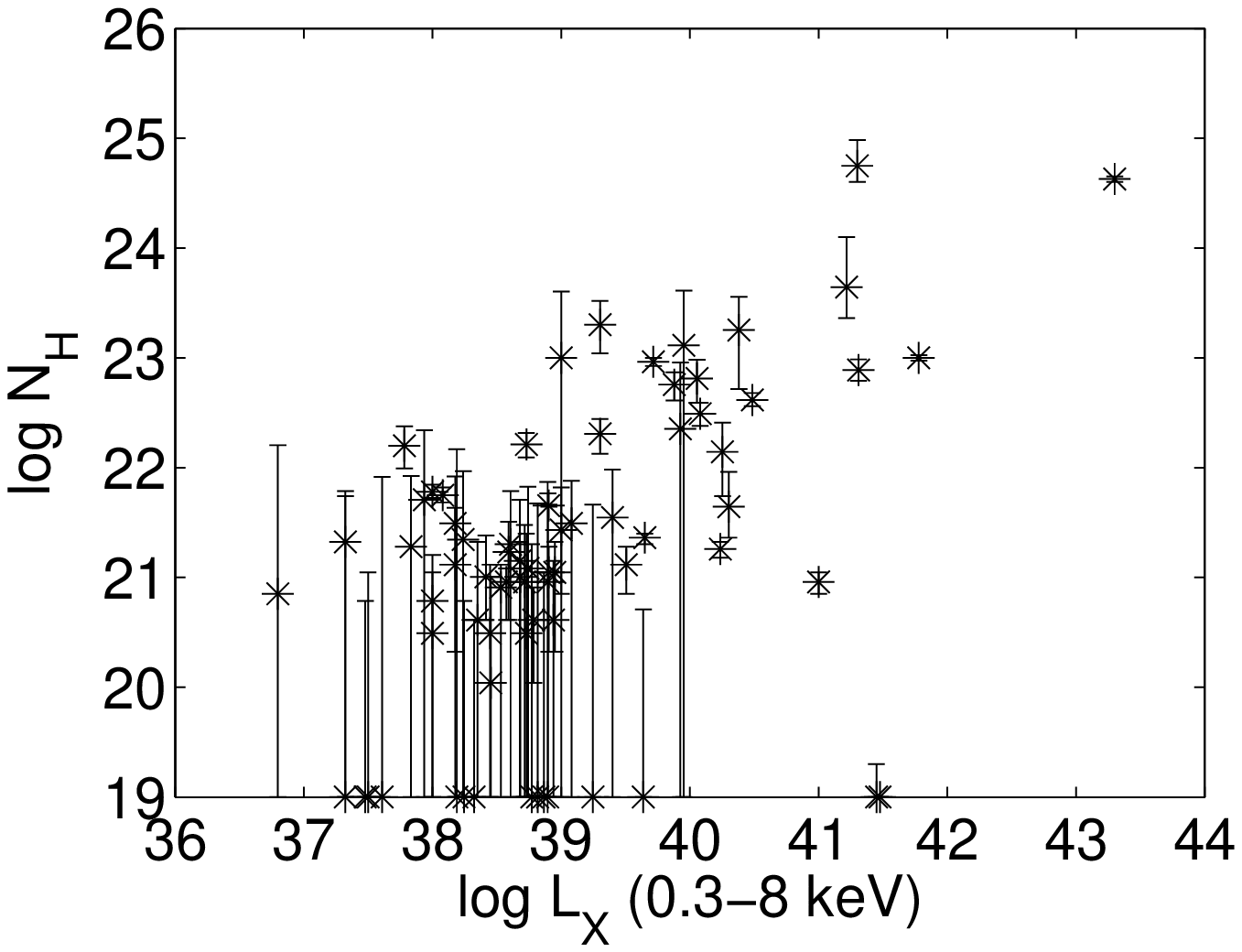}
\includegraphics[width=8.6cm]{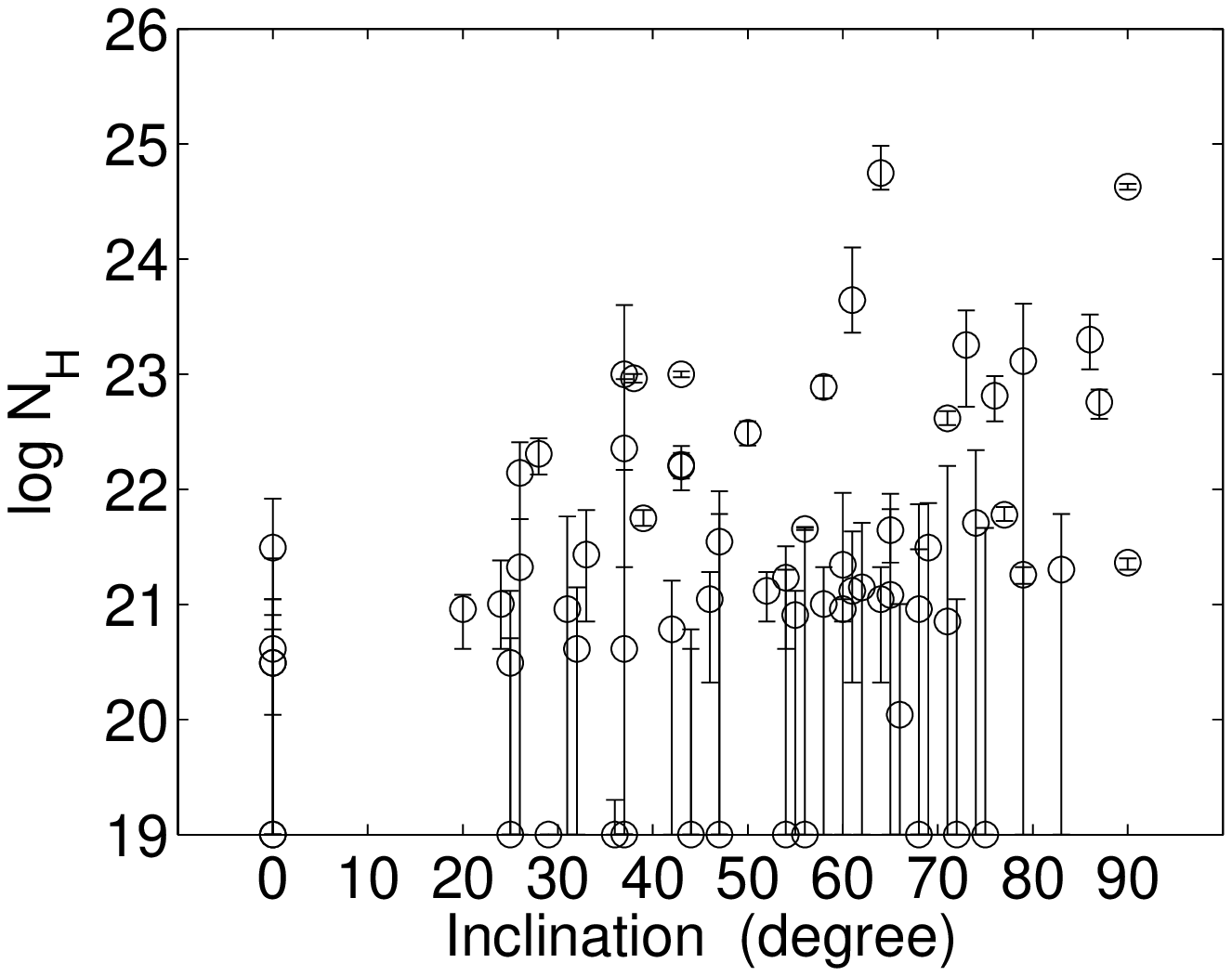}
%\plottwo{LxNH.eps}{InNH.eps} 
\caption{Left panel: relation between 
emitted X-ray luminosity and intrinsic column density. 
The error bars for $N_{\rm H}$ come
directly from spectral fitting, while we have chosen to plot
only the luminosity values (inferred from the best-fitting parameters)
without error bars. When the best-fitting value of the intrinsic
$N_{\rm H} \to 0$, we have arbitrarily assigned a value of $10^{19}$ cm$^{-2}$.
Right panel: relation between inclination angle of the host galaxy 
and intrinsic column density.
\label{fig6}}
\end{figure*}

Another possibility we may consider is that each of our nuclear
sources could instead be the integrated emission from many, fainter
sources in a very compact star-forming nucleus. After all, many of
our sample galaxies are classified as H{\footnotesize{II}} nuclei.
This scenario is already implausible given the predominant detection
of X-ray cores in earlier-type galaxies, as noted earlier. But we
can test this possibility more quantitatively by comparing the
nuclear X-ray and H$\alpha$ luminosities. Fifty-three of the 86
galaxies with an X-ray core in our sample are also included in the
optical spectroscopic Palomar survey of \citet{hfs97a}, which
provides nuclear H$\alpha$ fluxes. Both the H$\alpha$ and the
$0.3$--$8$ keV luminosities of a star-forming region are
proportional to the current or recent star formation rate (assuming
that it has stayed approximately constant over the last $\sim 10$
Myr): $L_{\rm H\alpha} \approx 1.3 \times 10^{41}$ SFR($M_{\odot}$
yr$^{-1}$) erg s$^{-1}$ \citep{ken98}, and $L_{0.3-8} \approx 1.0
\times 10^{40}$ SFR($M_{\odot}$ yr$^{-1}$) erg s$^{-1}$
\citep{ggs03}. Thus, we expect $L_{0.3-8} \sim 0.1 L_{\rm H\alpha}$
for a starburst-dominated compact nucleus. Instead, we expect
$L_{0.3-8} \sim 1$--$25 L_{\rm H\alpha}$ for typical low-luminosity
AGN \citep{ho08,flet06}. The results for our sample (Figure 5)
are more consistent with the AGN scenario.

\begin{figure}[t]
\includegraphics[width=7.7cm]{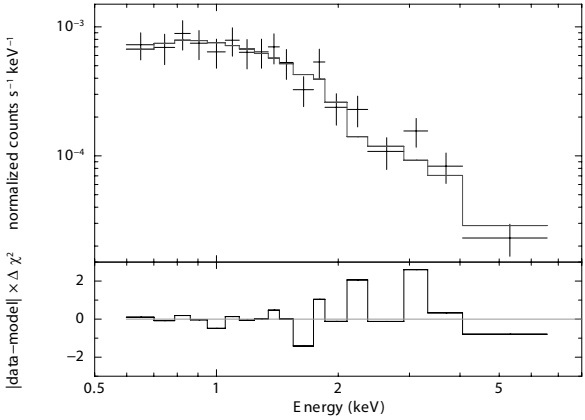} 
\caption{Coadded spectrum for all nuclear
sources with $< 50$ net counts. The spectrum is best fitted by an
absorbed power-law with photon index $\Gamma = 1.68^{+0.22}_{-0.27}$
and $N_{\rm H} = 4.8^{+8.1}_{-4.8} \times 10^{20} \rm cm^{-2}$.
\label{fig7}}
\end{figure}

\subsection{Intrinsic absorption and luminosity}

From our spectral fitting of individual sources, we found that the
intrinsic column density $N_{\rm H}$ appears to be correlated with
the unabsorbed X-ray luminosity (Figure 6, left panel): more luminous sources
tend to be more obscured, while most of the fainter sources are
consistent with column densities $\la 10^{21}$ cm$^{-2}$, or with
only line-of-sight Galactic absorption. There is also a dependence 
on the viewing angle of the host galaxy (Figure 6, right panel), 
as expected, but it is less strong than the luminosity dependence.
In terms of the unified AGN classification scheme, 
we found few or no ``Type-2'' nuclear BHs below an
X-ray luminosity $\approx 10^{40}$ erg s$^{-1}$. 
%If we label as Type-1 obscured all sources 
%with $N_{\rm H} > 10^{22}$ cm$^{-2}$ \citep{gil07,hop09},
%only $\la 10$ per cent of the nuclear sources with 
%$L_{\rm X} \la$ a few $10^{39}$ erg s$^{-1}$ are obscured. 
Only $\la 10$ per cent of the nuclear sources with
$L_{\rm X} \la$ a few $10^{39}$ erg s$^{-1}$ are obscured
by a neutral column density $N_{\rm H} > 10^{22}$ cm$^{-2}$.
But obscured sources represent two thirds (10 out of 15) 
of those with $L_{\rm X} \ga 10^{40}$ erg s$^{-1}$.
This apparent trend of higher absorption 
at higher luminosities is opposite to what seems to happens in more luminous AGN, 
with unabsorbed X-ray luminosities $\sim 10^{42}$--$10^{46}$ erg s$^{-1}$ 
\citep{has08,gil07,ued03}; see also \citet{hop09} for an alternative explanation 
of the luminosity dependence.

\begin{figure*}
\includegraphics[width=8.6cm]{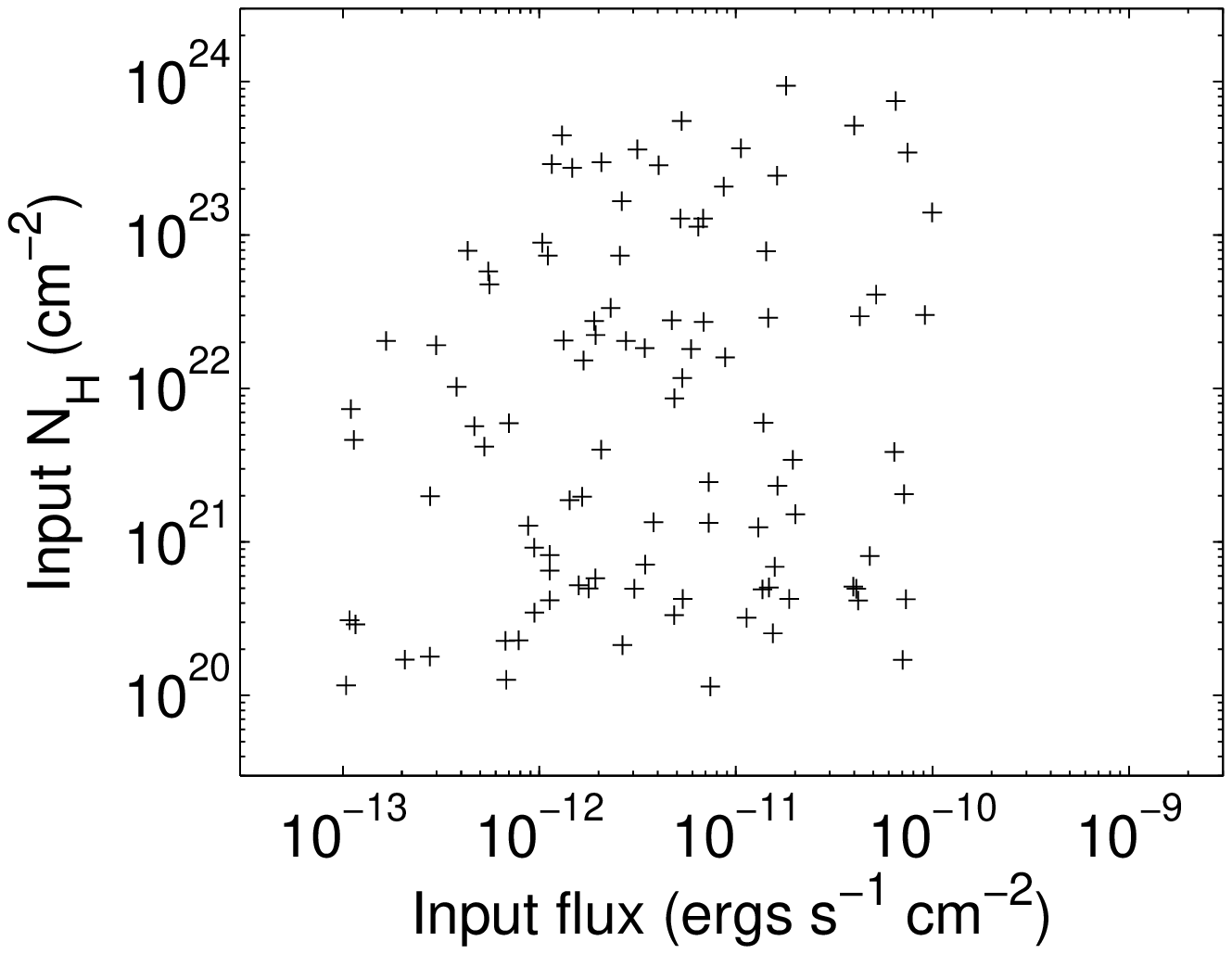}
\includegraphics[width=8.6cm]{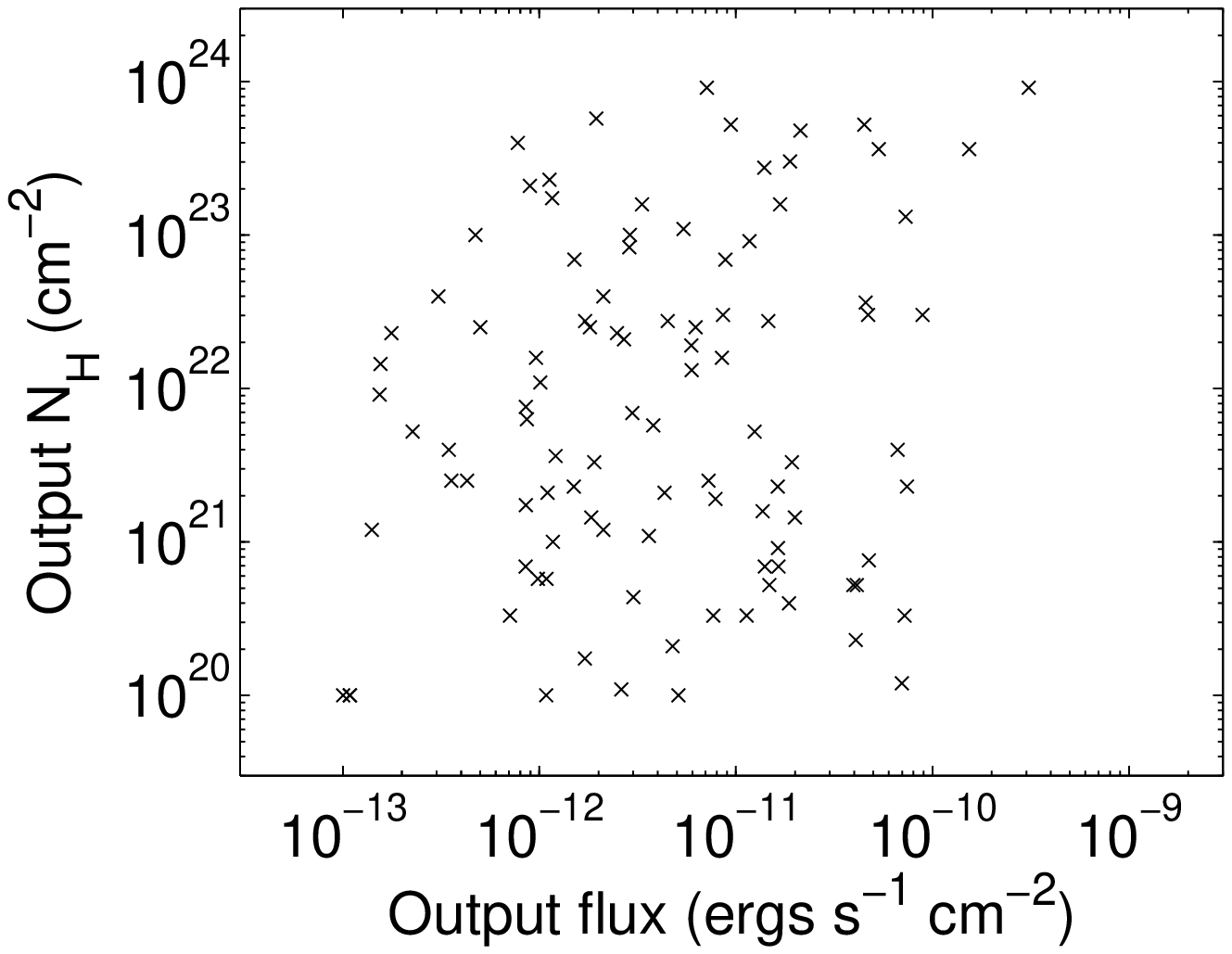} 
\caption{
Results of a MonteCarlo simulation to test a possible fitting bias
in the relation between emitted flux and intrinsic column density
(Section 3.4). Left panel: input distribution of unabsorbed fluxes
and column densities for a simulated sample of 100 sources.
The input distribution is uniform in log scale and uncorrelated.
Right panel: distribution of the fitted fluxes and column
densities for the same simulated spectra. The output distribution
is still uniform and uncorrelated, unlike the observed distribution
of the real sources.  \label{fig8}}
\end{figure*}

Before attempting a physical explanation for the observed trend, we
need to check that it is not the result of observational bias. We
have already excluded from the absorption--luminosity plot (Figure
6) all sources detected with $\le 10$ counts. One possibility might
be that we lose all sources with $N_{\rm H} \ga 10^{22}$ cm$^{-2}$
and $L_{\rm X} \la 10^{39}$ erg s$^{-1}$ because they would not have
enough counts to be detected. However, using simulated X-ray data
and spectral models, we verified that this is not the case: even for
$N_{\rm H} \approx 10^{23}$ cm$^{-2}$, most sources with a typical
photon index $\Gamma \approx 1.7$ and $L_{\rm X} \approx 10^{39}$
erg s$^{-1}$ would have enough counts in the $2$--$8$ keV band to be
detected. An alternative possibility is that most of the fitted
values of $N_{\rm H}$ in faint sources have been underestimated
(correspondingly, their photon indices and intrinsic luminosities would
also have been underestimated). To test this scenario, we stacked
the spectra of all sources with $\le 50$ counts; this is roughly
equivalent to stacking all spectra of nuclear sources with $L_{\rm
X} \la 10^{39}$ erg s$^{-1}$ (most of the galaxies in our sample are
at distances $\sim 10$--$15$ Mpc). We then fitted the co-added
spectrum with an absorbed power-law model; we obtain (Figure 7) that
the best-fitting $\Gamma \approx 1.7$, and $N_{\rm H, tot} \approx 5
\times 10^{20}$ cm$^{-2}$, which is only slightly higher than the average Galactic
absorption for the sources in the stacked sample. We conclude that
the co-added spectrum confirms a very low intrinsic absorption for
the least luminous nuclear BHs.

Furthermore, we conducted a MonteCarlo simulation to test whether
the apparent absorption--luminosity correlation could have been
spuriously introduced during spectral fitting (for example, due to
an overestimate of the intrinsic luminosity for sources with higher
$N_{\rm H}$). We assumed that the unabsorbed flux for a simulated
sample of 100 sources has a uniform distribution (in log scale) from
$10^{-13}$ to $10^{-10}$ erg cm$^{-2}$ s$^{-1}$ in the $0.3$--$8$
keV band (similar to the range of fluxes typical of our real sample
of nuclear sources), and their intrinsic $N_{\rm H}$ has a uniform
distribution (also in log scale) from $10^{20}$ to $10^{24}$
cm$^{-2}$ (Figure 8). In other words, we assumed no intrinsic
correlation between luminosity and $N_{\rm H}$. We also assumed an
absorbed power-law spectrum with $\Gamma = 1.7$ for every source. We
then fitted the simulated spectra (using the Cash statistics) with
absorbed power-law models, leaving $\Gamma$ and  $N_{\rm H}$ as free
parameters, and we estimated the unabsorbed fluxes implied by 
the best-fitting model for each source. The fitted
parameters do not show any correlation between unabsorbed 
fluxes and $N_{\rm H}$, or any significant bias in the
inferred fluxes (Figure 8).
%, nor is there any significant bias in the
%inferred flux (Figure 9).

\begin{figure}[t]
\hspace{-0.2cm}\includegraphics[width=8.4cm]{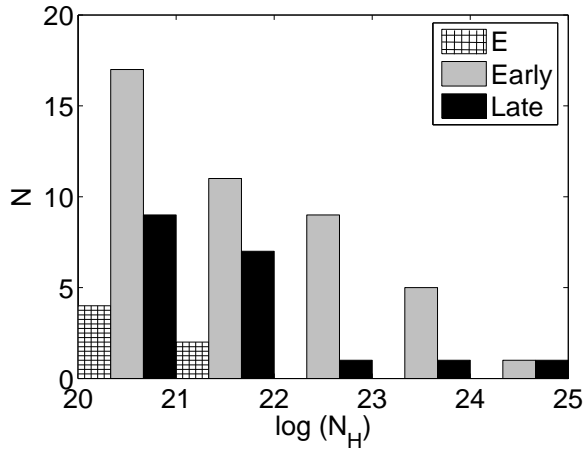} 
\caption{Relation between Hubble type and 
intrinsic column density $N_{\rm H}$. We binned the column density 
into five logarithmic ranges, and we grouped the galaxies 
into three morphological bins (ellipticals E, early-type spirals S0--Sb, 
late-type spirals Sc--Sm). Most of the obscured X-ray nuclei  
($N_{\rm H} \geq 10^{20}$ cm$^{-2}$) belong to early-type spirals, 
but a larger sample of galaxies is needed before we can 
determine any significant trend.  
\label{fig9}}
\end{figure}

We conclude that the absence of Type-2 low-luminosity nuclei,
and the positive correlation between
intrinsic absorption and luminosity up to $L_{\rm X} \approx 10^{42}$
erg s$^{-1}$ are probably real. This is opposite to the negative
correlation between absorption and luminosity known to exist
at higher luminosities \citep{has08,gil07,ued03}.
We also searched for a possible galaxy-morphology dependence 
of the intrinsic column density; for example, whether late-type 
spirals have systematically higher absorption than ellipticals.
We find that all Hubble types appear dominated by unobscured sources, 
although the small number of sources in each class does not permit us 
to draw stronger conclusions (Figure 9). 
The nuclei of early-type spirals in our X-ray detected 
sample seem to be the most obscured: 15 of the 42 nuclear sources
detected in early-type spirals have $N_{\rm H} \geq 10^{22}$ cm$^{-2}$; 
only 3 of the 20 nuclear sources in late-type spirals, and none of the 
6 elliptical nuclei have $N_{\rm H} \geq 10^{22}$ cm$^{-2}$.
However, this may still be due to small-number statistics 
in each class. A larger X-ray sample of galaxies will be needed, 
to study the Hubble-type dependence for a given range of luminosities 
and BH masses.

\begin{figure}
\vspace{-0.15cm}\hspace{0.8cm}\includegraphics[scale=0.45]{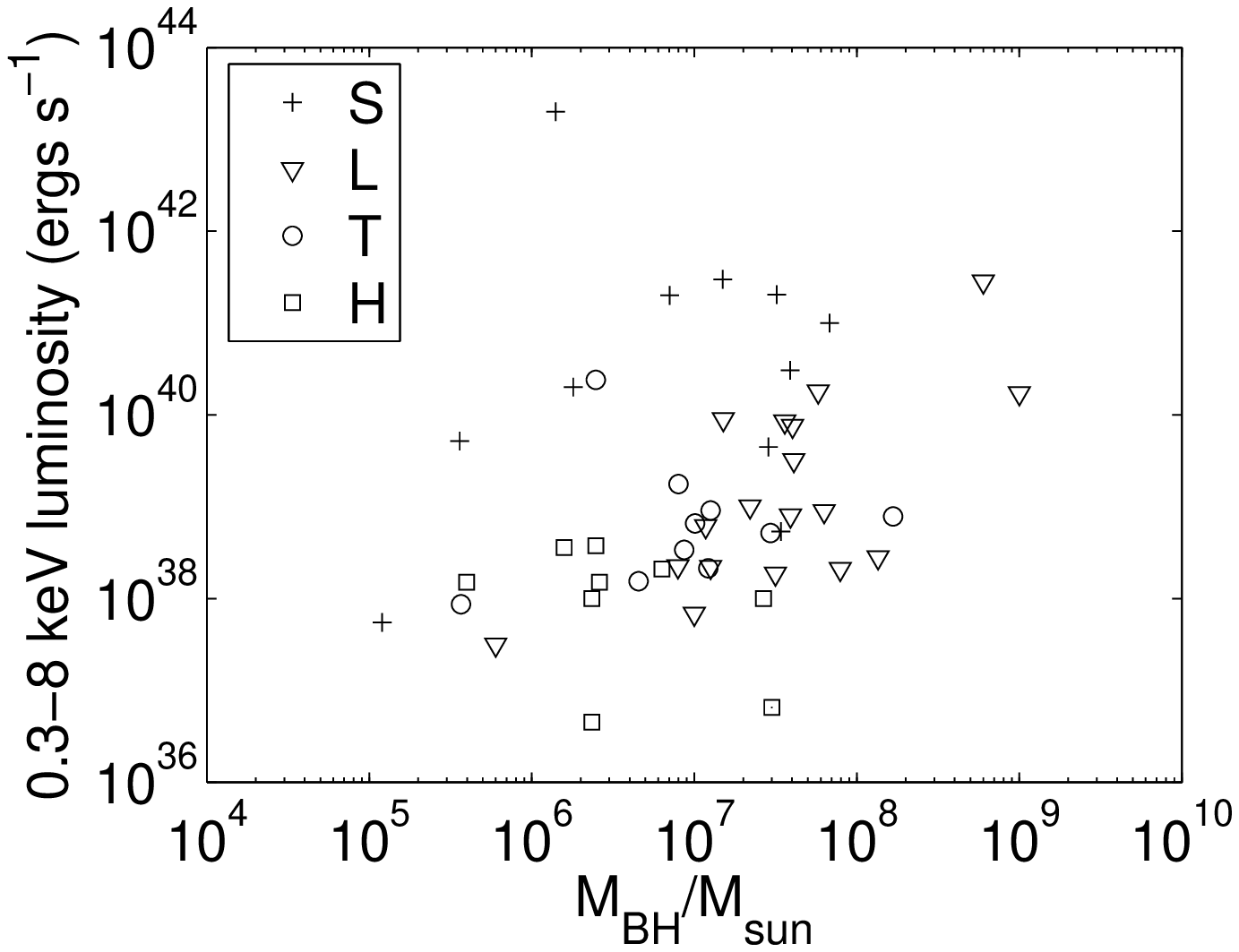}\\
\vspace{-0.0cm}\hspace{0.8cm}\includegraphics[scale=0.45]{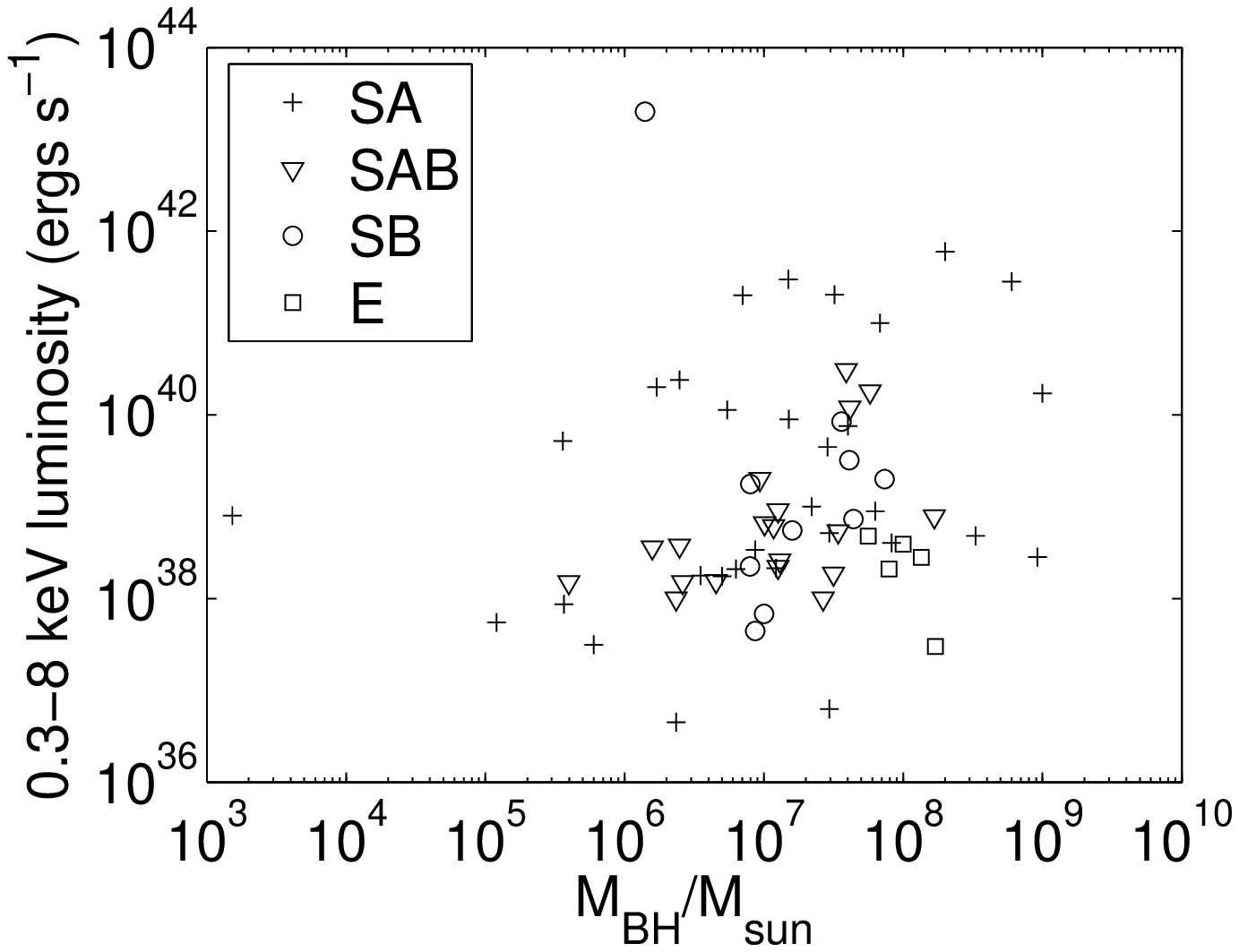} \\
\vspace{-0.15cm}\hspace{0.8cm}\includegraphics[scale=0.45]{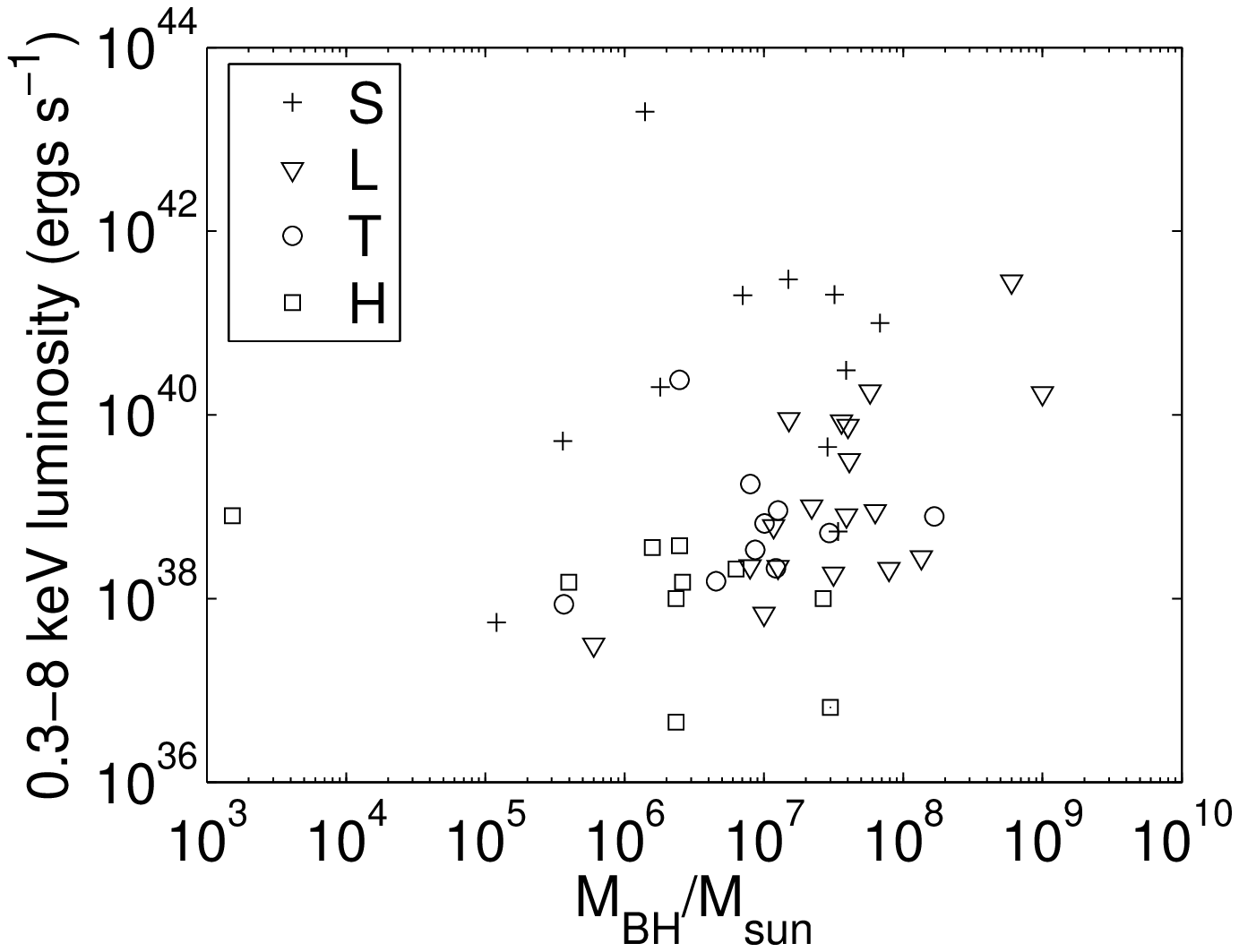}
\caption{Top panel: relation
between nuclear BH mass, X-ray luminosity and Hubble type of the
host galaxy (elliptical, early-type spiral or late-type spiral). The
nuclear X-ray luminosity appears almost independent of BH mass.
Middle panel: relation between nuclear BH mass, X-ray luminosity and
bar morphology (for the purpose of this and following plots, S0
galaxies have been included in the non-barred spiral class SA).
Barred galaxies (SB) do not appear to have brighter X-ray nuclei.
Bottom panel: relation between nuclear BH mass, X-ray luminosity and
optical spectroscopic nuclear class (S = Seyfert, L = LINER, T =
transition object, H = H{\footnotesize II} nucleus).  \label{fig10}}
\end{figure}

\subsection{Eddington ratios}

Among the 86 galaxies with an X-ray core in our extended sample, 31
already have a measurememnt of their nuclear BH mass in the
literature, based on stellar kinematics, gas kinematics, water
masers, or reverberation mapping (see detailed references in Table
4). For another 31 galaxies, stellar velocity dispersions of their
cores are available from Hyperleda\footnote{http://leda.univ-lyon1.fr},
with typical uncertainties $\sim 10$--$20$\%; in those cases, 
we used the $M_{\rm BH}$--$\sigma$ relation \citep{tre02} to estimate their BH masses.
For 3 other cases, only photometric observations are available: we
constrained their BH masses via the $M_{\rm BH}$--S$\acute{\rm
e}$rsic index relation \citep{gra07}. Thus, we have 65 nuclear BHs
in our sample with a mass estimate, out of 86 candidate X-ray cores
(Table 4).

\begin{figure}
\includegraphics[width=8.1cm]{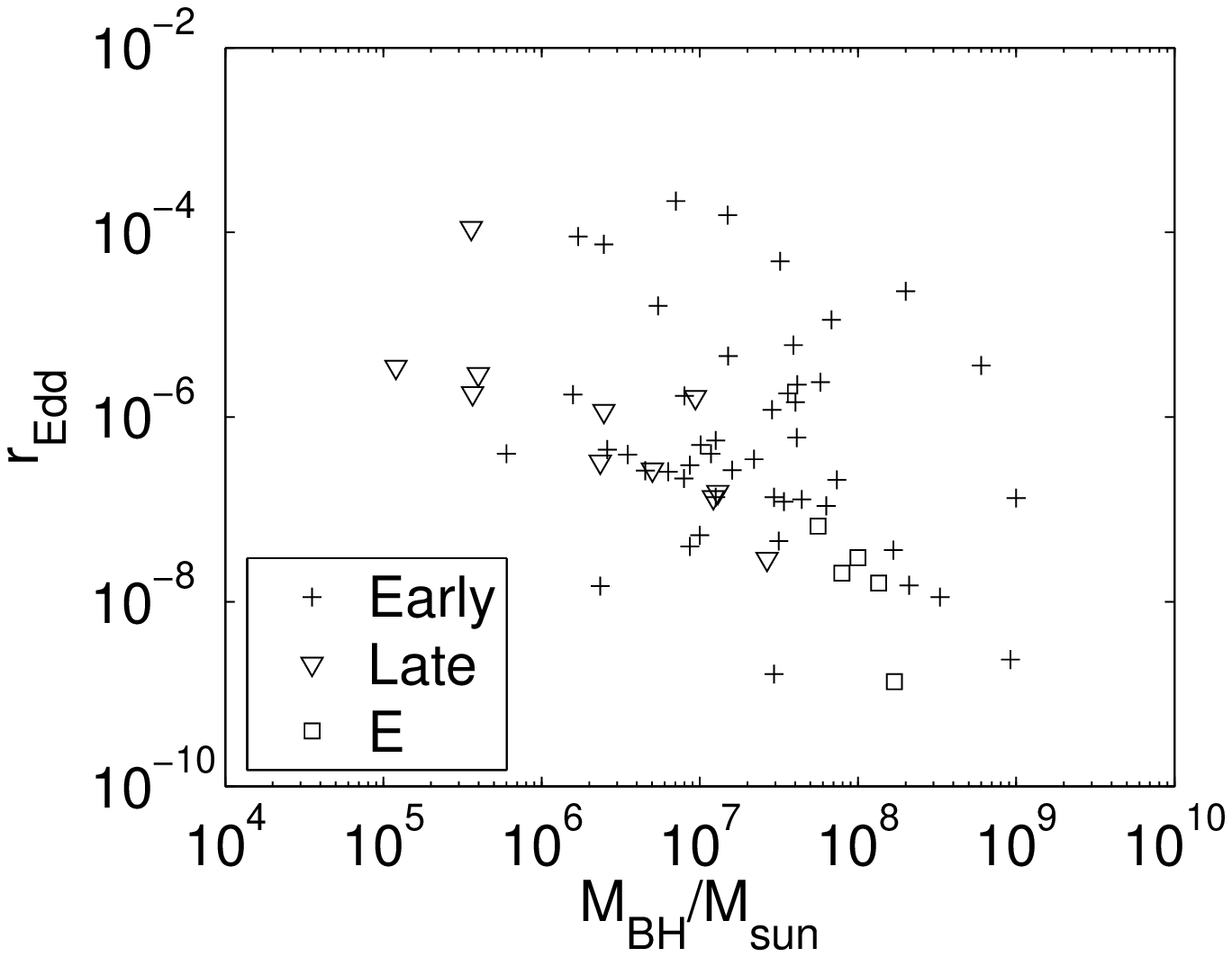}
\includegraphics[width=8.1cm]{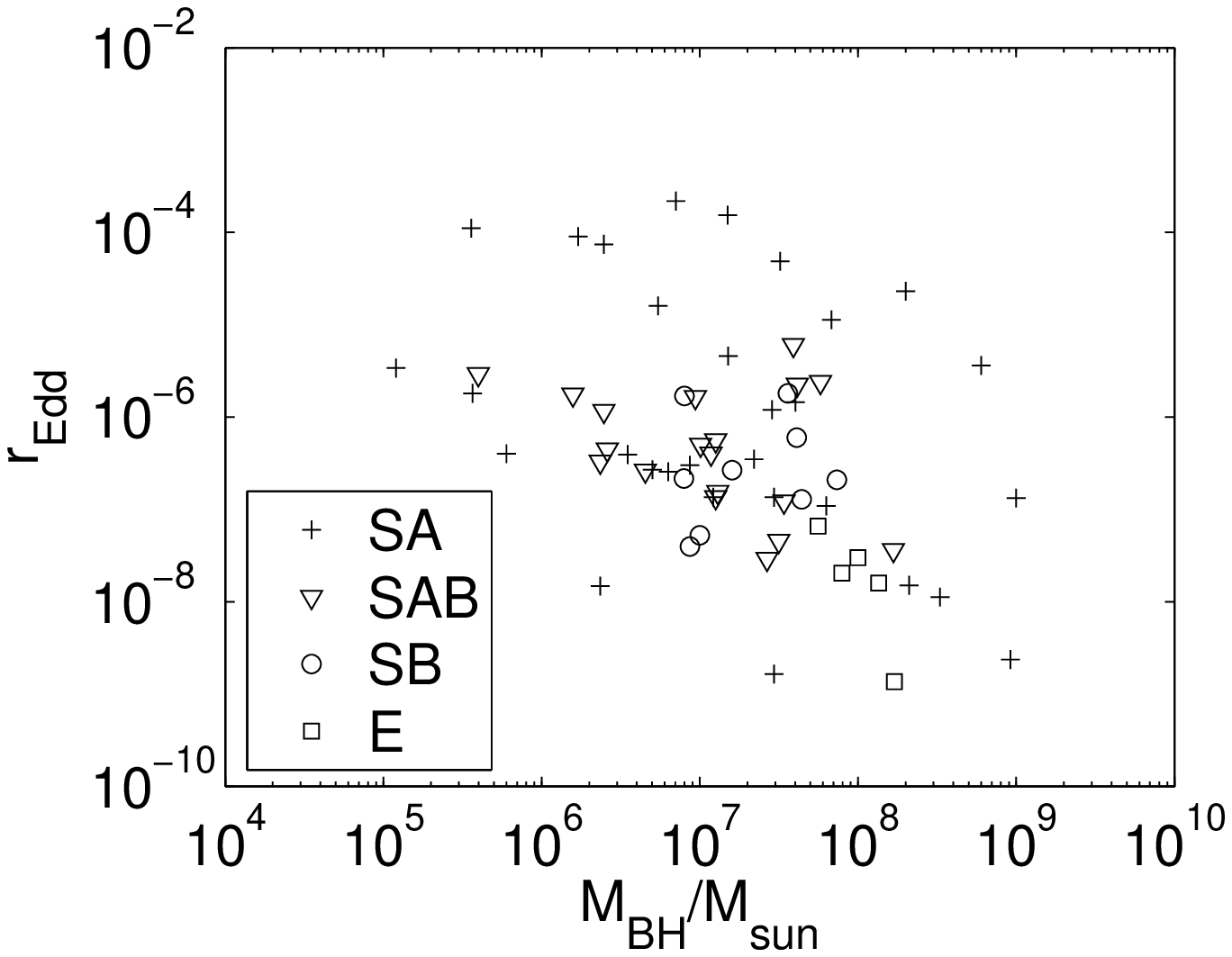} 
\caption{Left panel:
relation between BH mass, Eddington ratio and Hubble type.
Right panel: relation between BH mass, Eddington ratio and bar morphology.
Here and in the following figure, for clarity of presentation, 
our plots do not include the two very discrepant datapoints of 
M\,33 (very low mass nuclear BH) and NGC\,4945 (very high accretion rate 
and luminosity); see Table 4 for details. \label{fig11}}
\end{figure}

The elliptical galaxies in our sample have larger inferred BH masses
($M_{\rm BH} \sim 10^8 M_{\odot}$) and low X-ray luminosities
($L_{\rm X} \la 10^{39}$ erg s$^{-1}$), as expected. Late-type
spirals also have low X-ray luminosities ($L_{\rm X} \la 10^{39}$
erg s$^{-1}$) but the lowest BH masses ($M_{\rm BH} \sim
10^5$--$10^7 M_{\odot}$). The most luminous nuclear BHs in the local
universe ($L_{\rm X} \sim 10^{41}$--$10^{42}$ erg s$^{-1}$)
are found in early-type spirals (Figure 10, top panel).
X-ray nuclei of barred galaxies are less luminous than those 
of non-barred galaxies, for a given range of BH masses (Figure 10, middle panel). 
This is partly due to the higher fraction of strongly-barred galaxies 
in late-type spirals, which tend to have weaker nuclear emission.
Optically-classified Seyferts have the most luminous X-ray nuclei, 
as expected (Figure 10, bottom panel). There is only a
very weak positive correlation (slope $\approx 0.26$ and correlation
coefficient $\approx 0.19$) between BH mass and X-ray luminosity. If
the X-ray luminosity scaled with the Bondi accretion rate \citep{hoy39} onto the
nuclear BH, we would expect $L_{\rm X} \sim M_{\rm BH}^2
\rho(\infty)/c_s^3(\infty)$, where $\rho(\infty)$ and
$c_s^3(\infty)$ are the gas density and sound speed outside the
accretion radius. The higher gas density in the nuclear region of
spiral galaxies compensates for their lower BH masses.
In contrast, the lower abundance of gas available for accretion 
in elliptical galaxies is offset by higher BH masses.
Because the luminosity is almost independent of BH mass, 
there is a negative correlation (slope $\approx
-0.41$ and correlation coefficient $\approx -0.49$) between BH
masses and X-ray Eddington ratios 
$r_{\rm Edd} \equiv L_{0.3-8}/L_{\rm Edd}$ (Figure 11). 
In local-universe ellipticals, $L_{0.3-8} \sim
10^{-9}$--$10^{-8} L_{\rm Edd}$; in late-type spirals, $L_{0.3-8}
\sim 10^{-5}$--$10^{-4} L_{\rm Edd}$. 
%Figure 12 shows the positive
%correlation between X-ray luminosities and X-ray Eddington ratios
%(slope of $\approx 0.86$ and correlation coefficient $\approx 0.77$).
%The correlation coefficients are
%0.7, 0.97 and 0.6 for non-barred spiral galaxies (SA), barred
%spiral galaxies (SB) and mixed spiral galaxies (SAB) respectively.
%The correlation coefficients are 0.8, 0.87 and 0.98 for early type
%spiral galaxies, late type spiral galaxies and ellipse spiral
%galaxies respectively.
%The right-hand panel of Figure 12 is also another way to show that
%the majority of X-ray luminous nuclei (those with $L_{\rm X} \sim
%10^{41}$--$10^{42}$ erg s$^{-1}$) belong to non-barred, early-type
%spirals. 
The only outlier is the late-type, barred (SBcd)
Seyfert-2 galaxy NGC\,4945, perhaps the only ``true'' AGN in our
sample ($L_{\rm 0.3-8} \approx 2 \times 10^{43}$ erg s$^{-1}$ 
$\sim 0.1 L_{\rm Edd}$). 
This is also the galaxy with the highest intrinsic
absoprtion ($N_{\rm H} \sim 10^{24}$ cm$^{-2}$); see \citet{doet03,ito08} 
for detailed X-ray studies of this AGN.

%\section{ACCRETION RATE, X-RAY LUMINOSITY \& BOLOMETRIC LUMINOSITY}

%% The displaymath environment will produce the same sort of equation as
%% the equation environment, except that the equation will not be numbered
%% by LaTeX.

Estimating the mass accretion rate from the observed luminosities
requires the assumption of a model for the radiative and total
efficiency. In our local-universe sample, even the most luminous
nuclei (except for NGC\,4945) are likely to be in the
low-radiative-efficiency regime, which is thought to set 
in for $L_{\rm bol} \sim 10 L_{\rm X} \la 0.01 L_{\rm Edd}$, 
by analogy with stellar-mass BHs \citep{esi97,jes05}. 
We adopted the ADAF scenario, which includes contributions
to the emitted flux from disk-blackbody (truncated outer disk),
synchrotron, bremsstrahlung and inverse-Compton \citep{nbc97,nar98}.
ADAF spectral models scale with the BH mass  
and the dimensionless accretion parameter
$\dot{m} \equiv L_{\rm bol}/\left(\eta L_{\rm Edd}\right) 
= \dot{M} c^2/L_{\rm Edd} 
\propto \dot{M}/\dot{M}_{\rm Edd}$, where $\eta$ is the radiative efficiency 
($\eta \sim 0.1$ for efficient accretion)\footnote{With this definition, 
the transition between ADAF and standard-disk accretion occurs at  
$\dot{m}\sim 0.1$, and $\dot{m} \approx 10$ for $L_{\rm bol} \approx L_{\rm Edd}$.
The accretion parameter $\dot{m}$ is sometimes alternatively defined 
in the literature as $\equiv 0.1 \dot{M} c^2/L_{\rm Edd}$, 
so that $\dot{m} \approx 1$ for $L_{\rm bol} \approx L_{\rm Edd}$.}.
%$\dot{m} \equiv \dot{M}/\dot{M}_{\rm Edd} \equiv 0.1
%\dot{M} c^2/L_{\rm Edd}$. 
Other physical information  
is included in three (dimensionless) parameters:
the ratio of gas to magnetic pressure; the viscosity parameter;
and the fraction of viscous heating that goes into the electrons.
A useful table of band-limited X-ray luminosities for a grid
of ADAF spectral models, as a function of BH mass and accretion
parameter (with standard assumptions for the other three
parameters), was calculated by \citet{mel03}.
We used their grid values and interpolations
to estimate $\dot{m}$ of our sample nuclei, from their
observed X-ray luminosities and indirectly-inferred BH masses.
We obtain a range of accretion parameters $\dot{m} \sim
10^{-5}$--$10^{-2}$, and we have plotted them as a function of BH
mass, Hubble type and optical spectroscopic classification (Figure 12). 
%Finally, we show (Figure 14)
%the distribution of accretion parameters $\dot{m}$ (inferred from
%the X-ray luminosity with the assumption of an ADAF-like radiative
%efficiency) broken down for the different optical-spectroscopic
%classes of galactic nuclei in our sample (Seyferts, LINERs,
%transition objects, H{\footnotesize{II}} nuclei, in order of
%decreasing accretion rate).

\begin{figure*}
\includegraphics[width=8.4cm]{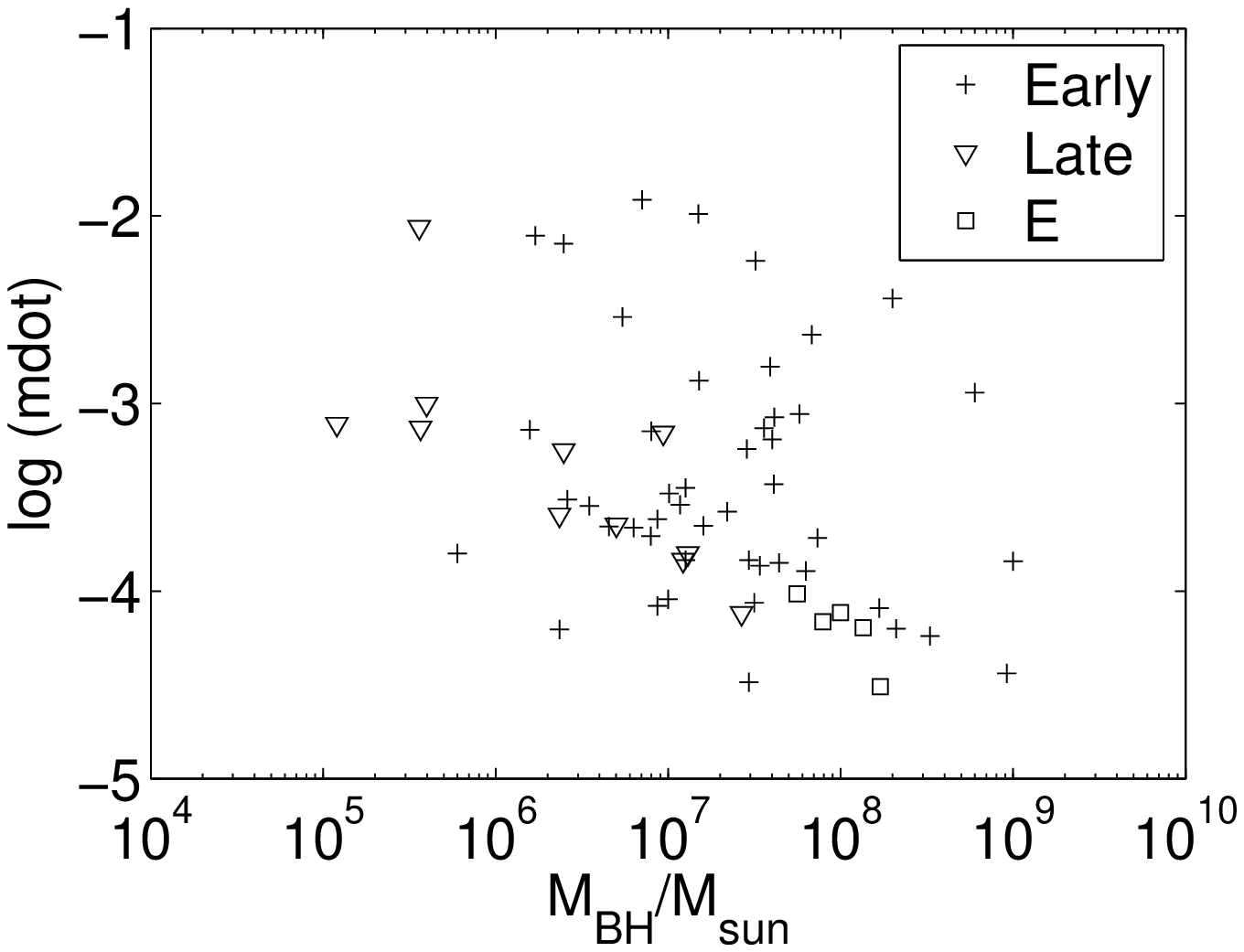}
\includegraphics[width=8.4cm]{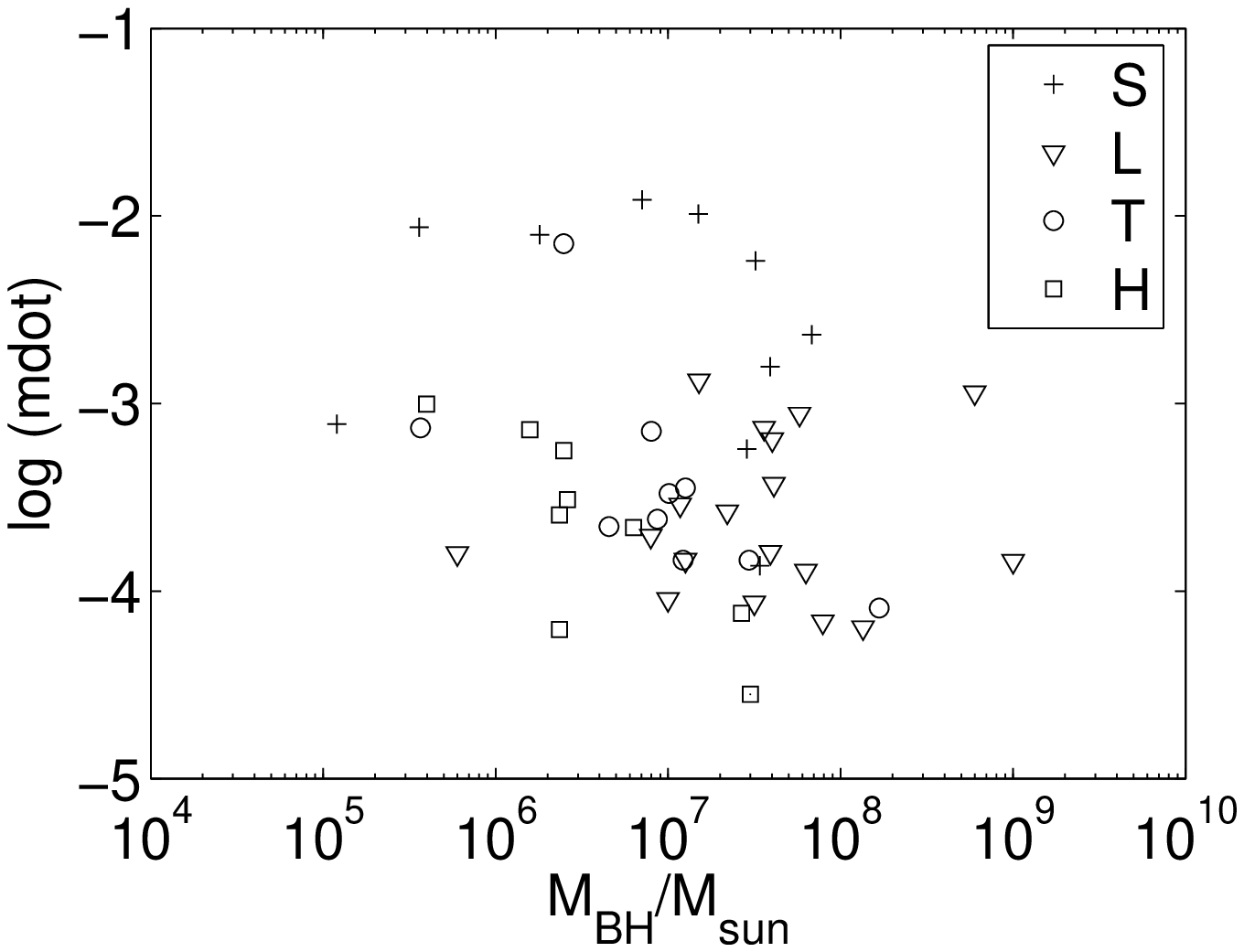}
%{LxMdEaLaE.eps}
\caption{Left panel: relation between BH mass,
accretion parameter $\dot{m}$ and Hubble type.
We estimated $\dot{m}$ from the emitted luminosities,
assuming the accretion rate/efficiency relations interpolated 
by \citet{mel03}.
Right panel: relation between BH mass,
accretion parameter $\dot{m}$ and optical spectroscopic nuclear
class (S = Seyfert, L = LINER, T = transition object, H = H{\footnotesize II} nucleus).
%$Right$: The relation between X-ray luminosity and $\dot{m}$.
\label{fig12}}
\end{figure*}

\section{DISCUSSION AND CONCLUSIONS}

We studied the X-ray nuclear activity of nearby galaxies (distance
$< 15$ Mpc), in a range of luminosities intermediate between
low-luminosity AGN and ``normal'' non-active galaxies. More
specifically, we chose a complete sample of optically/IR-selected
Northern galaxies, as defined in \citet{swa08}; most of the galaxies
in this sample were observed by {\it Chandra}/ACIS for a ULX survey.
We then extended that sample with another $\sim 70$ galaxies (also
at distances $< 15$ Mpc) with publicly-available {\it Chandra}/ACIS
data; the extended sample contains 187 galaxies. The main results
presented in this paper are:

1. We made a census of weakly active nuclei in the local
neighbourhood, down to a completeness limit $L_{\rm X} \approx 2
\times 10^{38}$ erg s$^{-1}$ in the $0.3$--$8$ keV band. Eighty-six
out of 187 galaxies have a point-like nuclear
X-ray source, coincident (within $1\arcsec$) with the radio or
infrared nuclear position. The presence of an X-ray core depends
strongly on Hubble type: $\approx 60$ per cent of early-type galaxies
(E to Sb) contain an X-ray core, but only $\approx 30$ per cent of
later-type galaxies. About 90 per cent of optically-classified
Seyfert and LINERs have a nuclear X-ray source with $L_{\rm X} \ga 2
\times 10^{38}$ erg s$^{-1}$; this fraction drops to $\approx 60$
per cent for transition objects, and $\approx 30$ per cent for
H{\footnotesize{II}} nuclei. The AGN demographics from our {\it
Chandra} survey is consistent with the AGN demographics inferred
from optical spectroscopic studies, for example the Palomar survey
\citep[][and references therein]{ho08}. Our {\it Chandra} survey
suggests that point-like nuclear X-ray emission is a reliable
indicator of BH activity in normal galaxies, especially in cases
where optical signatures of BH accretion may be swamped by the
surrounding stellar emission.

2. Spiral and elliptical galaxies within 15 Mpc are detected with
a continuous range of nuclear luminosities from $\sim 10^{38}$ erg s$^{-1}$
to $\sim 10^{42}$ erg s$^{-1}$. There is no gap between
low-luminosity AGN and BH activity in normal galaxies.
The cumulative luminosity distribution can be fitted
by a power-law with a slope of $\approx -0.5$.
Because of the relatively small number of galaxies in our sample,
we cannot yet accurately determine whether the luminosity
distribution of those faint nuclei matches the slope and normalization
of the luminosity distribution of fully-fledged AGN
($L_{\rm X} > \sim 10^{42}$ erg s$^{-1}$).
However, we are currently studying a larger {\it Chandra}
sample of galaxies (within 40 Mpc) and we will address
this issue in a follow-up paper.

3. For each individual X-ray core with luminosities $\sim
10^{38}$--$10^{39}$ erg s$^{-1}$, it is always very difficult to
distinguish between a nuclear BH and an unrelated, luminous X-ray
binary in the nuclear region. However, on a statistical basis, we
have discussed various reasons why we think that the majority of
detected sources are nuclear BHs rather than X-ray binaries (Section
3.1). Most of the X-ray sources are exactly coincident with the
nuclear position, and there are much fewer sources in the annulus
between $1\arcsec$ and $5\arcsec$ from the radio/optical nucleus.
Besides, the luminosity distribution has no hint of a break at
$L_{\rm X} \approx 10^{40}$ erg s$^{-1}$, as we would expect from a
population of high-mass X-ray binaries. The ratio between nuclear
H$\alpha$ luminosity (when available, from the Palomar survey) and
X-ray luminosity is more typical of low-luminosity AGN than of
star-forming regions with young X-ray binaries.

4. We fitted the spectra of each source, assuming absorbed
power-law models with two free parameters: the photon index
and the intrinsic absorption. The photon index is consistent
with the expected value for AGN ($\Gamma \approx 1.5$--$2$).
The intrinsic column density $N_{\rm H}$
is positively correlated with the emitted X-ray luminosity:
among the population of fainter nuclear BHs, very few or none
can be classified as Type-2 (highly obscured). This is the opposite
of the trend known for luminous AGN, with $L_{\rm X} > 10^{42}$ erg s$^{-1}$
\citep{has08,gil07,ued03}. It is still not clear what produces the absorption
(a geometrically-thick parsec-scale torus? An optically-thick disk wind?),
and hence we cannot determine from the data available
what causes the fraction of obscured sources to be highest for  
sources with $L_{\rm X} \sim 10^{42}$--$10^{43}$ erg s$^{-1}$,
and to decrease at both lower and higher nuclear BH luminosities.
At high luminosities, it was suggested that the thick torus gets
progressively evaporated or ablated by the radiation flux
from the central object \citep{has08,men08}. On the other hand,
X-ray faint nuclei have less gas available for accretion.
They may not be surrounded by a torus at all; or the torus may
collapse and become geometrically thin, so that only galaxies
seen perfectly edge-on would be classified as Type-2; or, instead,
the dramatic decrease in absorption may be caused by the suppression
of the optically-thick disk wind.
Alternatively, it was suggested \citep{hop09} that 
a large fraction of sources with 
$L_{\rm X} \sim 10^{42}$--$10^{44}$ erg s$^{-1}$ 
have been erroneously classified as obscured, because of optical dilution 
effects, and because of the transition from a standard-disk accretion 
geometry to a radiatively-inefficient flow. Both effects would make 
those sources less prominent or invisible in the UV/optical band, 
and slightly harder in the X-ray band, thus making them appear ``obscured'' 
even though they are not.
Based on those arguments, it was proposed that the fraction of truly obscured 
sources can be as low as 20 per cent, independent of luminosity \citep{hop09}.
However, an opposite result was obtained by \citet{rey08},
based on an optically-selected sample of quasars from
the Sloan Digital Sky Survey; they find that at least half
of the quasars in the nearby Universe ($z \la 0.8$) are
truly obscured, particularly in the high-luminosity population. 
In our sample of sources, the absorbing column densities 
are estimated directly from the fitted X-ray spectra, and 
do not rely on hardness ratios or optical/X-ray flux ratios, 
thus reducing the bias discussed by \citet{hop09}. Thus, 
we suggest that the high fraction (10 out of 15) of obscured sources 
at X-ray luminosities $> 10^{40}$ erg s$^{-1}$ ($0.3$--$8$ keV band) 
may be really due to a higher density of absorbing gas or dust  
around the nuclear BH (whatever its geometry) 
in that moderate luminosity range.
The fraction of obscured nuclei seems to depend 
more directly on luminosity rather than Hubble type.
In our sample, slightly more nuclei detected in early-type spirals 
seem to be obscured ($N_{\rm H} \geq 10^{22}$ cm$^{-2}$), 
compared with the obscured fractions in 
ellipticals and late-type spirals; however, we do not have 
enough galaxies to draw strong conclusions. 
We are planning further work with a larger sample of galaxies 
to address this issue.

5. Having argued that the apparent luminosity dependence of the obscured 
fraction of nuclei is truly due to changes in the column density 
of the absorbing medium, we can still look for a causal link between 
such changes and the disappearance of the standard disk at low accretion 
rates (or other transitions in the geometry of the accretion flow).
For example, the disk may disappear when it is
no longer fed by a large torus; or, vice versa, optically-thick winds
may be suppressed when the standard disk turns into an optically-thin
ADAF. It was already noted \citep{ho08} that the decrease or suppression 
of the intrinsic absorption at low luminosities often coincides 
with the disappearance of optical signatures of a standard accretion disk. 
If the apparent or real changes in the obscuration fraction 
are due to a sharp standard-disk/ADAF transition at $\dot{m} \sim 0.1$, 
we should not expect a trend in our sample, because all but one 
of our sources are almost certainly below this threshold, 
in the radiatively-inefficient regime. Instead, 
we see changes in the fraction of obscured sources 
over the X-ray luminosity range $\sim 10^{39}$--$10^{42}$ erg s$^{-1}$ 
and $\dot{m} \sim 10^{-4}$--$10^{-2}$. This suggests a more 
gradual evolution in the amount of absorbing material 
below the radiatively-inefficient threshold; or perhaps 
a more gradual disappearance of the outer accretion disk 
at low accretion rates.
It was also recently suggested \citep{zzt08,wan07} that a dusty torus
(produced during major galaxy mergers, or via secular evolution 
processes) can provide the main source
of fuel in a self-regulated way: when it is completely evaporated
by the radiation from the central source, the AGN phase turns off
and the supermassive BH becomes inactive; this scenario
is among those consistent with our observational findings.

6. We confirm that there is no positive correlation between
the presence of a bar and the X-ray luminosity of the nucleus.
Large-scale bars are highly
effective in delivering gas to the central few hundred parsecs of
a spiral galaxy and therefore enhance the probability and rate of
star formation in the nuclear region \citep{hs94}. However,
it was observed from optical/IR surveys \citep{hfs97d,hun99,lau04} 
that the presence of a bar seems to have no impact on either 
the frequency or strength of AGN activity---with the possible 
exception of Narrow-Line Seyfert 1 galaxies, which may be 
preferentially associated with barred spirals \citep{oht07}. 
In fact, our results suggest a negative correlation,
with strongly barred galaxies having a less active nuclear BH 
than non-barred/weakly barred galaxies, for a given range 
of BH masses and Hubble types (we do not have Narrow-Line Seyfert 1s 
in our sample). This preliminary but intriguing result will have to be tested
more strongly on our larger {\it Chandra} sample we are currently
studying. If confirmed, it may suggest that there is less gas reaching
the supermassive BH when there is circumnuclear star formation
(i.e., in most strongly-barred galaxies). In this scenario,
galaxies may cycle through phases of dominant nuclear
star formation (when a bar is present) and phases dominated
by nuclear BH accretion. One possible explanation could be that 
nuclear star formation in spiral galaxies actively disfavours 
gas inflows towards the nuclear BH---most of the gas not used for star 
formation may be blown away by supernova-powered outflows. 
In addition, there may be a mismatch between the characteristic 
timescales of bar formation and disruption 
($\approx$ a few $10^8$ yrs: \citet{com93}), and the longer timescales 
of nuclear BH accretion. In this scenario \citep{hun99}, 
bars and associated nuclear star formation would appear first, 
and nuclear activity later, after the bar-driven star formation 
has subsided.

7. There is an anticorrelation between the BH mass and the
Eddington ratio: elliptical galaxies in the local universe
have higher BH masses but much lower accretion parameters,
corresponding to X-ray luminosities $\sim 10^{-8} L_{\rm Edd}$.
When late-type spirals have an X-ray nuclear source,
their luminosities are $\sim 10^{-4} L_{\rm Edd}$,
because they have a larger supply of gas.
In any case, all classes of nuclear sources in our sample
are expected to be in the radiatively-inefficient regime,
dominated either by energy advection (e.g., ADAF)
or by non-radiative output channels (jets).
We are planning further work to determine the radio core luminosity
of the nuclear X-ray sources detected in our sample.

\acknowledgments

We thank Luis Ho for stimulating discussions,
particularly during his visit to Beijing, and the anonymous referee 
for his/her insightful suggestions.
SNZ acknowledges partial funding support by the Yangtze Endowment from
the Ministry of Education at Tsinghua University, Directional
Research Project of the Chinese Academy of Sciences under project
No. KJCX2-YW-T03 and by the National Natural Science Foundation of
China under grant Nos. 10521001, 10733010, 10725313, and by 973
Program of China under grant 2009CB824800.
RS acknowledges a UK-China Fellowship for excellence
and a Leverhulme Trust Fellowship (through University College London);
he also thanks the School of Physics at the University of Sydney 
for their hospitality and support during the completion of this work.

\clearpage

\begin{deluxetable}{lrrrrrrrrrrr}
\tabletypesize{\scriptsize} \tablecaption{Galactic properties and X-ray core morphology
 \label{tbl-1}} \tablewidth{0pt} \tablehead{
\colhead{Galaxy}   &
\colhead{Distance\tablenotemark{a}}& \colhead{Morphol.}  & \colhead{Class\tablenotemark{b}}&
\colhead{X-Ray Core\tablenotemark{c}} & \colhead{Exp.~Time\tablenotemark{d}} &
\colhead{Counts\tablenotemark{e}} & \colhead{Galactic $N_{\rm H}$\tablenotemark{f}} &
 \colhead{Sample\tablenotemark{g}}\\
\colhead{}  & \colhead{(Mpc)} & \colhead{Type} &
\colhead{}& \colhead{}& \colhead{(ks)} & \colhead{} & \colhead{(cm$^{-2}$)} &}
\startdata
IC10 &0.7[T88] &IBm &H &IV  &117.1  &$<3$ &5.3E+21  &S\\
IC239  &14.2[T88] &SAB(rs)cd &L2::&IV &4.5 &$<3$ &5.3E+20  &S\\
IC342  &3.9[T88] &SAB(rs)cd &H  &I &57.8 &585 &3.0E+21&S\\
IC396  &14.4[T88] &S &\nodata &I &4.9 &24 &1.1E+21 &S\\
IC1473  &11.5[T88] &S0 &\nodata &IV &3.3 &$<3$ &6.0E+20  &S\\
IC1613  &0.7[T88] &IB(s)m &\nodata & IV &49.9 &$<3$  &3.0E+20  &S\\
IC1727  &6.4[T88] &SB(s)m &T2/L2 &IV&3.9 &$<3$  &7.2E+20 &\\
%should have been in the S sample
IC5332  &8.4[T88] &SA(s)d &\nodata &IV&107.9 &$<5$ &1.4E+20 &\\
%southern
IC2574  &2.7[T88] &SAB(s)m &H  &IV&10.1 &$<3$ &2.4E+20&S\\
IC3521  &8.2[T88] &SBm &\nodata  &IV&1.9 &$<3$ &1.5E+20&S\\
IC3647  &8.5[NED] &Im &\nodata  &IV&5.1 &$<3$ &2.0E+20 &\\
IC3773  &14.5[NED] &E &\nodata  &IV&5.3 &$<3$ &1.7E+20 &\\
%no iras
%no iras
NGC14  &12.8[T88] &IB(s)m &\nodata  &IV&4.0  &$<3$  &4.1E+20 & S\\
NGC45  &8.1[T92] &SA(s)dm &\nodata  &I&65.9 &17 &2.2E+20 &\\
%southern
NGC55  &1.5[MM] &SB(s)m &\nodata  &IV&69.8  &$<4$ &1.7E+20 &\\
%southern
NGC205 (M110) &0.7[T88] &E5 &\nodata  &IV&10.0 &$<5$ &9.0E+20 &S\\
NGC253  &3.0[T92] &SAB(s)c &T &II &14.1 &580  &1.4E+20 &\\
NGC278  &11.8[T88] &SAB(rs)b &H &III &76.4 &$<12$  &1.3E+21 &S\\
NGC404  &2.4[T88] &SA(s)0- &L2 &I &25.9 &160  &5.3E+20 &S\\
NGC598 (M33) &0.7[T88] &SA(s)cd &H &II &93.9 &140000  &5.6E+20 &S\\
NGC625  &3.9[T88] &SB(s)m &H  &IV&61.1 &$<4$  &2.2E+20 &\\
%southern
NGC628 (M74) &9.7[T88] &SA(s)c &\nodata &I &46.4 &97 &4.8E+20 &S\\
NGC660  &12.8[T88] &SB(s)a &T2/H: &III &5.0 &$<8$  &4.9E+20 &S\\
NGC672  &7.5[T88] &SB(s)cd &H &IV&2.1 &$<3$  &7.2E+20 &S\\
NGC855  &8.2[T88] &E &\nodata &IV&1.7 &$<3$  &6.4E+20 &S\\
NGC891  &9.6[T88] &SA(s)b &H &I &50.8 &7 &7.6E+20 &S\\
NGC925  &9.4[T88] &SAB(s)d &H &I &2.2 &13  &6.3E+20 &S\\
NGC949  &10.3[T88] &SA(rs)b &\nodata &IV&2.7 &$<3$  &5.1E+20 &S\\
NGC959  &10.1[T88] &Sdm &H &IV&2.2 &$<3$  &5.7E+20 &S\\
NGC1003 &10.7[T88] &SA(s)cd &\nodata &IV&2.7 &$<3$ &7.9E+20 &S\\
NGC1012  &14.4[T88] &S0/a &\nodata &I &5.1 &24 &9.0E+20 &S\\
NGC1023  &11.4[SBF] &SB(rs)0- &\nodata &I &10.3 &63 &7.2E+20 &\\
%no iras detection
NGC1023A  &9.9[NED] &IB &\nodata &IV&10.3 &$<3$ &7.2E+20 &\\
%dwarf
NGC1023D  &9.3[NED] &dwarf &\nodata &IV&10.3 &$<3$ &7.0E+20 &\\
%dwarf PGC10133
NGC1036  &11.2[T88] &peculiar &\nodata &IV&3.1 &$<4$ &8.7E+20 &S\\
NGC1055  &12.6[T88] &SBb &T2/L2:: &IV&5.0 &$<3$ &3.4E+20  &S\\
NGC1058  &9.1[T88] &SA(rs)c &S2 &I &2.4 &3 &6.7E+20 &S\\
NGC1068 (M77) &14.4[T88] &SA(rs)b &S1.8 &II &12.8 &25000 &3.5E+20 &S\\
NGC1156  &6.4[T88] &IB(s)m &H &IV&1.9 &$<3$ &1.1E+21 &S\\
NGC1291  &8.6[T88] &SB(s)0/a &\nodata &II &60.4 &1228 &2.1E+20 &\\
%southern (-41) also identified as ngc1269
NGC1313  &3.7[T88] &SB(s)d &H &IV&49.9 &$<5$ &3.9E+20 &\\
%southern
NGC1396  &10.8[NED] &SAB0- &\nodata &IV&3.6 &$<3$ &1.4E+20 &\\
%southern
NGC1493  &11.3[T88] &SB(r)cd &\nodata &I &10.1 &47 &1.4E+20 &\\
%southern
NGC1507  &10.6[T88] &SB(s)m &\nodata &IV&2.8 &$<5$ &1.0E+21 &S\\
NGC1569  &1.6[T88] &IBm &H &III &96.8 &$<11$ &2.2E+21 &S\\
NGC1637  &8.9[T88] &SAB(rs)c &\nodata &I &168.1 &450 &4.4E+20 &\\
%southern
NGC1672  &14.5[T88] &SB(s)b &S2 &II &40.1 &30 &2.3E+20 &\\
%southern
NGC1705  &6.0[T88] &SA0- &H &IV&57.6 &$<6$ &4.2E+20 &\\
%southern
NGC1800  &7.4[T88] &IB(s)m &H &IV&46.7 &$<6$ &1.6E+20 &\\
%southern
NGC1808  &10.8[T88] &SAB(s)a &S2 &II &43.4 &550 &2.7E+20 &\\
%southern
NGC2337  &8.2[T88] &IBm &\nodata &IV&1.9 &$<3$ &8.4E+20 &S\\
NGC2403  &4.2[T88] &SAB(s)cd &H &IV&36.0 &$<5$ &4.1E+20 &S\\
NGC2500  &10.1[T88] &SB(rs)d &H &I &2.6 &7 &4.7E+20 &S\\
NGC2541  &10.6[T88] &SA(s)cd &T2/H: &IV&1.9 &$<3$ &4.6E+20 &S\\
NGC2552  &10.0[T88] &SA(s)m &\nodata &IV&7.9 &$<3$ &4.4E+20 &\\
%should have been in the S sample
NGC2681  &13.3[T88] &SAB(rs)0/a &L1.9 &I &80.9 &635 &2.5E+20 &S\\
NGC2683  &5.7[T88] &SA(rs)b &L2/S2 &I &1.7 &15 &3.0E+20 &S\\
NGC2787  &13.0[T88] &SB(r)0+ &L1.9 &I &30.9 &480 &4.3E+20 &S\\
NGC2841  &12.0[T88] &SA(r)b &L2 &II &28.2 &128 &1.5E+20 &S\\
NGC3031 (M81) &3.6[T88] &SA(s)ab &S1.5 &II &49.9 &2000 &4.2E+20 &S\\
NGC3034 (M82) &5.2[T88] &I0 &H &III &33.6 &$<80$ &4.0E+20 &S\\
NGC3077  &2.1[T88] &I0 &H &I &54.1 &254 &3.9E+20 &S\\
NGC3115  &9.7[SBF] &S0- &\nodata &II &37.4 &137 &4.3E+20 &\\
%southern
NGC3125  &11.5[NED] &E &\nodata &II &57.6 &17 &5.7E+20 &\\
%southern
NGC3184  &8.7[T88] &SAB(rs)cd &H &I &65 &28 &1.1E+20 &S\\
NGC3239  &8.1[T88] &IB(s)m &\nodata &IV&1.9 &$<3$ &2.7E+20 &S\\
NGC3274  &5.9[T88] &SABd &\nodata &IV&1.7 &$<5$ &1.9E+20 &S\\
NGC3344  &6.1[T88] &SAB(r)bc &H &I &1.7 &7 &2.2E+20 &S\\
NGC3351 (M95) &8.1[T88] &SB(r)b &H &III &40.0 &$<28$ &2.9E+20 &S\\
NGC3368 (M96) &8.1[T88] &SAB(rs)ab &L2 &II &1.9 &6 &2.8E+20 &S\\
NGC3377  &11.2[SBF] &E5 &\nodata &I &40.1 &110 &2.9E+20 &\\
%no iras
NGC3379 (M105) &10.6[SBF] &E1 &L2/T2:: &II &341.1 &858 &2.8E+20 &\\
%no iras
NGC3384  &11.6[SBF] &SB(s)0- &\nodata &II &10.0 &29 &2.7E+20 &\\
%no iras
NGC3412  &11.3[SBF] &SB(s)0o &\nodata &I &10.0 &3 &2.6E+20 &\\
%no iras
NGC3413  &8.8[T88] &S0 &\nodata &IV&1.7 &$<3$ &2.0E+20 &S\\
NGC3432  &7.8[T88] &SB(s)m &H &IV&1.9 &$<3$ &1.8E+20 &S\\
NGC3486  &7.4[T88] &SAB(r)c &S2 &IV&1.7 &$<6$ &1.9E+20 &S\\
NGC3489  &12.1[SBF] &SAB(rs)0+ &T2/S2 &I &1.7 &11 &1.9E+20 &\\
%no iras
NGC3495  &12.8[T88] &Sd &H: &IV&4.1 &$<3$ &4.2E+20 &S\\
NGC3507  &11.8[T92] &SB(s)b &L2 &I &39.7 &288 &1.6E+20 &\\
%no iras
NGC3521  &7.2[T88] &SAB(rs)bc &H/L2:: &II &10.0 &24 &4.1E+20 &S\\
NGC3556 (M108) &14.1[T88] &SB(s)cd &H &II &60.1 &6 &7.9E+19 &S\\
NGC3593  &5.5[T88] &SA(s)0/a &H &I &1.9 &4 &1.8E+20 &S\\
NGC3600  &10.5[T88] &Sa &H &IV&2.6 &$<3$ &1.9E+20 &S\\
NGC3623 (M65) &7.3[T88] &SAB(rs)a &L2: &I &1.7 &8 &2.2E+20 &S\\
NGC3627 (M66) &6.6[T88] &SAB(s)b &T2/S2 &I &1.7 &9 &2.4E+20 &S\\
NGC3628  &7.7[T88] &Sb &T2 &III &58.7 &$<25$ &2.2E+20 &S\\
NGC3675  &12.8[T88] &SA(s)b &T2 &IV&1.7 &$<5$ &2.2E+20 &S\\
NGC3985  &8.3[T88] &SB(s)m &\nodata &IV&1.7 &$<5$ &2.1E+20 &S\\
NGC3998  &14.1[SBF] &SA(r)0o &L1.9 &I &15.0 &19000 &1.2E+20 &\\
%PGC35286, too faint in iras
NGC4020  &8.0[T88] &SBd &\nodata &IV&1.7 &$<3$ &1.6E+20 &S\\
NGC4026  &13.6[SBF] &S0 &\nodata &I &15.1 &24 &2.0E+20 &\\
%no iras
NGC4062  &9.7[T88] &SA(s)c &H &IV&2.2 &$<3$ &1.6E+20 &S\\
NGC4096  &8.8[T88] &SAB(rs)c &H &IV&1.7 &$<3$ &1.7E+20 &S\\
NGC4111  &15.0[SBF] &SA(r)0+ &L2 &II &15.1 &269 &1.4E+20 &\\
%no iras
NGC4136  &9.7[T88] &SAB(r)c &H &I &18.5 &18 &1.6E+20 &S\\
NGC4138  &13.8[SBF] &SA(r)0+ &S1.9 &I &6.1 &817 &1.4E+20 &\\
%no iras
NGC4150  &9.7[T88] &SA(r)0o &T2 &IV&1.7 &$<5$ &1.6E+20 &S\\
NGC4203  &9.7[T88] &SAB0- &L1.9 &I &1.7 &310 &1.2E+20 &S\\
NGC4204  &7.9[T88] &SB(s)dm &\nodata &IV&2.0 &$<3$ &2.4E+20 &S\\
NGC4207  &8.3[T88] &Scd &\nodata &IV&1.7 &$<5$ &1.8E+20 &S\\
NGC4214  &3.5[T88] &IAB(s)m &H &IV&29.0 &$<5$ &1.5E+20 &S\\
NGC4244  &3.1[T88] &SA(s)cd &H &IV&49.8 &$<5$ &1.7E+20 &S\\
NGC4245  &9.7[T88] &SB(r)0/a &H &IV&7.1 &$<3$ &1.7E+20 &S\\
NGC4258 (M106) &9.6[T88] &SAB(s)bc &S1.9 &II &21.2 &3000 &1.2E+20 &S\\
NGC4274  &9.7[T88] &SB(r)ab &H &IV&1.9 &$<3$ &1.8E+20 &S\\
NGC4286  &8.6[NED] &SA(r)0/a &\nodata &IV&37.9 &$<5$ &1.8E+20 &\\
%no iras
NGC4309  &11.9[T88] &SAB(r)0+ &\nodata &IV&3.3 &$<5$ &1.6E+20 &S\\
NGC4310  &9.7[T88] &SAB(r)0+ &\nodata &IV&2.4 &$<3$ &1.8E+20 &S\\
NGC4312  &2.1[T88] &SA(rs)ab &\nodata &IV&1.9 &$<3$ &2.5E+20 &S\\
NGC4314  &9.7[T88] &SB(rs)a &L2 &II &16.1 &17 &1.8E+20 &S\\
NGC4321 (M100) &14.1[KP] &SAB(s)bc &T2 &II &38.3 &68 &2.4E+20 &\\
%M100, should have been in the S sample
NGC4341  &12.5[NED] &SAB(s)0o &\nodata &IV&38.7 &$<3$ &1.6E+20 &\\
%IC3260, no iras
NGC4342  &10.0[NED] &S0- &\nodata &I &38.7 &171 &1.6E+20 &\\
%IC3256, no iras
NGC4343  &13.9[T88] &SA(rs)b &\nodata &I &4.7 &4 &1.6E+20 &S\\
NGC4370  &10.7[T88] &Sa &\nodata &IV&40.9 &$<5$ &1.6E+20 &S\\
NGC4395  &3.6[T88] &SA(s)m &S1.8 &I &72.9 &8000 &1.4E+20 &S\\
NGC4414  &9.7[T88] &SA(rs)c &T2: &I &1.7 &8 &1.4E+20 &S\\
NGC4419  &13.5[SBF] &SB(s)a &T2 &I &5.1 &40 &2.7E+20 &\\
%should have been in the S sample
NGC4448  &9.7[T88] &SB(r)ab &H &IV&2.0 &$<5$ &1.8E+20 &S\\
NGC4449  &3.0[T88] &IBm &H &IV&26.9 &$<9$ &1.4E+20 &S\\
NGC4471  &10.8[NED] &E &S2:: &IV&40.0 &$<6$ &1.7E+20 &\\
%PGC41185 no iras
NGC4485  &9.3[T88] &IB(s)m &H &IV&98.7 &$<3$ &1.8E+20 &\\
%no iras
NGC4490  &7.8[T88] &SB(s)d &H &IV&19.5 &$<5$ &1.8E+20 &S\\
NGC4491  &6.8[T88] &SB(s)a &\nodata &IV&2.1 &$<5$ &2.3E+20 &S\\
NGC4509  &12.8[T88] &Sab &H &IV&3.9 &$<3$ &1.5E+20 &S\\
NGC4527  &13.5[T88] &SAB(s)bc &T2 &II &4.9 &17 &1.9E+20 &\\
%should have been in the S sample
NGC4548 (M91) &15.0[KP] &SB(rs)b &L2 &I &3.0 &27 &2.4E+20 &\\
%should have been in the S sample
NGC4559  &9.7[T88] &SAB(rs)cd &H &I &11.9 &60 &1.5E+20 &S\\
NGC4561  &12.3[T88] &SB(rs)dm &\nodata &I &3.5 &92 &2.1E+20 &S\\
NGC4564  &15.0[SBF] &E &\nodata &II &18.3 &33 &2.4E+20 &\\
%no iras
NGC4565  &9.7[T88] &SA(s)b &S1.9 &I &59.2 &2100 &1.3E+20 &S\\
NGC4592  &9.6[T88] &SA(s)dm &\nodata &IV&2.1 &$<3$ &1.8E+20 &S\\
NGC4594 (M104) &9.8[SBF] &SA(s)a &L2 &II &18.7 &2700 &3.8E+20 &\\
%Sombrero, southern
NGC4618  &7.3[T88] &SB(rs)m &H &IV&9.4 &$<3$ &1.9E+20 &S\\
NGC4625  &8.2[T88] &SAB(rs)m &\nodata &IV&1.7 &$<5$ &1.5E+20 &S\\
NGC4627  &7.4[T88] &E4 &\nodata &IV&59.9 &$<6$ &1.3E+20 &S\\
NGC4631  &6.9[T88] &SB(s)d &H &IV&59.9 &$<5$ &1.3E+20 &S\\
NGC4636  &14.7[SBF] &E0 &L1.9 &II&210.0 &202 &1.8E+20 &\\
%no iras
NGC4670  &11.0[T88] &SB(s)d &\nodata &I&2.6 &12 &1.1E+20 &S\\
NGC4697  &11.8[SBF] &E6 &\nodata &II&39.7 &120 &2.1E+20 &\\
%southern
NGC4713  &10.9[T92] &SAB(rs)d &T2 &I&4.9 &10 &2.0E+20 &\\
%should have been in the S sample
NGC4725  &12.4[T88] &SAB(r)ab &S2: &I&2.4 &196 &1.0E+20 &S\\
NGC4736 (M94) &4.3[T88] &SA(r)ab &L2 &II&47.3 &90 &1.4E+20 &S\\
NGC4826 (M64) &4.1[T88] &SA(rs)ab &T2 &III&1.8 &$<9$ &2.6E+20 &S\\
NGC4945  &5.2[T88] &SB(s)cd &S2 &II&49.7 &1000 &1.6E+21 &\\
%southern
NGC5055 (M63) &7.2[T88] &SA(rs)bc &T2 &II&28.3 &200 &1.3E+20 &S\\
NGC5068  &6.7[T88] &SAB(rs)cd &\nodata &IV&28.3 &$<3$ &7.8E+20 &\\
%southern
NGC5102  &4.0[SBF] &SA0- &H &I&34.6 &14 &4.3E+20 &\\
%southern
NGC5128 (Cen A) &4.2[SBF] &S0 &S2 &II&99.5 &134000 &8.6E+20 &\\
%southern
NGC5194 (M51a) &7.7[T88] &SA(s)bc &S2 &II&14.9 &200 &1.6E+20 &S\\
NGC5195 (M51b) &7.7[T88] &I0 &L2: &II&41.7 &522 &1.6E+20 &S\\
NGC5204  &4.8[T88] &SA(s)m &H &IV&48.9 &$<4$ &1.4E+20 &S\\
NGC5236 (M83) &4.7[T88] &SAB(s)c &\nodata &II&49.5 &690 &3.8E+20 &\\
%southern
NGC5253  &3.2[KP] &peculiar &H &II&57.3 &70 &3.9E+20 &\\
%southern
NGC5457 (M101) &5.4[T88] &SAB(rs)cd &H &I&98.2 &310 &1.2E+20 &S\\
NGC5474  &6.0[T88] &SA(s)cd &H &IV&1.7 &$<3$ &1.2E+20 &S\\
NGC5585  &7.0[T88] &SAB(s)d &H &IV&5.3 &$<5$ &1.4E+20 &S\\
NGC5879  &12.3[T92] &SA(rs)bc &T2/L2 &I&90.1 &159 &1.5E+20 &\\
%should have been in the S sample
NGC5949  &11.2[T88] &SA(r)bc &\nodata &IV&3.0 &$<3$ &2.0E+20 &S\\
NGC6503  &6.1[T88] &SA(s)cd &T2/S2: &I&13.2 &15 &4.1E+20 &S\\
NGC6690  &12.2[T88] &Sd &\nodata &IV&3.5 &$<3$ &5.9E+20 &S\\
NGC6822  &0.5[MM] &IB(s)m &\nodata &IV&28.4 &$<3$ &9.5E+20 &\\
%southern
NGC6946  &5.5[T88] &SAB(rs)cd &H &II&58.3 &159 &2.1E+21 &S\\
NGC7013  &14.2[T88] &SA(r)0a &L &I&4.4 &49 &1.7E+21 &S\\
NGC7090  &6.6[T92] &SBc &\nodata &IV&57.4 &$<5$ &2.8E+20 &\\
%southern
NGC7320  &13.8[T88] &SA(s)d &H &I&19.9 &13 &8.0E+20 &S\\
NGC7331  &14.3[T88] &SA(s)b &T2 &II&30.1 &78 &8.6E+20 &S\\
NGC7424  &11.5[T88] &SAB(rs)cd &\nodata &IV&47.8 &$<6$ &1.3E+20 &\\
%southern
NGC7457  &13.2[SBF] &SA(rs)0- &\nodata &I&9.1 &10 &5.6E+20 &\\
%no iras
NGC7640  &8.6[T88] &SB(s)c &H &IV&1.9 &$<3$ &1.0E+21 &S\\
NGC7741  &12.3[T88] &SB(s)cd &H &IV&3.5 &$<3$ &4.7E+20 &S\\
NGC7793  &3.7[T92] &SA(s)d &H &IV&49.5 &$<6$ &1.2E+20 &\\
PGC3589  &0.08[MM] &E &\nodata &IV &6.1 &$<3$ & 2.0E+20  &\\
%SCULPTOR DWARF ELLIPTICAL, southern
PGC13449 &11.0[NED] &SAB(s)0o &\nodata &IV&44.1 &$<5$ &1.3E+20 &\\
%southern
PGC13452  &12.0[NED] &E0 &\nodata &IV&63.7 &$<5$ &1.4E+20 &\\
%southern
PGC16744  &9.1[NED] &SB0 &\nodata &IV&46.7 &$<3$ &1.6E+20 &\\
%southern
PGC24175  &10.5[T88] &I0 &\nodata &II&20.0 &188 &9.8E+20 &\\
%southern
PGC40512  &10.3[NED] &E &\nodata &IV&40.2 &$<3$ &2.5E+20 &\\
%no iras
PGC46093  &5.2[NED] &Im &\nodata &IV&28.3 &$<3$ &1.3E+20 &\\
%no iras
PGC50779 (Circinus) &4.2[T88] &SA(s)b &S2 &II&24.6 &6600 &5.6E+21 &\\
%southern
UGC2126  &9.5[NED] &SABdm &\nodata &IV&2.7 &$<4$  &7.8E+20 &\\
%PGC10025, no iras
UGC5336 (Ho IX) &3.42[KP] &Im &\nodata &IV&5.1 &$<3$ &4.1E+20 &\\
%PGC28757 = Holmberg IX, southern
UGC6456  &1.4[T88] &peculiar &\nodata &IV&10.6 &$<3$ &3.8E+20 &\\
%PGC35286, too faint in iras
UGC7636  &3.7[NED] &Im &\nodata &IV&10.4 &$<3$ &1.7E+20 &\\
%PGC41258 no iras
UGC8041  &14.2[T88] &SB(s)d &\nodata &IV&4.7 &$<3$ &1.6E+20 &S\\
%PGC44014
UGC11466  &11.2[T88] &Sab &\nodata &IV&2.7 &$<5$ &1.3E+21 &S\\
%PGC63552

\enddata

%% Text for table notes should follow after the \enddata but before
%% the \end{deluxetable}. Make sure there is at least one \tablenotemark
%% in the table for each \tablenotetext.

\tablenotetext{a}{References for the distances are given in brackets,
as follows: KP = Key Project \citep{fre01};
SBF = surface brightness fluctuations \citep{ton01};
T92 = nearby galaxy flow model \citep{tul92};
T88 = Nearby Galaxy Catalog \citep{tul88};
NED = distances computed from recessional velocities
relative to the cosmic microwave background.}
\tablenotetext{b}{Optical classification
of the nuclear spectrum, from \citet{hfs97a} and NED:
H = H{\tiny II} nucleus;
S = Seyfert;
L = LINER;
T = transition object.
The number attached to the class letter designates
the type (1.0, 1.2, 1.5, 1.8, 1.9, and 2); quality ratings
are given by ":" and "::" for uncertain and highly uncertain
classifications, respectively.}
\tablenotetext{c}{X-ray classification
of the nuclear region, from our study:
I = dominant point-like X-ray nucleus;
II = point-like nuclear source embedded in diffuse emission;
III = diffuse X-ray emission in the nuclear region without
a point-like core;
IV = no detectable X-ray emission at the nuclear position. See examples
in Figure 1.}
\tablenotetext{d}{{\it Chandra}/ACIS exposure time for the datasets used
in our analysis.}
\tablenotetext{e}{{\it Chandra}/ACIS X-ray counts or upper limits
for a point-like nuclear X-ray source; upper limits are at the
$95$\% confidence level, estimated using a Bayesian method \citep{kbn91}.}
\tablenotetext{f}{Line-of-sight
column density, from \citet{dil90}.}
\tablenotetext{g}{S denotes galaxies included in the optically/IR-selected
sample (see Section 2).}

\end{deluxetable}

\begin{deluxetable}{lrrrrrrrrrrrrr}
\tabletypesize{\scriptsize} \tablecolumns{14} \tablewidth{0pc}
\tablecaption{Fraction of X-ray core detections for different classes of galaxies}
\tablehead{ \colhead{Sample} & \multicolumn{4}{c}{Morphological
Class} & \colhead{} & \multicolumn{3}{c}{Bar
Structure Class\tablenotemark{a}}&
\colhead{}   & \multicolumn{4}{c}{Spectroscopic Class\tablenotemark{b}}\\
\cline{2-5} \cline{7-9} \cline{11-14}\\
\colhead{} & \colhead{E} & \colhead{S0--Sb}  & \colhead{Sc--Sm}& \colhead{Irr/pec}&
\colhead{} & \colhead{SA} & \colhead{SAB}   & \colhead{SB}  &
\colhead{}    & \colhead{S} & \colhead{L}& \colhead{T}& \colhead{H}}
\startdata
Optical/IR & 0/3 & 27/50 & 18/50 & 2/13 & & 21/34 & 16/27 & 6/26 & & 7/9 & 12/13 & 5/12 & 16/46\\
Extended & 6/14 & 51/80 & 25/68 & 4/25 & & 30/48 & 24/41 & 16/42 & & 14/16 & 19/20 & 12/20 & 18/54\\
\enddata
%\tablenotetext{a}{Early denotes the a and b type spiral galaxies;
%Late denotes the c, d, and m type spiral galaxies.}
\tablenotetext{a}{SA: non-barred spirals; SAB: weakly barred spirals; SB: strongly barred spirals.}\\[-5pt]
\tablenotetext{b}{S: Seyferts; L: LINERs; T: transition objects; H: H{\tiny II} nuclei.}
%\tablecomments{The numbers in the table denotes X-ray core
%detection numbers and the total subclass source numbers, e.g.,
%21/67 in the first column means there are 21 galaxies have X-ray
%nuclear emission in all 67 early spiral galaxies.}
\end{deluxetable}

\begin{deluxetable}{lrrrrrrr}
\tabletypesize{\scriptsize} \tablecolumns{8} \tablewidth{0pc}
\tablecaption{Dependence of nuclear X-ray detections on the bar class}
\tablehead{ \colhead{Sample} & \multicolumn{3}{c}{Early-type spirals (S0--Sb)}
& \colhead{} & \multicolumn{3}{c}{Late-type spirals (Sc--Sm)}\\
\cline{2-4} \cline{6-8} \\
\colhead{~} & \colhead{SA} & \colhead{SAB}  & \colhead{SB} &
\colhead{~} & \colhead{SA} & \colhead{SAB}   & \colhead{SB}}
\startdata
Optical/IR & 14/20 & 9/12 & 2/9 & ~ & 7/14 & 7/15 & 4/17\\
Extended & 22/30 & 13/19 & 10/18 & ~ & 8/18 & 11/22 & 6/24\\
\enddata
\end{deluxetable}

\begin{deluxetable}{lrrrrrrrrrrr}
\tabletypesize{\scriptsize} \tablecaption{Galactic nuclei with point-like X-ray emission\label{tbl-2}} \tablewidth{0pt}
\tablehead{ \colhead{Galaxy} & \colhead{$L_{\rm 0.3-8}$\tablenotemark{a}} &
\colhead{$N_{\rm H}$\tablenotemark{b}} & \colhead{Model\tablenotemark{c}} &
\colhead{$i$\tablenotemark{d}}& \colhead{$M_{\rm BH}$} &
\colhead{Method\tablenotemark{e}}& \colhead{$r_{\rm Edd}$\tablenotemark{f}} &
\colhead{log($\dot{m}$)\tablenotemark{g}} \\
\colhead{} & \colhead{($10^{39}$ $\rm ergs$ $\rm s^{-1}$)} &
\colhead{($10^{22}$ cm$^{-2}$)}& \colhead{}&
\colhead{(\degr)} & \colhead{($10^{7}$ $M_{\sun}$)} &
\colhead{} & \colhead{} } \startdata
IC342   &   0.37   &   $0.09^{+0.03}_{-0.05}$  &ARP  &   20  &   0.25   &MS &   1.2E$-$06    &   $-3.3$   &  \\[4pt]
IC396  &   2.5    &   $0.35^{+0.61}_{-0.35}$   &AP &  47  &   \nodata &\nodata &  \nodata &   \nodata &  \\[4pt]
NGC45   &   0.021 &   $<0.6$  &AP &   47  &   \nodata &\nodata &   \nodata &   \nodata &  \\[4pt]
NGC253  &   2.0   &   $20^{+13}_{-9}$  &[1] &   86  &   0.94   &MS &   1.6E$-$06    &   $-3.2$   &  \\[4pt]
NGC404  &   0.031    &   $<0.11$  &AP &   0   &   0.06    &MS &  4.0E$-$07    &   $-3.8$   &  \\[4pt]
NGC598 (M33)  &   0.8   &   $0.45^{+0.01}_{-0.01}$    &AP,[2] &  56  &   $< 1.5 \times 10^{-4}$    &S,[3] & $>0.004$ & $> -1.4$   &   \\[4pt]
NGC628 (M74)  &   0.18    &   $<0.06$   &AP &  0   &   0.5    &BB,[4]  & 2.7E$-$07    &   $-3.7$   &    \\[4pt]
NGC891  &   0.0045    &   \nodata &P &  84  &   0.23 &MS &  1.5E$-$08 &   $-4.2$ &   \\[4pt]
NGC925 &   0.59    &   $<0.2$  &AP &   54  &   \nodata &\nodata &  \nodata &   \nodata &  \\[4pt]
NGC1012 &   165    &   $44^{+82}_{-21}$    &AP &  61  &   \nodata &\nodata &  \nodata &   \nodata &  \\[4pt]
NGC1023 &   0.73    &   $<0.11$  &AP &   72  &   4.4  &S,[5]  &   1.3E$-$07    &   $-3.9$   &    \\[4pt]
NGC1058 &   0.055 &   \nodata &P &  16  &   0.012   &MS &   3.4E$-$06    &   $-2.7$   &  \\[4pt]
NGC1068 (M77) &   300    &  $\la 0.03$   &[6] &  29  &   1.50    &M,[7] &  1.6E$-$04    &   $-2.0$   &   \\[4pt]
NGC1291 &   2.0   &   $2.0^{+0.7}_{-0.7}$   &AP,[8]  &   28  &   7.4  &MS  &   2.1E$-$07    &   $-3.7$   &   \\[4pt]
NGC1493 &   0.54    &  $0.03^{+0.22}_{-0.03}$  &AP  &   0   &   \nodata &\nodata &  \nodata &   \nodata &   \\[4pt]
NGC1637 &   0.12 &  $0.56^{+0.10}_{-0.08}$   &[9] &   39  &   \nodata &\nodata &  \nodata &   \nodata &   \\[4pt]
NGC1672 &   1.0   &   $10^{+30}_{-10}$    &[10] &  37  &   \nodata &\nodata &  \nodata &   \nodata &   \\[4pt]
NGC1808 &   12 &   $3.1^{+0.8}_{-0.7}$  &[11]  &   50  &   4.1  &MS  &   1.7E$-$06    &   $-3.2$   &   \\[4pt]
NGC2500 &   0.35    &   \nodata &P &  0   &   \nodata &\nodata &  \nodata &   \nodata &  \\[4pt]
NGC2681 &   0.61   &   $0.04^{+0.04}_{-0.03}$    &ARP &  0   &   1.2   &MS &   4.0E$-$07    &   $-3.5$   &  \\[4pt]
NGC2683 &   9.0  &   $13^{+28}_{-13}$   &AHP &   79  &   1.5   &MS &   4.6E$-$06    &   $-2.9$   &   \\[4pt]
NGC2787 &   3.2 &   $0.13^{+0.06}_{-0.06}$    &AP &  52  &   4.1    &G,[12] &  6.0E$-$07    &   $-3.4$   &  \\[4pt]
NGC2841 &   0.89  &  $0.11^{+0.10}_{-0.09}$  &AP  &   64  &   6.3   &BB,[13] &   1.1E$-$07    &   $-3.9$   &  \\[4pt]
NGC3031 (M81) &  100 & $0.09^{+0.02}_{-0.02}$   &AP,[14] &   60  &   6.8   &BB,[15] &   1.1E$-$05    &   $-2.7$   &   \\[4pt]
NGC3077 &   0.060  &  $1.6^{+0.8}_{-0.6}$  &AP  &   43  &   \nodata &\nodata &  \nodata &   \nodata &    \\[4pt]
NGC3115 &   0.28    &  $0.01^{+0.09}_{-0.01}$   &AP &  66  &   92   &S,[16] &   2.4E$-$09    &   $-4.4$   &   \\[4pt]
NGC3125 &   0.04    &   $<0.8$   &AP &  \nodata &   \nodata &\nodata &  \nodata &   \nodata &    \\[4pt]
NGC3184 &   0.02 &  $0.21^{+0.34}_{-0.21}$   &AP &  26  &   \nodata &\nodata &  \nodata &   \nodata &   \\[4pt]
NGC3344 &   0.36 &   \nodata &P &  23  &   0.16    &BB,[4] &  1.8E$-$06    &   $-3.1$   &   \\[4pt]
NGC3368 (M96) &   0.19    &   \nodata &P &  50  &   3.2    &BB,[4] &  4.5E$-$08    &   $-4.1$   &    \\[4pt]
NGC3377 &   0.39 & $0.17^{+0.15}_{-0.13}$   &AP &   54  &   10    &S,[17,18] &  3.0E$-$08    &   $-4.1$   &    \\[4pt]
NGC3379 (M105) & 0.28 & $0.03^{+0.1}_{-0.03}$  &AP &   25  &   13.5    &S,[19] &  1.4E$-$08    &   $-4.2$   &   \\[4pt]
NGC3384 &   0.55 & $0.12^{+0.55}_{-0.12}$ &AP &   65  &   1.6   &S,[17,20] &   2.4E$-$07    &   $-3.7$   &    \\[4pt]
NGC3412 &   0.044    &   \nodata &P &  59  &   0.87   &MS &   4.0E$-$08    &   $-4.1$   &    \\[4pt]
NGC3489 &   0.66    &  $<0.5$ &AP &  56  &   1.0  &MS  &   5.0E$-$07    &   $-3.5$   &    \\[4pt]
NGC3507 &   0.22   &  $0.04^{+0.17}_{-0.04}$   &ARP &  37  &   0.79  &MN  &   2.2E$-$07    &   $-3.7$   &   \\[4pt]
NGC3521 &   0.15 &  $0.13^{+0.30}_{-0.11}$   &AP &  61  &   0.26    &MN &  4.4E$-$07    &   $-3.5$   &  \\[4pt]
NGC3556 (M108) &   0.023    &   \nodata &P &  81  &   \nodata &\nodata &  \nodata &   \nodata &  \\[4pt]
NGC3593 &   0.21 &   \nodata &P &  69  &   0.63    &BB,[4] &  2.6E$-$07    &   $-3.7$   &   \\[4pt]
NGC3623 (M65) &   0.22    &   \nodata &P &  81  &   1.3   &BB,[4] &   1.4E$-$07    &   $-3.8$   &   \\[4pt]
NGC3627 (M66) &   0.912    &   \nodata   &P &   65  &   1.3 &BB,[4] &  5.6E$-$07 &   $-3.5$ &    \\[4pt]
NGC3998 &   280    &  $<0.01$   &AP &  36  &   60   &S,[21] &   3.6E$-$06    &   $-2.7$   &    \\[4pt]
%NGC4026 &   0.4    &  $0.20^{+0.41}_{-0.20}$  &AP  &   83  &   8.25   &MS &   3.8E$-$08    &   $-4.09$   &   \\[4pt]
NGC4026 &   0.4    &  $0.20^{+0.41}_{-0.20}$  &AP  &   83  &   21   &S,[22] &   1.5E$-$08    &   $-4.2$   &   \\[4pt]
NGC4111 &   7.6  &  $5.7^{+1.7}_{-1.6}$  &ARP  &   87  &   4.0   &MS &   1.4E$-$06    &   $-3.2$   &    \\[4pt]
NGC4136 &   0.15 &  $0.31^{+0.52}_{-0.31}$    &AP &  0   &   0.04    &BB,[4] &  2.9E$-$06    &   $-3.0$   &    \\[4pt]
NGC4138 &   200    & $7.7^{+1.9}_{-1.6}$   &AP &   58  &   3.2   &MS &   4.9E$-$05    &   $-2.2$   &   \\[4pt]
NGC4203 &   18 &   $1.4^{+1.2}_{-0.8}$    &AHRP &  26  &   5.8    &MS &  2.4E$-$06    &   $-3.1$   &   \\[4pt]
NGC4258 (M106) & 31 & $4.1^{+0.6}_{-0.5}$   &AP &   71  &   3.9    &M,[23] &  6.1E$-$06    &   $-2.7$   &   \\[4pt]
NGC4314 &   0.068    &  $0.19^{+0.65}_{-0.19}$   &AP &  15  &   1.0    &BB,[4] &  5.2E$-$08    &   $-4.1$   &   \\[4pt]
NGC4321 (M100) &   0.15 & $<1.5$   &ARP &  37  &   0.45    &MS &  2.6E$-$07    &   $-3.7$   &     \\[4pt]
NGC4342 &   0.48  &  $0.10^{+0.11}_{-0.10}$  &AP  &   58  &   33  &S,[13]  &   1.1E$-$08    &   $-4.2$   &   \\[4pt]
NGC4343 &   0.25    &   \nodata &P &  \nodata &   \nodata &\nodata &  \nodata &   \nodata &  \\[4pt]
NGC4395 &   5.2  & $9.2^{+0.8}_{-0.8}$   &AHP &   38  &   0.036    &R,[24] &  1.3E$-$04    &   $-2.0$   &   \\[4pt]
NGC4414 &   0.21  &   \nodata &P &  50  &   1.2 &MS &  1.4E$-$07 &  $-3.8$ &   \\[4pt]
NGC4419 &   1.8    &  $<0.5$   &AP &  75  &   0.8    &MS &  1.7E$-$06    &   $-3.2$   &    \\[4pt]
NGC4527 &   0.78    &  $<0.7$   &AP &  68  &   16.7    &MS &  3.6E$-$08    &   $-4.1$   &   \\[4pt]
NGC4548 (M91) & 8.4 &  $2.3^{+6.8}_{-2.3}$   &AP &   37  &   3.6  &MS  &   1.8E$-$06    &   $-3.1$   &   \\[4pt]
NGC4559 &   1.2   & $0.31^{+0.45}_{-0.31}$    &AP &  69  &   \nodata &\nodata &  \nodata &   \nodata &   \\[4pt]
NGC4561 &   4.3    &  $<0.05$   &AP &  25  &   \nodata &\nodata &  \nodata &   \nodata & \\[4pt]
NGC4564 &   0.48 & $0.14^{+0.37}_{-0.14}$   &AP &   62  &   5.6  &S,[17,20]  &   6.4E$-$08    &   $-4.0$   &   \\[4pt]
NGC4565 &   4.5   &  $0.23^{+0.02}_{-0.03}$  &AP  &   90  &   2.9   &MS &   1.2E$-$06    &   $-3.2$   &  \\[4pt]
NGC4594 (M104) &  17 & $0.18^{+0.03}_{-0.03}$  &AP  &   79  &   100   &BB,[4] &   1.3E$-$07    &   $-3.8$   &  \\[4pt]
NGC4636 &   0.21 & $<0.04$   &[25] &  44  &   7.9    &MS,[15] &  2.0E$-$08    &   $-4.2$   &   \\[4pt]
NGC4670 &   0.79  & $0.09^{+0.49}_{-0.09}$   &AP &  31  &   \nodata &\nodata &  \nodata &   \nodata &  \\[4pt]
NGC4697 &   0.03 & $<0.06$  &[26]  &   44  &   17    &S,[17,20] &  1.0E$-$09    &   $-4.6$   &  \\[4pt]
NGC4713 &   0.18    &   \nodata &P  &  53  &   \nodata &\nodata &  \nodata &   \nodata &  \\[4pt]
NGC4725 &   0.54   & $1.6^{+0.4}_{-0.4}$  &AHBP  &   43  &   3.4  &MS  &   1.2E$-$07    &   $-3.9 $  &  \\[4pt]
NGC4736 (M94) & 1.0 & $0.27^{+0.39}_{-0.20}$  &AP, [27]  &   33  &   2.2    &MS &  3.5E$-$07  &   $-3.7$   &   \\[4pt]
NGC4945 &   20000  & $425^{+25}_{-25}$   &[28] &   90  &   0.14    &M,[29] &  0.11    &   $\sim 10$   &  \\[4pt]
NGC5055 (M63) & 0.34 & $0.08^{+0.05}_{-0.08}$   &AP &   55  &   0.87   &MS &   3.0E$-$07    &   $-3.6$   &    \\[4pt]
NGC5102 &   0.006   &  $<1.5$ &AP &   71  &   3.0    &MS &  1.6E$-$09    &   $-4.5$   &   \\[4pt]
NGC5128 (Cen A) &  600 & $10.0^{+0.6}_{-0.6}$   &AP,[30] &   43  &   20    &G,[31] &  1.9E$-$05    &   $-2.5 $  &   \\[4pt]
NGC5194 (M51a) &  200 & $560^{+400}_{-160}$   &[32,33] &   64  &   0.71  &MS  &   2.2E$-$04    &   $-2.1$   &    \\[4pt]
NGC5195 (M51b) &  0.8 & $0.11^{+0.08}_{-0.09}$ &AP,[34] &  46  &   3.9    &MS &  1.6E$-$07    &   $-3.9$   &     \\[4pt]
NGC5236 (M83) &   0.26 & $0.10^{+0.14}_{-0.06}$   &AP,[35] &   24  &   1.3    &S,[36] &  1.5E$-$07    &   $-3.8$   &   \\[4pt]
NGC5253 &   0.1  & $0.6^{+0.1}_{-0.1}$  &[37]   &   77  &   \nodata &\nodata &  \nodata &   \nodata &   \\[4pt]
NGC5457 (M101) &   0.1  & $0.03^{+0.08}_{-0.03}$   &AP,[38] &  0   &   0.24    &MS &  3.2E$-$07    &   $-3.8$   &   \\[4pt]
NGC5879 &   24  & $18^{+18}_{-13}$  &ARP  &   73  &   0.25  &MS  &   7.4E$-$06    &   -$2.9$   &  \\[4pt]
NGC6503 &   0.086  &  $0.5^{+1.7}_{-0.5}$  &AP  &   74  &   0.037   &MS &   1.8E$-$06    &   $-3.1$   &   \\[4pt]
NGC6946 &   0.1   &  $0.06^{+0.10}_{-0.06}$   &AP &  42  &   2.7    &MN &  2.9E$-$08    &   $-4.1$   &   \\[4pt]
%\enddata
%\end{deluxetable}
%\begin{deluxetable}{lrrrrrrrrrrr}
%\tabletypesize{\scriptsize} \tablecaption{Continues Table 4}
%\startdata
NGC7013 &   11.4 &  $6.5^{+3.1}_{-2.6}$    &AHP &  76  &   0.54  &MS  &   1.6E$-$05    &   $-2.5$   &  \\[4pt]
NGC7320 &   0.17    &  $0.22^{+0.71}_{-0.22}$  &AP  &   60  &   \nodata &\nodata &  \nodata &   \nodata &  \\[4pt]
NGC7331 &   0.52    &  $0.09^{+0.21}_{-0.09}$  &AP  &   68  &   3.0  &MS  &   1.4E$-$07    &   $-3.8$   &  \\[4pt]
NGC7457 &   0.18    &   \nodata &P &  56  &   0.35    &S,[17,20] &  4.0E$-$07    &   $-3.5$   &   \\[4pt]
PGC24175    &   0.87  &  $0.04^{+0.10}_{-0.04}$  &AP &   32  &   \nodata &\nodata &  \nodata &   \nodata &  \\[4pt]
PGC50779    & 20 & $0.44^{+0.47}_{-0.21}$  &[39]  &   65  &   0.17  &M,[40]  &   9.1E$-$05    &   $-2.1$   &   \\[4pt]
\enddata
\tablenotetext{a}{Emitted luminosity in the $0.3$--$8$ keV band,
inferred from the best-fitting spectral model.}
\tablenotetext{b}{Intrinsic neutral-hydrogen column density,
inferred from the best-fitting spectral model. For sources
with $< 10$ net counts, such estimates are not meaningful,
and no value is listed; the luminosity of those sources is estimated
assuming Galactic line-of-sight absorption only.}
\tablenotetext{c}{Our spectral models are coded as follows.
P: power-law with Galactic absorption;
AP: power-law with Galactic and intrinsic absorption;
ARP: two-component model, with an absorbed optically-thin
thermal plasma and an absorbed power-law component;
AHP: power-law absorbed by both neutral hydrogen and
an ionized absorber;
AHRP: two-component model with an optically-thin
thermal plasma component and a power-law component,
absorbed by both neutral hydrogen and an ionized absorber.
AHBP: two-component model with a disk-blackbody and
a power-law component, absorbed by both
neutral hydrogen and an ionized absorber. When a reference
number is given, we used spectral-analysis results
from the literature.}
%weet02:  \citet{weet02}
%isb02: \citet{isb02}. imet03: \citet{imet03} jen08: \citet{jen08}
%baet05: \citet{baet05} poet06: \citet{poet06} wth08: \citet{wth08}
%doet03: \citet{doet03} evet04: \citet{evet04} sw02: \citet{sw02}
%suet04: \citet{suet04} sw01: \citet{sw01} tw04: \citet{tw04}
%swet03: \citet{swet03} psm01: \citet{psm01} yws01: \citet{yws01}
%plet08: \citet{plet08} peet02: \citet{peet02} tw01: \citet{tw01}
%fuet01: \citet{fuet01}}
\tablenotetext{d}{Inclination angle of the host galaxy, from \citet{dev91}}
\tablenotetext{e}{Methods used for estimating the nuclear
BH masses (assuming that late-type galaxies contain BHs), as follows.
S: stellar kinematics;
%, using axisymmetric dynamical models with three-integral distribution
%functions;
G: gas kinematics;
M: kinematics of water-maser clumps;
R: reverberation mapping;
MS: $M$--$\sigma$ relation \citep{teet02};
MN: $M$--Sersi\'{c} index relation \citep{gra07};
BB: BH mass--bulge mass relation \citep{dr06}.}
\tablenotetext{f}{X-ray Eddington ratio, defined as $L_{0.3-8}/L_{\rm Edd}$.}
\tablenotetext{g}{Accretion parameter required for the inferred
X-ray luminosity, assuming the radiatively-inefficient ADAF model; 
we used the grid of ADAF solutions plotted in \citet{mel03}. 
From our definition of $\dot{m} \equiv \dot{M}c^2/L_{\rm Edd}$, 
the Eddington luminosity corresponds to $\dot{m} \sim 10$.}

\tablerefs{
[1]: \citet{weet02}; [2]: \citet{plet08}; [3]: \citet{geb01}; [4]: \citet{dr06};
[5]: \citet{boet01}; [6]: \citet{yws01};  [7]: \citet{gg97};
[8]: \citet{isb02};  [9]: \citet{imet03}; [10]: \citet{jen08};
[11]: \citet{baet05}; [12]: \citet{saret01}; [13]: \citet{cv99};
[14]: \citet{swet03}; [15]: \citet{mf01a}; [16]: \citet{edb99};
[17]: \citet{gebet03}; [18]: \citet{koet98}; [19]: \citet{geb00b};
[20]: \citet{pinet03}; [21]: \citet{boet00}; [22]: \citet{gul09};
[23]: \citet{heet99};
[24]: \citet{petet05}; [25]: \citet{poet06}; [26]: \citet{wth08};
[27]: \citet{peet02}; [28]: \citet{doet03}; [29]: \citet{gmh97};
[30]: \citet{evet04}; [31]: \citet{ttg00}; [32]: \citet{tw01};
[33]: \citet{fuet01}; [34]: \citet{tw04}; [35]: \citet{sw02};
[36]: \citet{maret01}; [37]: \citet{suet04}; [38]: \citet{psm01};
[39]: \citet{sw01}; [40]: \citet{greet03}}
%(1) \citet{boet01}  (2) \citet{edb99} (3)
%\citet{gebet02} (4) \citet{koet98} (5) \citet{geb00b} (6)
%\citet{pinet02} (7) \citet{boet00} (8) \citet{cv99} (9)
%\citet{dr06} (10) \citet{mf01a} (11) \citet{gmh97}
%(12)\citet{ttg00} (13) \citet{maret01} (14) \citet{greet03} (15)
%\citet{petet05}(16) \citet{heet99} (17) \citet{saret01} (18)
%\citet{mf01a} (19) \citet{gg97}}
\end{deluxetable}


\begin{thebibliography}{}
%\bibitem[Anderson \& Ulvestad(2005)]{au05} Anderson, J. M.,
%\& Ulvestad, J. S. 2005, \apj, 627, 674
%\bibitem[Anderson et al.(2004)]{auh04} Anderson, J. M.,
%Ulvestad, J. S., \& Ho, L. C. 2004, \apj, 603, 42
\bibitem[Antonucci(1993)]{ant93} Antonucci, R. 1993, ARA\&A, 31, 473
\bibitem[Arnaud(1996)]{arn96} Arnaud, K. A. 1996, 
Astronomical Data Analysis Software and Systems V, ASP Conference Series, 
Vol. 101, G. H. Jacoby and J. Barnes, eds., p. 17
\bibitem[Barth et al.(2005)]{bgh05} Barth, A. J., Greene,
J. E., \& Ho, L. C. 2005, \apj, 619, L151
%\bibitem[Barth et al.(2004)]{baet04} Barth, A. J., Ho, L. C.,
%Rutledge, R. E., \& Sargent, W. L. W. 2004, \apj, 607, 90
%\bibitem[Beichman et al.(1988)]{bei88} Beichman, C. A., Neugebauer, G., Habing, H. J., Clegg, P. E., 
%\& Chester, T. J. 1988, NASA RP-1190 (Washington: GPO)
\bibitem[Bower et al.(2000)]{boet00} Bower, G. A., Wilson, A. S.,
Heckman, T. M., Magorrian, J., Gebhardt, K., Richstone, D. O.,
Peterson, B. M., \& Green, R. F. 2000, \baas, 32, 1566
\bibitem[Bower et al.(2001)]{boet01} Bower, G. A., et al. 2001,
\apj, 550, 75
\bibitem[Cappi et al.(2006)]{caet06} Cappi, M., et al. 2006, \aap, 446, 459
\bibitem[Cash(1979)]{cas79} Cash, W. 1979, ApJ, 228, 939
\bibitem[Combes \& Elmegreen(1993)]{com93} Combes, F., \& Elmegreen, B. G. 1993, A\&A, 271, 391
\bibitem[Cretton \& van den Bosch(1999)]{cv99} Cretton, N., \& van
den Bosch, F. C. 1999, \apj, 514, 704
\bibitem[de Vaucouleurs et al.(1991)]{dev91} de Vaucouleurs, G., 
de Vaucouleurs, A., Corwin, H. G., Buta, R. J.,
     Paturel, G., \& Fouque, P. 1991, Third Reference Catalogue of Bright Galaxies 
    (Springer-Verlag: New York)
\bibitem[Desroches \& Ho(2009)]{des09} Desroches, L.-B., \& Ho, L. C. 2009, ApJ, 690, 267
\bibitem[Dickey \& Lockman(1990)]{dil90} Dickey, J. M., \&
Lockman, F. J. 1990, \araa, 28, 215
%\bibitem[Djorgovski \& King(1984)]{djo84} Djorgovski, S.,
%    \& King, I. R. 1984, \apjl, 277, L49
%\bibitem[Doi et al.(2005)]{doet05} Doi, A., Kameno, S., Kohno, K.,
Nakanishi, K., \& Inous, M. 2005, \mnras, 363, 692
\bibitem[Done et al.(2003)]{doet03} Done, C., Madejski, G. M.,
Zycki, P. T., \& Greenhill, L. J. 2003, \apj, 588, 763
\bibitem[Dong et al.(2007)]{doet07} Dong, X. B., et al. 2007. \apj, 657, 700
\bibitem[Dong \& De Robertis(2006)]{dr06} Dong, X. Y., \& De
Robertis, M. M. 2006, \apj, 131, 1236
\bibitem[Dubus, Charles \& Long(2004)]{dcl04} Dubus, G., Charles, P. A., \& Long, K. S.
2004, A\&A, 425, 95
\bibitem[Dudik et al.(2005)]{duet05} Dudik, R. P., Satyapal, S.,
Gliozzi, M., \& Sambruna, R. M. 2005, \apj, 620, 113
\bibitem[Elvis(2000)]{elv00} Elvis, M. 2000, ApJ, 545, 63
\bibitem[Elvis et al.(2004)]{elv04} Elvis, M., Risaliti, G., Nicastro, F.,
Miller, J. M., Fiore, F., \& Puccetti, S. 2004, ApJ, 615, L25
\bibitem[Emsellem et al.(1999)]{edb99} Emsellem, E.,
Dejonghe, H., \& Bacon, R. 1999 \mnras, 303, 495
\bibitem[Esin et al.(1997)]{esi97} Esin, A. A., McClintock, J. E., 
\& Narayan, R. 1997, ApJ, 489, 865
\bibitem[Evans et al.(2004)]{evet04} Evans, D. A., Kraft, R. P.,
Worrall, D. M., Hardcastle, M. J., Jones, C., Forman, W. R., \&
Murray, S. S. 2004, \apj, 612, 786
%\bibitem[Falcke et al.(2000)]{faet00} Falcke, H., Nagar, N. M.,
%Wilson, A. S., \& Ulvestad, J. S. 2000, \apj, 542, 197
\bibitem[Fender, Belloni \& Gallo(2004)]{fen04} Fender, R. P., Belloni, T. M.,
\& Gallo, E. 2004, MNRAS, 355, 1105
%\bibitem[Ferrarese \& Merritt(2000)]{fm00} Ferrarese, L.,
%\& Merritt, D. 2000, \apj, 9, 539
\bibitem[Ferrarese et al.(2006)]{feet06} Ferrarese, L., et al.
2006, \apj, 644, L21
\bibitem[Filho, Barthel \& Ho(2006)]{fbh06} Filho, M. E., Barthel, P. D.,
\& Ho, L. C. 2006, A\&A, 451, 71
\bibitem[Filippenko \& Ho(2003)]{fh03} Filippenko, A. V., \& Ho, L. C.
2003, \apj, 588, L13
%\bibitem[Filippenko \& Sargent(1985)]{fs85} Filippenko, A. V., \& Sargent, W. L. W.
%1985, \apjs, 57, 503
%\bibitem[Fiore et al.(2001)]{fiet01} Fiore, F., et al. 2001, \apj, 556,
%150
\bibitem[Flohic et al.(2006)]{flet06} Flohic, H. M. L. G.,
Eracleous, M., Chartas, G., Shields, J. C., \& Moran, E. C. 2006,
\apj, 647, 140
\bibitem[Freedman et al.(2001)]{fre01} Freedman, W., et al. 2001, \apj, 553, 47
\bibitem[Fukazawa et al.(2001)]{fuet01} Fukazawa, Y., Iyomoto, N.,
Kubota, A., Matsumoto, Y., \& Makishima, K. 2001, \aap, 374, 73
%\bibitem[Gallimore et al.(2006)]{gaet06} Gallimore, J. F., Axon, D.
%J., O$\arcmin$Dea, C. P., Baum, S. A., \& Pedlar, A. 2006, \aj,
%132, 546
%\bibitem[Gebhardt et al.(2000a)]{geb00a} Gebhardt, K., et al. 2000a, \apj, 539, L13
\bibitem[Fullmer \& Londsdale(1989)]{ful89} Fullmer, L. \& Londsdale, C. J. 1989, 
Cataloged galaxies and quasars observed in the IRAS survey 
(Pasadena: Jet Propulsion Laboratory).
\bibitem[Garcia et al.(2005)]{gar05} Garcia, M. R., Williams, B. F., Yuan, F., 
Kong, A. K. H., Primini, F. A., Barmby, P., Kaaret, P., \& Murray, S. S. 2005, ApJ, 632, 1042 
\bibitem[Gebhardt et al.(2000b)]{geb00b} Gebhardt, K., et al. 2000b, \aj, 119,
1157
\bibitem[Gebhardt et al.(2001)]{geb01} Gebhardt, K., et al. 2001, \aj, 122, 2469
\bibitem[Gebhardt et al.(2003)]{gebet03} Gebhardt, K., et al.
2003, \apj, 583, 92
%\bibitem[Georgantopoulos et al.(2002)]{geet02} Georgantopoulos, I.,
%Panessa, F., Akylas, A., Zezas, A., Cappi, M., \& Comastri, A.
%2002, \aap, 386, 60
%\bibitem[George \& Fabian(1991)]{gf91} George, I. M., \& Fabian,
%A. C. 1991, \mnras, 249, 352
\bibitem[Ghosh et al.(2008)]{gho08} Ghosh, H., Mathur, S., Fiore, F., 
\& Ferrarese, L. 2008, ApJ, 687, 216
\bibitem[Gilfanov(2004)]{gil04} Gilfanov, M. 2004, \mnras, 349,
146
%\bibitem[Gilfanov et al.(2004)]{ggs04} Gilfanov, M., Grimm, H.-J.,
%\& Sunyaev, R. 2004, Nucl. Phys. B Proc. Suppl., 132, 369
\bibitem[Gilli et al.(2007)]{gil07} Gilli, R., Comastri, A., \& Hasinger, G. 2007, A\&A, 463, 79
\bibitem[Gonz\'{a}lez-Mart\'{i}n et al.(2006)]{goet06} Gonz\'{a}lez-Mart\'{i}n,
O., Masegosa, J., Marquez, I., Guerrero, M. A., \& Dultzin-Hacyan,
D. 2006, \aap, 460, 45
\bibitem[Graham \& Driver(2007)]{gra07} Graham, A. W., \& Driver,
S. P. 2007, \apj, 655, 77
\bibitem[Greene \& Ho(2004)]{gh04} Greene, J. E., \& Ho, L. C.
2004, \apj, 610, 722
%\bibitem[Greene \& Ho(2006)]{gh06} Greene, J. E., \& Ho, L. C.
%2006, \apj, 641, L21
\bibitem[Greene \& Ho(2007)]{gh07} Greene, J. E., \& Ho, L. C.
2007, \apj, 670, 92
\bibitem[Greene et al.(2008)]{ghb08} Greene, J. E., Ho, L.
C., \& Barth, A. J. 2008, \apj, 688, 159
\bibitem[Greenhill \& Gwinn(1997)]{gg97} Greenhill, L. J., \&
Gwinn, C. R. 1997, \apss, 248, 261
\bibitem[Greenhill et al.(1997)]{gmh97} Greenhill,
L. J., Moran, J. M., \& Herrnstein, J. R. 1997, \apj, 481, L23
\bibitem[Greenhill et al.(2003)]{greet03} Greenhill, L., et al.
2003, \apj, 590, 162
\bibitem[Grimm et al.(2003)]{ggs03} Grimm, H. J., Gilfanov, M., 
\& Sunyaev, R. 2003, \mnras, 339, 793
%\bibitem[Guainazzi et al.(2000)]{guet00} Guainazzi, M.,
%Oosterbroek, T., Antonelli, L. A., \& Matt, G. 2000, \aap, 364,
%L80
\bibitem[Gultekin et al.(2009)]{gul09} Gultekin, K., et al. 2009, ApJ, in press (arXiv:0901.4162)
\bibitem[Hao et al.(2005a)]{hao05a} Hao, L., et al. 2005, 129, 1783
\bibitem[Hao et al.(2005b)]{hao05b} Hao, L., et al. 2005, 129, 1795
\bibitem[Hasinger(2008)]{has08} Hasinger, G. 2008, \aap, 490, 905
\bibitem[Heckman(1980)]{hec80} Heckman, T. M. 1980, \aap, 87, 142
\bibitem[Heller \& Shlosman(1994)]{hs94} Heller, C. H., \& Shlosman, I. 1994, \apj, 424, 84
\bibitem[Herrnstein et al.(1999)]{heet99} Herrnstein, J. R., et al.
1999, \nat, 400, 539
%\bibitem[Ho(in preparation)]{hopp} Ho, L. C., in preparation
%\bibitem[Ho(1999a)]{ho99a} Ho, L. C. 1999a, \apj, 510, 631
%\bibitem[Ho(1999b)]{ho99b} Ho, L. C. 1999b, \apj, 516, 672
\bibitem[Ho(2008)]{ho08} Ho, Luis C. 2008, \araa, 46, 475
%\bibitem[Ho et al.(1995)]{hfs95} Ho, Luis C.,
%Filippenko, A. V., \& Sargent, W. L. W. 1995, \apjs, 98, 477
\bibitem[Ho et al.(1997a)]{hfs97a} Ho, Luis C.,
Filippenko, A. V., \& Sargent, W. L. W. 1997a, \apjs, 112, 315
\bibitem[Ho et al.(1997b)]{hfs97b} Ho, Luis C.,
Filippenko, A. V., \& Sargent, W. L. W. 1997b, \apj, 487, 568
\bibitem[Ho et al.(1997c)]{hfs97c} Ho, Luis C.,
Filippenko, A. V., \& Sargent, W. L. W. 1997c, \apj, 487, 579
\bibitem[Ho et al.(1997d)]{hfs97d} Ho, L. C.,
Filippenko, A. V., \& Sargent, W. L. W. 1997d, \apj, 487, 591
\bibitem[Ho et al.(2001)]{hoet01} Ho, L. C., et al. 2001,
\apj, 549, L51
%\bibitem[Ho et al.(2003)]{hfs03} Ho, Luis C.,
%Filippenko, A. V., \& Sargent, W. L. W. 2003, \apj, 583, 159
%\bibitem[Ho et al.(1997e)]{hfsp97e} Ho, Luis C.,
%Filippenko, A. V., Sargent, W. L. W., \& Peng, C.Y. 1997e, \apjs,
%112, 391
%\bibitem[Ho et al.(1999)]{hoet99a} Ho, L. C., Ptak, A.,
%Terashima, Y., Kunieda, H., Serlemitsos, P. J., Yaqoob, T., \&
%Koratkar, A. P. 1999, \apj, 525, 168
%\bibitem[Ho \& Ulvestad(2001)]{hu01} Ho, L. C., \& Ulvestad, J. S.
%2001, \apjs, 133, 77
\bibitem[Hopkins et al.(2009)]{hop09} Hopkins, P. F., Hickox, R., Quataert, E., 
\& Hernquist, L. 2009, MNRAS, in press (arXiv:0901.2936)
\bibitem[Hoyle \& Lyttleton(1939)]{hoy39} Hoyle, F., \& Lyttleton, R. A. 1939, PCPS, 35, 405
\bibitem[Hunt \& Malkan(1999)]{hun99} Hunt, L. K., \& Malkan, M. A. 1999, ApJ, 516, 660
\bibitem[Immler et al.(2003)]{imet03} Immler, S., Wang, Q. D.,
Leonard, D. C., \& Schlegel, E. M. 2003, \apj, 595, 727
\bibitem[Irwin et al.(2002)]{isb02} Irwin, J. A.,
Sarazin, C. L., \& Bergman, J. N. 2002, \apj, 570, 152
%\bibitem[Ishisaki et al.(1996)]{iset96} Ishisaki, Y., et al. 1996. \pasj, 48, 237
\bibitem[Itoh et al.(2008)]{ito08} Itoh, T., et al. 2008, PASJ, 60, 251
%\bibitem[Iyomoto et al.(2001)]{iyet01} Iyomoto, N., Fukazawa, Y.,
%Nakai, N., \& Ishihara, Y. 2001, \apj, 561, L69
%\bibitem[Iyomoto et al.(1997)]{iyet97} Iyomoto, N., Makishima, K., Fukazawa, Y.,
%Tashiro, M., \& Ishihara, Y. 1997, \pasj, 49, 425
%\bibitem[Iyomoto et al.(1996)]{iyet96} Iyomoto, N., Makishima, K., Fukazawa, Y.,
%Tashiro, M., Ishihara, Y., Nakai, N., \& Taniguchi, Y. 1996,
%\pasj, 48, 231
%\bibitem[Iyomoto et al.(1998a)]{iyet98a} Iyomoto, N., Makishima, K., Matsushita, K.,
%Fukazawa, Y., Tashiro, M., \& Ohashi, T. 1998a, \apj, 503, 168
%\bibitem[Iyomoto et al.(1998b)]{iyet98b} Iyomoto, N., Makishima,
%K., Tashiro, M., Inoue, S., Kaneda, H., Matsumoto, Y., \& Mizuno,
%T. 1998b, \apj, 503, L31
\bibitem[Jenkins et al.(2008)]{jen08} Jenkins, L. P., Brandt, W.
N., Colbert, E. J. M., Levan, A. J., Roberts, T. P., Ward, M. J.,
\& Zezas, A. 2008, proceedings of the ESAC workshop on X-rays from nearby galaxies, 
Madrid (Spain), September 2007, MPE Report 295, p.~65 (arXiv:0801.2356)
\bibitem[Jester(2005)]{jes05} Jester, S. 2005, ApJ, 625, 667
\bibitem[J\'{i}menez-Bail\'{o}n et al.(2005)]{baet05} Jim\'{e}nez-Bail\'{o}n, E., Santos-Lle\'{o}, M.,
Dahlem, M., Ehle, M., MasHesse, J. M., Guainazzi, M., Heckman, T.
M., \& Weaver, K. A. 2005, \aap, 442, 861
%\bibitem[Kauffmann et al.(2003)]{kau03} Kauffmann, G., et al. 2003, \mnras,
%346, 1055
%\bibitem[Keel(1983)]{ke83} Keel, W. C. 1983, \apj, 269, 466
\bibitem[Kennicutt(1998)]{ken98} Kennicutt, R. C. 1998, ARA\&A, 36, 189
%\bibitem[Kormendy(1993)]{ko93} Kormendy, J. 1993, in The
%Nearest Active Galaxies, ed. J. Beckman, L. Colina, \& H. Netzer
%(Madrid: Consejo Superior de Investigaciones Cient\'{i}ficas), 197
\bibitem[Kormendy(2004)]{kor04} Kormendy, J. 2004, in Coevolution
of Black Holes and Galaxies, ed. L. C. Ho (Cambridge: Cambridge
Univ. Press), 1
\bibitem[Kormendy et al.(1998)]{koet98} Kormendy, J., Bender, R.,
Evans, A. S., \& Richstone, D. 1998, \aj, 115, 1823
%\bibitem[Kormendy \& Richstone(1995)]{kr95} Kormendy, J., \&
%Richstone, D.O. 1995, \araa, 33, 581
\bibitem[Kraft et al.(1991)]{kbn91} Kraft, R. P.,
Burrows, D. N., \& Nousek, J. A. 1991, \apj, 374, 344
%\bibitem[Krips et al.(2007)]{kr07} Krips, M., et al. 2007, \aap, 464,
%553
%\bibitem[Kukula et al.(1995)]{kuet95} Kukula, M. J., Pedlar, A.,
%Baum, S. A., \& O$\arcmin$Dea, C. P. 1995, \mnras, 276, 1262
\bibitem[Lauer et al.(1995)]{lau95} Lauer, T. R., et al. 1995, AJ, 110, 2622
\bibitem[Laurikainen et al.(2004)]{lau04} Laurikainen, E., Salo, H., 
\& Buta, R. 2004, ApJ, 607, 103
%\bibitem[Lightman \& White(1988)]{lw88} Lightman, A. P., \& White,
%T. R. 1988, \apj, 335, 57
%\bibitem[Liu(2008)]{Liu08} Liu, J. F. 2008, in preparation
\bibitem[Lynden-Bell(1969)]{lyn69} Lynden-Bell, D. 1969, Nature, 223, 690
\bibitem[Magorrian et al.(1998)]{mago98} Magorrian, J., et al. 1998, \aj, 115, 2285
%\bibitem[Makishima et al.(1994)]{maet94} Makishima, K., et al. 1994, \pasj,
%46, L77
\bibitem[Marconi et al.(2001)]{maret01} Marconi, A., Capetti, A.,
Axon, D. J., Koekemoer, A., Macchetto, D., \& Schreier, E. J.
2001, \apj, 549, 915
%\bibitem[Matsumoto et al.(2001)]{maet01} Matsumoto, Y., Fukazawa,
%Y., Nakazawa, K., Iyomoto, N., \& Makishima, K. 2001, \pasj, 53,
%475
\bibitem[Menci et al.(2008)]{men08} Menci, N., Fiore, F., Puccetti, S., 
\& Cavaliere, A. 2008, ApJ, 686, 219
\bibitem[Merloni et al.(2003)]{mel03} Merloni, A., Heinz, S.,
    \& Di Matteo, T. 2003, \mnras, 345, 1057
\bibitem[Merritt \& Ferrarese(2001a)]{mf01a} Merritt, D., \&
Ferrarese, L. 2001a, ASPC, 249, 335
\bibitem[Merritt \& Ferrarese(2001b)]{mf01b} Merritt, D., \&
Ferrarese, L. 2001b, \mnras, 320, L30
%\bibitem[Miller et al.(2003)]{mi03} Miller, C. J., Nichol, R. C.,
%Gomez, P. L., Hopkins, A. M., \& Bernardi, M. 2003, \apj, 597, 142
\bibitem[Moshir et al.(1993)]{mos93} Moshir, M., et al. 1993, IRAS Faint Source Catalog, 
on Vizier. 
\bibitem[Murray \& Chiang(1998)]{mur98} Murray, N., \& Chiang, J. 1998, ApJ, 494, 125
\bibitem[Murray et al.(1995)]{mur95} Murray, N., Chiang, J., Grossman, S. A.,
   \& Voit, G. M. 1995, ApJ, 451, 498
\bibitem[Nagar et al.(2005)]{nfw05} Nagar, N. M.,
Falcke, H., \& Wilson, A. S. 2005, \apj, 435, 521
%\bibitem[Nagar et al.(2000)]{net00} Nagar, N. M.,
%Falcke, H., Wilson, A. S., \& Ho, L. C. 2000, \apj, 542, 186
%\bibitem[Nagar et al.(2002)]{net02} Nagar, N. M.,
%Falcke, H., Wilson, A. S., \& Ulvestad, J. S. 2002, \aap, 392, 53
%\bibitem[Nager et al.(2001)]{nwf01} Nagar, N. M.,
%Wilson, A. S., \& Falcke, H. 2001, \apj, 559, L87
\bibitem[Narayan(2002)]{nar02} Narayan, R. 2002, in Lighthouses
of the Universe, ed. M. Gilfanov, R. Sunyaev, \& E. Churazov (New York: Springer), 405
\bibitem[Narayan et al.(1997)]{nbc97} Narayan, R., Barrett, D., \& McClintock, 
J. E. 1997, ApJ, 482, 448
\bibitem[Narayan et al.(1998)]{nar98} Narayan, R., Mahadevan, R., 
Grindlay, J. E., Popham, R. G., \ Gammie, C. 1998, ApJ, 492, 554
\bibitem[Narayan \& Yi(1995)]{ny95a} Narayan, R., \& Yi, I. 1995,
\apj, 444, 231
%\bibitem[Narayan \& Yi(1995b)]{ny95b} Narayan, R., \& Yi, I. 1995b,
%\apj, 452, 710
\bibitem[Nilson(1973)]{nil73} Nilson, P. 1973, Uppsala General Catalogue 
of Galaxies (Uppsala: Astron. Obs.) 
\bibitem[Ohta et al.(2007)]{oht07} Ohta, K., Aoki, K., Kawaguchi, T., 
\& Kiuchi, G. 2007, ApJS, 169, 1
\bibitem[Panessa et al.(2006)]{paet06} Panessa, F., Bassani, L.,
Cappi, M., Dadina, M., Barcons, X., Carrera, F. J., Ho, L. C., \&
Iwasawa, K. 2006, \aap, 455, 173
\bibitem[Pellegrini(2005)]{pel05} Pellegrini, S. 2005, \apj, 624,
155
%\bibitem[Pellegrini et al.(2000a)]{peet00a} Pellegrini, S., Cappi, M.,
%Bassani, L., Della Ceca, R., \& Palumbo, G. G. C. 2000a, \aap,
%360, 878
%\bibitem[Pellegrini et al.(2000b)]{peet00b} Pellegrini, S., Cappi, M.,
%Bassani, L., Malaguti, G., Palumbo, G. G. C., \& Persic, M. 2000b,
%\aap, 353, 447
\bibitem[Pellegrini et al.(2002)]{peet02} Pellegrini, S.,
Fabbiano, G., Fiore, F., Trinchieri, G., \& Antonelli, A. 2002,
\aap, 383, 1
\bibitem[Pence et al.(2001)]{psm01} Pence, W. D.,
Snowden, S. L., \& Mukai, K. 2001, \apj, 561, 189
\bibitem[Peterson et al.(2005)]{petet05} Peterson, B. M., et al.
2005, \apj, 632, 799
%\bibitem[Phillips et al.(1986)]{pet86} Phillips, M. M., Jenkins,
%C. R., Dopita, M. A., Sadler, E. M., \& Binette, L. 1986, \aj, 91,
%1062
\bibitem[Pinkney et al.(2003)]{pinet03} Pinkney, J., et al. 2003, \apj,
596, 903
\bibitem[Plucinsky et al.(2008)]{plet08} Plucinsky, P. P, et al.
2008, \apj, 174, 366
\bibitem[Posson-Brown et al.(2006)]{poet06} Posson-Brown, J.,
Raychaudhury, S., Forman, W., Donnelly, R. H., \& Jones, C. 2009,
ApJ, in press (astro-ph/0605308)
%\bibitem[Ptak et al.(1999)]{ptet99} Ptak, A., Serlemitsos, P. J.,
%Yaqoob, T., \& Mushotzky, R. 1999, \apjs, 120, 179
%\bibitem[Ptak et al.(2004)]{ptet04} Ptak, A., Terashima, Y., Ho,
%L. C., \& Quataert, E. 2004, \apj, 606, 173
%\bibitem[Ptak et al.(1996)]{ptet96} Ptak, A., Yaqoob, T.,
%Serlemitsos, P. J., Kunieda, H., \& Terashima, Y. 1996, \apj, 459,
%542
\bibitem[Quataert \& Narayan(1999)]{qua99} Quataert, E., \& Narayan, R. 1999, ApJ, 520, 298
\bibitem[Reyes et al.(2008)]{rey08} Reyes, R., et al.~2008, AJ, 136, 2373
\bibitem[Rice et al.(1988)]{ric88} Rice, W., Lonsdale, C. J., Soifer, B. T.,
Neugebauer, G., Kopan, E. L., Lloyd, L. A., de Jong, T., \& Habing, H. J. 1988,
ApJS, 68, 91
%\bibitem[Roberts et al.(2001)]{rsw01} Roberts, T.
%P., Schurch, N. J., \& Warwick, R. S. 2001, \mnras, 324, 737
%\bibitem[Roberts \& Warwick(2000)]{rw00} Roberts, T. P., \& Warwick,
%R. S. 2000, \mnras, 315, 98
%\bibitem[Sadler et al.(1989)]{sjk89} Sadler, E. M.,
%Jenkins, C. R., \& Kotanyi, C. G. 1989, \mnras, 240, 591
\bibitem[Sakamoto et al.(1999)]{sak99} Sakamoto, K., Okumura, S. K., 
Ishizuki, S., \& Scoville, N. Z. 1999, ApJ, 525, 691
\bibitem[Salpeter(1964)]{sal64} Salpeter, E. E. 1964, ApJ, 140, 796
\bibitem[Sarzi et al.(2001)]{saret01} Sarzi, M., Rox. H. W.,
Shields, J. C., Rudnick, G., Ho, L. C., McIntosh, D. H.,
Filippenko, A. V., \& Sargent, W. L. W. 2001, \apj, 550, 65
\bibitem[Satyapal et al.(2005)]{saet05} Satyapal, S., Dudik, R.
P., O$\arcmin$Halloran, B., \& Gliozzi, M. 2005, \apj, 633, 86
\bibitem[Satyapal et al.(2004)]{ssd04} Satyapal, S.,
Sambruna, R. M., \& Dudik, R. P. 2004, \aap, 414, 825
\bibitem[Satyapal et al.(2008)]{saty08} Satyapal, S., Vega, D., Dudik, R. P., 
Abel, N. P., \& Heckman, T. 2008, ApJ, 677, 926
\bibitem[Seth et al.(2008)]{set08} Seth, A., Ag\"{u}eros, M., Lee, D., 
\& Basu-Zych, A. 2008, ApJ, 678, 116
\bibitem[Skrutskie et al.(2006)]{skr06} Skrutskie, M. F., et al. 2006, AJ, 131, 1163.
\bibitem[Smith \& Wilson(2001)]{sw01} Smith, D. A., \& Wilson, A.
S. 2001, \apj, 557, 180
\bibitem[Soria \& Wu(2002)]{sw02} Soria, R., \& Wu, K. 2002, \aap,
384, 99
\bibitem[Summers et al.(2004)]{suet04} Summers, L. K., Stevens, I.
R., Strickland, D. K., \& Heckman, T. M. 2004, \mnras, 351, 1
\bibitem[Swartz et al.(2003)]{swet03} Swartz, D. A., Ghosh, K. K.,
McCollough, M. L., Pannuti, T. G., Tennant, A. F., \& Wu, K. 2003,
\apj, 144, 213
\bibitem[Swartz et al.(2004)]{swartz04} Swartz, D. A., Ghosh, K.
K., Tennant, A. F., \& Wu, K. 2004, ApJS, 154, 519
\bibitem[Swartz et al.(2008)]{swa08} Swartz, D. A., Soria, R.,
\& Tennant, A. F. 2008, ApJ, 684, 282
%\bibitem[Terashima et al.(2000a)]{thp00} Terashima, Y., Ho,
%L. C., \& Ptak, A. F. 2000a, \apj, 539, 161
%\bibitem[Terashima et al.(2000b)]{teet00a} Terashima, Y., Ho, L.
%C., Ptak, A. F., Mushotzky, R. F., Serlemitsos, P. J., Yaqoob, T.,
%\& Kunieda, H. 2000b. \apj, 533, 729
%\bibitem[Terashima et al.(2000c)]{teet00b} Terashima, Y., Ho, L.
%C., Ptak, A. F., Yaqoob, T., Kunieda, H., Misaki, K., \&
%Serlemitsos, P. J. 2000c, \apj, 535, L79
\bibitem[Terashima et al.(2002)]{teet02} Terashima, Y., Iyomoto,
N., Ho, L. C., \& Ptak, A. F. 2002, \apjs, 139, 1
%\bibitem[Terashima et al.(1998a)]{teet98a} Terashima, Y., Kunieda,
%H., Misaki, K., Mushotzky, R. F., Ptak, A. F., \& Reichert, G. A.
%1998a, \apj, 503, 212
%\bibitem[Terashima et al.(1998b)]{teet98b} Terashima, Y., Ptak, A.,
%Fujimoto, R., Itoh, M., Kunieda, H., Makishima, K., \&
%Serlemitsos, P. J. 1998b, \apj, 496, 210
\bibitem[Terashima \& Wilson(2001)]{tw01} Terashima, Y., \&
Wilson, A. S. 2001, \apj, 560, 139
%\bibitem[Terashima \& Wilson(2003)]{tw03} Terashima, Y., \&
%Wilson, A. S. 2003, \apj, 583, 145
\bibitem[Terashima \& Wilson(2004)]{tw04} Terashima, Y., \&
Wilson, A. S. 2004, \apj, 601, 735
\bibitem[Thatte et al.(2000)]{ttg00} Thatte, N., Tecza,
M., \& Genzel, R. 2000, \aap, 364, L47
\bibitem[Tonry et al.(2001)]{ton01} Tonry, J. L., Dressler, A., Blakeslee,
J. P., Ajhar, E. A., Fletcher, A. B., Luppino, G. A., Metzger, M.
R., \& Moore, C. B. 2001, \apj, 546, 681
\bibitem[Tremaine et al.(2002)]{tre02} Tremaine. S., et al. 2002, \apj, 574,
740
\bibitem[Tully(1988)]{tul88} Tully, R. B. 1988, Nearby
Galaxies Catalogue (Cambridge: Cambridge Univ.Press)
\bibitem[Tully et al.(1992)]{tul92} Tully, R. B., Shaya, E. J. \& Pierce, M. J. 1992, \apjs, 80, 479
\bibitem[Ueda et al.(2003)]{ued03} Ueda, Y., Akiyama, M., Ohta, K., \& Miyaji, T. 2003, ApJ, 598, 886
%\bibitem[Ulvestad \& Ho(2001a)]{uh01a} Ulvestad, J. S., \& Ho, L.
%C. 2001a, \apj, 558, 561
%\bibitem[Ulvestad \& Ho(2001b)]{uh01b} Ulvestad, J. S., \& Ho, L.
%C. 2001b, \apj, 562, L133
%\bibitem[Ulvestad \& Wilson(1989)]{uw89} Ulvestad, J. S., \& Wilson,
%A. S. 1989, \apj, 343, 659
\bibitem[Wang \& Kauffmann(2008)]{wan08} Wang, L., \& Kauffmann, G. 2008, MNRAS, 391, 785
\bibitem[Wang \& Zhang(2007)]{wan07} Wang, J.-M., \& Zhang, E.-P. 2007, ApJ, 660, 1072
\bibitem[Weaver et al.(2002)]{weet02} Weaver, K. A., Heckman, T.
M., Strickland, D. K., \& Dahlem, M. 2002, \apj, 576, L19
%\bibitem[Weaver et al.(1999)]{weet99} Weaver, K. A., Wilson, A. S.,
%Henkel, C., \& Braatz, J. A. 1999, \apj, 520, 130
\bibitem[Wrobel \& Heeschen(1991)]{wh91} Wrobel, J. M., \& Heeschen
D. S. 1991, \aj, 101, 148
\bibitem[Wrobel et al.(2008)]{wth08} Wrobel, J. M.,
Terashima, Y., \& Ho, L. C. 2008, \apj, 675, 1041
%\bibitem[Yaqoob et al.(1995)]{yaet95} Yaqoob, T., Serlemitsos, P.
%J., Park, A., Mushotzky, R. F., Kunieda, H., \& Terashima, Y.
%1995, \apj, 455, 508
\bibitem[Young et al.(2001)]{yws01} Young, A. J.,
Wilson, A. S., \& Shopbell, P. L. 2001, \apj, 556, 6
\bibitem[Yuan \& Narayan(2004)]{yua04} Yuan, F., \& Narayan, R. 2004, ApJ, 612, 724
\bibitem[Zhu et al.(2008)]{zzt08} Zhu, L., Zhang, S. N.,
\& Tang, S. M. 2008, arXiv:0807.3992
\end{thebibliography}
\end{document}